\documentclass[
reprint,
aps,
prd,
twocolumn,
superscriptaddress,
nofootinbib,
preprintnumbers,
floatfix
]{revtex4-1}

\include{notations}
\usepackage[utf8]{inputenc}
\usepackage[T1]{fontenc}
\usepackage{graphicx}
\graphicspath{ {figures/} }
\usepackage[export]{adjustbox}
\usepackage{dcolumn}
\usepackage{bm}
\usepackage{hyperref}
\usepackage{amsmath}	
\usepackage{amssymb}	
\usepackage{slashed}
\usepackage[normalem]{ulem}
\usepackage[dvipsnames]{xcolor}
\usepackage{xspace}
\usepackage{multirow}
\usepackage{booktabs}
\usepackage{aas_macros}
\usepackage{multirow}
\usepackage{caption}
\usepackage{array}
\usepackage{subfigure}
\usepackage{rotating}
\usepackage{adjustbox}
\newcommand{\bthis}[1]{\textcolor{black}{#1}}
\hypersetup{colorlinks,linkcolor={red},citecolor={blue},urlcolor={blue}}

\newcommand{\pinta}{\texttt{pinta}}

\newcommand{\psrfits}{\texttt{PSRFITS}}

\newcommand{\tempotwo}{\texttt{TEMPO2}}
\newcommand{\dmcalc}{\texttt{DMcalc}}

\newcommand{\enterprise}{\texttt{ENTERPRISE}}
\newcommand{\dynesty}{\texttt{DYNESTY}}
\newcommand{\ptmcmc}{\texttt{PTMCMCSAMPLER}}
\allowdisplaybreaks

\begin{document}

\preprint{APS/123-QED}

\title{Noise analysis of the Indian Pulsar Timing Array data release I}

\author{Aman Srivastava}
 \email{amnsrv1@gmail.com}
 \affiliation{Department of Physics, IIT Hyderabad, Kandi, Telangana 502284, India}

\author{Shantanu Desai}
\affiliation{Department of Physics, IIT Hyderabad, Kandi, Telangana 502284, India}

\author{Neel Kolhe}
\affiliation{Department of Physics, St. Xavier’s College (Autonomous), Mumbai 400001, Maharashtra, India}

\author{Mayuresh Surnis}
\affiliation{Department of Physics, IISER Bhopal, Bhopal Bypass Road, Bhauri, Bhopal 462066, Madhya Pradesh, India}

\author{Bhal Chandra Joshi}
\affiliation{National Centre for Radio Astrophysics, Pune University Campus, Pune 411007, India}

\author{Abhimanyu Susobhanan}
\affiliation{ Center for Gravitation, Cosmology, and Astrophysics, University of Wisconsin-Milwaukee, Milwaukee, WI 53211, USA}

\author{Aur\'elien Chalumeau}
\affiliation{Dipartimento di Fisica “G. Occhialini", Universit\'a degli Studi di Milano-Bicocca, Piazza della Scienza 3, 20126 Milano, Italy}

\author{Shinnosuke Hisano}
\affiliation{Kumamoto University, Graduate School of Science and Technology, Kumamoto, 860-8555, Japan}

\author{Nobleson K.}
\affiliation{Department of Physics, BITS Pilani Hyderabad Campus, Hyderabad 500078, Telangana, India}

\author{Swetha Arumugam}
\affiliation{Department of Electrical Engineering, IIT Hyderabad, Kandi, Telangana 502284, India}

\author{Divyansh Kharbanda}
\affiliation{Department of Physics, IIT Hyderabad, Kandi, Telangana 502284, India}

\author{Jaikhomba Singha}
\affiliation{Department of Physics, Indian Institute of Technology Roorkee, Roorkee-247667, India}

\author{Pratik Tarafdar}
\affiliation{The Institute of Mathematical Sciences, C. I. T. Campus, Taramani, Chennai 600113, India}

\author{P Arumugam}
\affiliation{Department of Physics, Indian Institute of Technology Roorkee, Roorkee-247667, India}

\author{Manjari Bagchi}
\affiliation{The Institute of Mathematical Sciences, C. I. T. Campus, Taramani, Chennai 600113, India}
\affiliation{Homi Bhabha National Institute, Training School Complex, Anushakti Nagar, Mumbai 400094, India}

\author{Adarsh Bathula}
\affiliation{Department of Physical Sciences, Indian Institute of Science Education and Research, Mohali, Punjab, India -140306.}

\author{Subhajit Dandapat}
\affiliation{Department of Astronomy and Astrophysics, Tata Institute of Fundamental Research, Homi Bhabha Road, Navy Nagar, Colaba, Mumbai 400005, India}

\author{Lankeswar Dey}
\affiliation{Department of Physics and Astronomy, West Virginia University, P.O. Box 6315, Morgantown, WV 26505, USA}
\affiliation{Center for Gravitational Waves and Cosmology, West Virginia University, Chestnut Ridge Research Building, Morgantown, WV 26505, USA}

\author{Churchil Dwivedi}
\affiliation{Department of Earth and Space Sciences, Indian Institute of Space Science and Technology, Valiamala, Thiruvananthapuram, Kerala, India (695547)}

\author{Raghav Girgaonkar}
\affiliation{Department of Physics and Astronomy, University of Texas Rio Grande Valley, One West University Blvd., Brownsville, Texas 78520, USA}

\author{A. Gopakumar}
\affiliation{Department of Astronomy and Astrophysics, Tata Institute of Fundamental Research, Homi Bhabha Road, Navy Nagar, Colaba, Mumbai 400005, India}

\author{Yashwant Gupta}
\affiliation{National Centre for Radio Astrophysics, Pune University Campus, Pune 411007, India}

\author{Tomonosuke Kikunaga}
\affiliation{Kumamoto University, Graduate School of Science and Technology, Kumamoto, 860-8555, Japan}

\author{M. A. Krishnakumar}
\affiliation{Max-Planck-Institut f\"ur Radioastronomie, Auf dem H\"ugel 69, 53121 Bonn, Germany}
\affiliation{Fakult\"at f\"ur Physik, Universit\"at Bielefeld, Postfach 100131, 33501 Bielefeld, Germany}

\author{Kuo Liu}
\affiliation{Max-Planck-Institut f\"{u}r Radioastronomie, Auf dem H\"{u}gel 69, 53121, Bonn, Germany}

\author{Yogesh Maan}
\affiliation{National Centre for Radio Astrophysics, Pune University Campus, Pune 411007, India}

\author{P K Manoharan}
\affiliation{Arecibo Observatory, University of Central Florida, PR 00612, USA}

\author{Avinash Kumar Paladi}
\affiliation{Joint Astronomy Programme, Indian Institute of Science, Bengaluru, Karnataka, 560012, India}

\author{Prerna Rana}
\affiliation{Department of Astronomy and Astrophysics, Tata Institute of Fundamental Research, Homi Bhabha Road, Navy Nagar, Colaba, Mumbai 400005, India}

\author{Golam M. Shaifullah}
\affiliation{Dipartimento di Fisica “G. Occhialini", Universit\'a degli Studi di Milano-Bicocca, Piazza della Scienza 3, 20126 Milano, Italy}
\affiliation{INFN, Sezione di Milano-Bicocca, Piazza della Scienza 3, I-20126 Milano, Italy}
\affiliation{INAF, Osservatorio Astronomico di Cagliari, Via della Scienza 5, 09047 Selargius, Italy }

\author{Keitaro Takahashi}
\affiliation{Division of Natural Science, Faculty of Advanced Science and Technology, Kumamoto University,
2-39-1 Kurokami, Kumamoto 860-8555, Japan}
\affiliation{International Research Organization for Advanced Science and Technology, Kumamoto University, 2-39-1 Kurokami, Kumamoto 860-8555, Japan
}

\begin{abstract}
The Indian Pulsar Timing Array (InPTA) collaboration has recently made its first official data release (DR1) for a sample of 14 pulsars using 3.5  years of uGMRT observations.
We present the results of single-pulsar noise analysis for each of these 14 pulsars using the InPTA DR1. 
For this purpose, we consider  white noise, achromatic red noise, dispersion measure (DM) variations, and scattering variations in our analysis. 
We apply  Bayesian model selection to obtain the preferred noise models among these for each pulsar.  For PSR J1600$-$3053, we find no evidence of DM and scattering variations, while for PSR J1909$-$3744, we find no significant scattering variations. Properties vary dramatically among pulsars. For example, we find a strong chromatic noise with chromatic index $\sim$ 2.9 for PSR J1939+2134, indicating the possibility of a scattering index that doesn't agree with that expected for a Kolmogorov scattering medium consistent with similar results for millisecond pulsars in past studies. Despite the relatively short time baseline, the noise models broadly agree with the other PTAs and provide, at the same time, well-constrained DM and scattering variations.

\end{abstract}

\maketitle

\section{Introduction}
\label{sec:intro}

Millisecond pulsars (MSPs) are known for their exceptional rotational stability and accuracy comparable to atomic clocks. 
Pulsar Timing Array experiments (PTAs)~\cite{FosterBacker1990} aim to detect ultra-low frequency ($\sim$ 1-100 nHz) gravitational waves (GWs) by monitoring an ensemble of MSPs distributed across the Galaxy. 
This is possible because the GWs travelling across the line of sight to a pulsar perturb the null geodesics along which the pulsar electromagnetic signals propagate, thereby modulating their times of arrival (ToAs) radio pulses.
GW signals in the PTA frequency range are typically expected to originate from orbiting supermassive black hole binaries (SMBHBs) in the inspiral phase, both as a stochastic GW background (GWB) formed by the incoherent addition of GWs from a large number of SMBHBs, and as strong individual sources standing out above this background \cite{Burke-SpolaorTaylor+2019}.
Such a GWB induces spatially correlated ToA modulations in different pulsars, characterized by the Hellings-Downs overlap reduction function \cite{HellingsDowns1983}. 
Other proposed sources of nanohertz GWs include cosmological phase transitions \cite{Hogan1986}, cosmic strings \cite{DamourVilenkin2000}, and relic GWs emanating from cosmic fluctuations in the early universe \cite{Grishchuk2005}. 

PTA experiments working towards the goal of detection of nHz GWs include the European Pulsar Timing Array \cite[EPTA:][]{KramerChampion2013}, the Parkes Pulsar Timing Array \cite[PPTA:][]{Hobbs2013}, the North American Nanohertz Observatory for Gravitational Waves \cite[NANOGrav:][]{McLaughlin2013}, the Indian Pulsar Timing Array \cite[InPTA:][]{JoshiArumugasamy+2018}, and emerging PTAs such as the Chinese Pulsar Timing Array \cite[CPTA:][]{Lee2016}, and the MeerTime Pulsar Timing Array \cite{SpiewakBailes+2022}.
The International Pulsar Timing Array consortium \cite[IPTA:][]{HobbsArchibald+2010} aims to improve the prospects of nanohertz GW detection and post-detection science by combining the data and resources from different PTA experiments.
Over the last decade, the PTA experiments have put increasingly stringent constraints on the stochastic GWB, culminating in the recent detection of a common red noise process in multiple PTA datasets \cite{GonchorovReardon+2021,ChalumeauBabak+2022,ArzoumanianBaker+2020,AntoniadisArzoumanian+2022}.

The InPTA experiment~\cite{JoshiGopu+2022} aims to use the upgraded Giant Metre-wave Radio Telescope~\cite[uGMRT:][]{GuptaAjithkumar+2017} to complement the international PTA efforts via low-frequency observations of pulsars.
The uGMRT observations significantly improve the prospects of characterizing the interstellar medium effects, such as dispersion measure (DM)\footnote{Integrated free electron density along the line of sight to the pulsar.} and scatter-broadening variations, which are the strongest at low frequencies \cite{KrishnakumarManoharan+2021}.
The recently published InPTA Data Release 1 \cite[InPTA DR1:][]{TarafdarNobleson+2022} provided ToA measurements, timing analysis, and the characterization of DM variations for 14 pulsars over a time span of 3.5 years, estimated using both the traditional narrowband method \cite{Taylor1992} and the more recent wideband method \cite{PennucciDemorestRansom2014,Pennucci2019}.

The intrinsic wander of the rotation rate of the constituent pulsars, the variations in DM and scatter-broadening, as well as  the instrumental noise of radio telescopes are often covariant with the slowly varying GW signature in the data and act as sources of  chromatic and achromatic noise. The detection and characterization of GWs are strongly affected by the faithfulness of noise models and can be highly dependent on custom noise modelling for each pulsar \cite{caballero2016,lentati2016,ArzoumanianBaker+2020,chen2021,ChalumeauBabak+2022}.
Characterizing these single pulsar noise processes, which are uncorrelated across the constituent pulsars, is a crucial first step for extracting the weak GW signal, which is otherwise correlated across pulsars \cite{GonchorovReardon+2021,ChalumeauBabak+2022,ArzoumanianBaker+2020,AntoniadisArzoumanian+2022}.

This work presents the single pulsar noise analysis (SPNA) of the 14 pulsars present in the InPTA DR1 using the \enterprise{} package \cite{EllisVallisneri+2019}.
We perform Bayesian model selection among a finite set of noise models for each pulsar based on the Bayes factors estimated using the  \dynesty{} package \cite{Speagle2020}, which implements the dynamic nested sampling algorithm \cite{nestedsampling}. 
 Finally, we perform parameter estimation for the preferred noise model using \ptmcmc{}~\cite{EllisvanHaasteren2017}. We note that corresponding noise analyses have also been carried out with EPTA DR2 for six pulsars~\cite{ChalumeauBabak+2022} and the PPTA DR2 dataset for 26 pulsars~\cite{GonchorovReardon+2021}, and we shall also do a comparison with their results for the same  pulsar as appropriate.
\bthis{We should note that with a relatively modest timing baseline of 3.5 years, the red noise modeled using the InPTA dataset may be inconsistent with the other PTAs, which have a timing baseline of over a decade.}

The rest of this article is organized  as follows: Section~\ref{sec:inpta-dr1} briefly describes the InPTA DR1. 
Section~\ref{sec:noise-models} discusses the various noise sources incorporated into our analysis. 
Section~\ref{sec:anl-tech} discusses the Bayesian analysis methodology for noise model selection and parameter estimation.
Section~\ref{sec:discussion} discusses the noise modelling results for each pulsar.
We present our conclusions in Section~\ref{sec:conclusion}.

 \section{Brief description of the I\lowercase{n}PTA DR1}
\label{sec:inpta-dr1}

The InPTA DR1 \cite{TarafdarNobleson+2022} consists of observations of 14 MSPs conducted using the uGMRT \cite[][]{GuptaAjithkumar+2017} as part of the InPTA experiment from 2018 to 2021 typically with a bi-weekly cadence. 
These observations were carried out during observing cycles 34$-$35 and 37$-$40 of the uGMRT, where the 30 uGMRT antennae were divided into multiple phased subarrays, simultaneously observing the same source in multiple bands in total intensity mode \cite{JoshiGopu+2022}.
The channelized time series data generated by the uGMRT are recorded using the GMRT Wideband Backend~\cite{ReddyKudale+2017} in a binary raw data format, and RFI-mitigated and partially folded into \psrfits{} archives using the \pinta{} pipeline \cite[][]{SusobhananMaan+2021}.
The narrowband ToAs were measured using the Global Positioning System (GPS) and a local topocentric frequency standard was provided by the hydrogen maser clock at the GMRT.
The ToAs were fitted using \tempotwo{} \cite[][]{HobbsEdwardsManchester2006} to obtain the timing residuals. 
The timing procedure involved epoch-wise DM correction by incorporating the DM time series obtained using \dmcalc{} \cite{KrishnakumarManoharan+2021} from low-frequency simultaneous multi-band uGMRT data. 
Additionally, wideband timing residuals were generated using the wideband likelihood method described in Ref.~\cite[][]{AlamArzoumanian+2020b} and implemented in the \texttt{TEMPO} \cite[][]{NiceDemorest+2015} pulsar timing package.
For our analysis, we use only the narrowband data for all pulsars. More details of the InPTA DR1 can be found in ~\cite{TarafdarNobleson+2022}.

\section{Noise models for PTA data}
\label{sec:noise-models}

In this section, we discuss the various noise components used in our analysis. The noise analysis is critical in the search for gravitational waves to separate noise processes from the correlated GWB signal. \citet{hazboun2020} using simulations showed that improper noise models could cause bias in GWB estimates. Hence, it is critical to robustly model these noise sources to search and characterize any such correlated signals among pulsars. We model the noise processes as a stationary Gaussian processes (GP) \cite{vhaasteren+2014}. The details of the myriad achromatic and chromatic noise processes are described in this section, which will be used to obtain custom noise models for each pulsar.

\subsection{White noise}
White noise refers to the stochastic signal, where the power spectral density  is constant  across the whole frequency range and is uncorrelated across time. In PTA data, white noise dominates at high frequencies. It is modelled by re-scaling the initial ToA uncertainties ($\sigma_{ToA}$) as follows:
\begin{equation}
    \sigma^2 = \rm{EFAC}^2 \times (\sigma^2_{\text{ToA}} + \rm{EQUAD}^2) 
\end{equation}
where the \text{EFAC} accounts for radiometer noise and the \text{EQUAD} denotes the intrinsic scatter related to  the stochastic profile variations \cite{Liu.et.al,shannon2014,lam2016}. Hence, the white noise covariance matrix $C_W$, which is a diagonal matrix with diagonal elements as the re-scaled variances of ToAs, is given by:
\begin{equation}
    C_{W,i,j} = \sigma^2_{ij} \delta_{ij}.
\end{equation}
Note that the re-scaling is based on the {\it ansatz} that this uncorrelated ToA noise is Gaussian. Refs.~\cite{lentati2014,vallisneri2017} have discussed the non-Gaussian character of this noise, and its presence in a few MSPs has also been recently reported~\cite{GonchorovReardon+2021}. Although, we do not investigate the non-Gaussianity aspect in this work, modelling it may provide better ToA precision, which we plan to explore in the future.

\subsection{Red noise}
In pulsar timing, red noise refers to a  time-correlated noise, which is  stronger at lower frequencies compared to higher frequencies. 
As the GWB itself may appear as a correlated red noise signal that is spatially correlated across pulsars \cite{shannon2010}, it is of utmost importance to correctly model the pulsar-specific red noise in the data. The red noise is modelled as a stationary Gaussian process, and we adopt the ``Fourier space'' representation of the Gaussian process \cite{temponest2014}.
The timing residuals $t_i$ at each epoch due to the stochastic red signal (SRS) are approximated as:
\begin{equation}
    \delta t^{\text{SRS}}(t_i) =\sum_{l=1}^{N} X_l \cos (2\pi t_i f_l) + Y_l \sin(2\pi t_i f_l)
\end{equation}
where one can easily notice that $X_l$ and $Y_l$ appear as weights, and the basis functions are:
\begin{align}
    F_{2l-1} (t_i) = \cos (2\pi t_i f_l) \\
    F_{2l}(t_i) = \sin (2\pi t_i f_l)
\end{align}
where $l = 1,2,...,N$. If $f_l = \bthis{l}/T$ where $T$ is the total observing time span, and if the epochs are evenly spaced, then this would correspond to the discrete Fourier transform. Also, we typically truncate the set at a \bthis{ high frequency (beyond which the spectrum is dominated by white noise)} instead of using the entire set, using an evenly spaced set of frequencies, truncating at $N/T$, where $N$ is the number of Fourier modes. We use $N$ as a hyper-parameter in our noise model selection among four predefined values for each pulsar. The choice of the optimum number of Fourier modes is discussed in Ref. \cite{ChalumeauBabak+2022}. 
\\
The covariance matrix $\Sigma$ for Fourier coefficients $X_l$, $Y_l$ is defined by power spectral density (PSD), $S$. For our analysis, we will use the power law for fitting the red noises, which can be written as follows:
\begin{equation}
S(A, \gamma) = \frac{A^2}{12\pi^2}\left(\frac{f}{yr^{-1}}\right)^{-\gamma} yr^3   
\end{equation}
where $S(A,\gamma)$ is the power spectral density, $A$ is the amplitude with normalization at a frequency of (1 $\text{yr}^{-1}$), and $\gamma$ is the spectral index. The covariance matrix for red noise in the frequency domain (see \cite{ChalumeauBabak+2022} and references within) is given by
\begin{equation}
    \Sigma _{\kappa \alpha l\beta} = S(f_k:A_\alpha,\gamma_\alpha)\delta_{kl}\delta_{\alpha\beta}/T
\end{equation}
where $l,k = 1,2,...,N$, and $\alpha,\beta$ denote the indices of the pulsar. The Kronecker delta function has been introduced,  since we take into account the spatially uncorrelated red noise.

\subsubsection{Achromatic red noise}
Achromatic red noise (RN), also known as timing noise, is modelled in PTA data to account for the  spin irregularities in pulsars \cite{cordes1985, allessandro1995}. This noise might not be significant in MSPs compared to younger pulsars but it can be detected with data over a long baseline (e.g. \cite{alam2021,GonchorovReardon+2021}). This is the observing frequency-independent noise originating from the pulsar. We model achromatic red noise using the power law described above for our analysis.

\subsubsection{Chromatic red noise}
There are delays in ToAs due to the interaction of pulse signals with matter along the path of propagation, such as the ionized interstellar medium (IISM), the ionosphere of the Earth, and the interplanetary medium. Delays in such signals are observing-frequency dependent in nature. One such dominating effect is due to the dispersion, which causes the frequency-dependent delay in the arrival time of pulses. The delay in ToAs due to the DM is related to the observing frequency according to  $\Delta t^{\text{DM}} \propto \nu^{-2}$, where $\nu$ is the observing frequency, and $DM$ is the dispersion measure \cite{LorimerKramer2004}. The timing model accounts for this effect by considering its value at reference epoch along with its first (\texttt{DM1}) and second derivatives (\texttt{DM2}). However, turbulence and inhomogeneity in the IISM coupled with the relative motion of the earth, pulsar and the IISM, may induce an additional time-correlated red noise due to these DM variations (DMv), which depends on the observing frequency \cite{you2007,keith2013}. Another such effect is the delay due to the scattering variations (Sv) caused by the signal's multi-path propagation in IISM due to refraction and diffraction, which occurs when the radio pulses from a pulsar pass through the interstellar medium (ISM), leading to delay, broadening, and other distortion of the pulses \cite{LorimerKramer2004}. The delay due to the scattering is given by $\Delta t^{\text{SC}} \propto \nu^{-4}$. It is crucial to have multi-band observations to disentangle the chromatic components of the red noise (see, for e.g. Ref.~\cite{caballero2016}). 

For the covariance matrix $F^{\text{chrom}}_i$ of chromatic noise, we use the same formula as that used for  red noise, with the additional dependency of induced ToA delays on the observing frequency, as given below:
\begin{equation}
    F^{\text{chrom}}_i = F_i\times\left(\frac{\nu_i}{1.4GHz}\right)^{-\chi}
\end{equation}
where $F_i$ is the Fourier transform of the time-domain red noise signal and contains the incomplete sine and cosine functions, $\nu_i$ is observing frequency, and $\chi$ is the chromatic index, which is 0, 2, and 4 for RN, DMv, and Sv, respectively.
 Apart from DMv and Sv, we also use the ``Free Chromatic Noise'' model (FCN), which has the chromatic index ($\chi_{FCN}$) as an additional free parameter along with the amplitude and spectral index (see Refs.~\cite{GoncharovShannon+2021,ChalumeauBabak+2022}). This model is used as a diagnostic tool for our selected noise models, where we fit for the ($\chi_{FCN}$) to look for the presence of achromatic and chromatic red noise. \bthis{The contributions of solar wind to DM and scattering variations are included in our GP model in our analysis, where we do not factorize between a small quasi-periodic variation due to inter-planetary medium and the stochastic variation due to ISM.}

\section{Analysis techniques}
\label{sec:anl-tech}

\subsection{Bayesian analysis}

We now provide a brief prelude to the Bayesian model comparison techniques for selecting the best noise model. In this work, Bayesian model comparison is used for selecting the best noise model and for selecting the optimum number of Fourier modes for the selected noise model. Bayesian regression is then used for estimating the optimum parameters of the selected noise model. More details on Bayesian inference and Bayesian model selection can be found in Refs.~\cite{Sanjib,Weller,Krishak} (and references therein). We follow the same notation as in Ref.~\cite{Sanjib}.

The starting point for Bayesian Model comparison is the Bayes Theorem, which states that for a model $M$ parameterized by the parameter vector $\theta$ and given the data $D$:
\begin{equation}
  P(\theta|D,M) = \frac{P(D|\theta,M)P(\theta|M)}{P(D|M)},
  \label{eq:bayesthm}
 \end{equation} 
where $P(\theta|M)$ is the prior on the parameter vector ($\theta$) for that model; $P(D|\theta,M)$ represents the likelihood; $P(\theta|D,M)$ represents the posterior probability; and $P(D|M)$ is the marginal likelihood, also known as the Bayesian Evidence. \bthis{The priors used in our analysis can be found  in Table \ref{priors}.} For Bayesian parameter estimation, one needs to evaluate the posterior $P(\theta|D,M)$. 

For model selection, we need to evaluate the Bayesian evidence, which can be defined as:
\begin{equation}
P(D|M) = \int P(D|\theta,M)P(\theta|M) \, d\theta
\label{eq:evid}
\end{equation}
 
To perform model selection between two models $M_1$ and $M_2$, we calculate the Bayes factor (BF), which is given by the ratio of the Bayesian evidence for the two models:
\begin{equation} 
B_{21} = \frac{\int P(D|M_2, \theta_2)P(\theta_2|M_2) \, d\theta_2}{\int P(D|M_1, \theta_1)P(\theta_1|M_1) \, d\theta_1} 
\end{equation}
The Bayes factor is then used for Bayesian model comparison. The model with the larger value of Bayesian evidence will be considered the favored model. We then use Jeffrey's scale to assess the significance of the favored model~\cite{Weller}. Based on this scale, a Bayes Factor $< 1$ indicates negative support for the model in the numerator ($M_2$), thereby favouring the model $M_1$. A value exceeding 10 points to ``substantial'' evidence for $M_2$, while a value greater than 100 points to decisive evidence. We choose 100 as the threshold Bayes factor above which a more complex model is chosen over another with a smaller number of free parameters.
In case the Bayes factor is greater than one but less than 100, we follow Occam's razor and choose the model with fewer free parameters. In the case where both models have the same number of free parameters, we chose the model based on prior information on the presence of the parameters in that dataset. 
We assume that the stochastic processes present in our data are Gaussian, and the data is represented by these Gaussian processes, whereas deterministic signals are included in the timing parameters. We apply Gaussian likelihood in our analysis following the previous studies \citep{haasteren2009, haasteren2012, taylor2017}, and \texttt{ENTERPRISE} is used to evaluate the likelihood function.

We now provide details of the model selection procedure for selecting the best noise model, followed by harmonic mode selection. Finally, we provide details of the parameter estimation procedure for the selected noise model.

\begin{table*}[t!]
 \caption{Priors used in our noise analysis. This table gives the distributions for priors used in the Bayesian analysis for model selection. Here $\cal U$ and $\log_{10} $$\cal U$ stand for uniform and log-uniform distributions, respectively.}
	\setlength{\tabcolsep}{10pt}
 {\renewcommand\arraystretch{1.5}
\begin{tabular}{ lcc }
 Noise (abrev.) & Parameters & Priors (or fixed val.) \\ \hline \hline
 White Noise & EFAC & $\cal U$(0.1,5) \\
 (WN) & EQUAD [s] & $\log_{10}$$\cal U$$(10^{-9},10^{-5})$ \\ \hline

 Achromatic red-noise & $A_{RN}$ & $\log_{10} $$\cal U$$(10^{-18},10^{-10})$ \\
 (RN) & $\gamma_{RN}$ & $\cal U$(0,7) \\ \hline

 DM variations & $A_{DM}$ & $\log_{10} $$\cal U$$(10^{-18},10^{-10})$ \\
 (DMv) & $\gamma_{DM}$ & $\cal U$(0,7) \\ \hline

 Scattering variations & $A_{Sv}$ & $\log_{10} $$\cal U$$(10^{-18},10^{-10})$ \\
 (Sv) & $\gamma_{Sv}$ & $\cal U$(0,7) \\ \hline

 Free chromatic noise & $A_{FCN}$ & $\log_{10} $$\cal U$$(10^{-18},10^{-10})$ \\
 (FCN) & $\gamma_{FCN}$ & $\cal U$(0,7) \\
  & $\chi_{FCN}$ & $\cal U$(0,7) \\ \hline
\end{tabular}
}
\label{priors}
\end{table*}

\begin{table}
\centering
\caption{Six pre-defined noise models were used for model selections.
\textbf{W}, \textbf{R}, \textbf{D}, and \textbf{S} stand for white noise, achromatic red noise, DM variations and scattering variations, respectively. \textit{Red noise parameters} column gives the number of red noise parameters for each model.}
\label{6models}
\setlength{\tabcolsep}{4.5pt}
 {\renewcommand\arraystretch{1.6}
\begin{tabular}{lll}
\textbf{Model name} & \textbf{Noise model} &\textbf{Red noise}\\ 
 & & \textbf{parameters} \\ \hline \hline
Model1(W)    & WN   &  0       \\ \hline
Model2(WR)    & WN + RN   &      2   \\ \hline
Model3(WRD)    & WN + RN + DMv  &    4   \\ \hline
Model4(WDS)    & WN + DMv + Sv  &    4   \\ \hline
Model5(WRDS)    & WN + RN + DMv + Sv &   6   \\ \hline
Model6(WD)    & WN + DMv    &   2   \\ \hline
\end{tabular}
}
\end{table}

\begin{table*}[!]
\centering
\caption{The table contains the $\ln(\text{BF})$ with respect to the selected model for each model for all 14 pulsars. The zeroes, which are in bold text in each row, represent the selected model based on the Bayes factor and the simplicity of the model. We used the maximum number of Fourier modes while performing the model selection for all pulsars.}
\label{model_select}
\setlength{\tabcolsep}{14pt}
 {\renewcommand\arraystretch{2.0}
\begin{tabular}{ccccccc}
\hline
\textbf{Pulsar} & \textbf{Model1} & \textbf{Model2} & \textbf{Model3} & \textbf{Model4} & \textbf{Model5} & \textbf{Model6} \\ 
& \textbf{(W)} & \textbf{(WR)} & \textbf{(WRD)} & \textbf{(WDS)} & \textbf{(WRDS)} & \textbf{(WD)} \\ \hline \hline
J0437$-$4715  &  -250.8  & \textbf{0}  & 1.6   & -94.8   &  0.7  & -186.5   \\ \hline
J0613$-$0200  & -77.5  & \textbf{0}  &  0   & -8.4   &  -0.7  & -14.1   \\ \hline
J0751+1807  & \textbf{0}  & -0.6   & -0.6   & -0.9   & -1.2   & -0.2     \\ \hline
J1012+5307  & -8.2   & -5.8   & -0.2   & 0.1   & -1.3   & \textbf{0}  \\ \hline
J1022+1001  & -246.8   & -85.6   & 2.0    & 0.2   & 1.1    & \textbf{0}  \\ \hline
J1600$-$3053  & \textbf{0}  & 1.9    & 2.3    & 0.8   & 2.0    & 1.1    \\ \hline
J1643$-$1224  & -164    & -150   & -137    & -16   & \textbf{0}  & -159    \\ \hline
J1713+0747  & -38    & -34    & -31    & \textbf{0}  & 1    & -30    \\ \hline
J1744$-$1134  & -48.4   & -19.8   & 0.7    & -0.6   & 0.2   & \textbf{0}  \\ \hline
J1857+0943  & -6.0    & \textbf{0} & 0.3    & -6.6   & -1     & -5.7    \\ \hline
J1909$-$3744  & -334.0  & -132.2   & \textbf{0}  & -18.8   & 4.4    & -43.5    \\ \hline
J1939+2134  & -1914.4   & -1498.8   & -557.5    & -56.8  & \textbf{0}  & -729.7    \\ \hline
J2124$-$3358  & -15.1   & \textbf{0}  & -0.2    & -4.8   & -2.0   & -4.4     \\ \hline
J2145$-$0750  & -150.4   & -56.0    & -30.8    & -12.0  & \textbf{0}  & -33.7    \\ \hline
\end{tabular}
}
\end{table*}

\subsection{Model selection}
For model selection, we calculate the Bayes factor by applying nested sampling using the \texttt{DYNESTY} package. In the first step, we perform Bayesian model selection to look for EQUAD in white noise for each pulsar. We use two white noise-only models, i.e. the first with only EFAC and the second with both EFAC and EQUAD. Once we finalize the white noise for each pulsar, we perform the Bayesian model selection using six pre-defined noise models listed in Table \ref{6models}. We construct these six models with different combinations of RN, DMv and Sv. As the $\Delta\text{DM}$ estimates are very precise, it is highly unlikely to have Sv without a discernible DMv in the data. Hence, we did not use a model with only RN and Sv (without DMv) for our analysis. For this model selection step, we use the highest number of Fourier modes for RN, DMv as well as Sv, i.e. the highest frequency mode equivalent to a frequency of once per month, which roughly corresponds to two observations (using Nyquist sampling theorem) per pulsar, as the cadence of InPTA observations is roughly 14 days. The priors for EQUAD and all types of red noise amplitudes are log-uniform priors ($\log_{10} $$\cal U$), which serve as reliable approximations of non-informative priors for scale-invariant parameters. The priors for EFACs and spectral index for the various red noise models are uniform distributions ($\cal U$). For the same prior sets used for SPNA by \citet{ChalumeauBabak+2022}, our tests show that this prior range is adequate for our dataset. The final selected models, along with the Bayes factors for all the pulsars, can be found in Table \ref{model_select}. \\

\subsection{Selection of number of Fourier modes}
As shown in Ref.~\cite{ChalumeauBabak+2022}, the noise models are sensitive to the number of Fourier modes (\bthis{N}) for all types of red noise. Hence, it is imperative to perform a model selection for different numbers of Fourier modes for each pulsar to obtain an optimum number. \bthis{One thing to note is that, in principle, these optimum numbers can be different for RN, DMv and Sv. Hence, we carry out the model selection   separately, for a selected number of Fourier modes for each noise component, as described below.} This is accomplished amongst the four values of $\bthis{N} = 2,5,8,12~(\times T_{span})$, where $T_{span}$ is in years. As the data spans differ for a few pulsars, we have created the number of Fourier modes as a function of $T_{span}$. These four values are chosen to evenly spread the number from the lowest to highest Fourier modes. \bthis{The lowest mode corresponds to the frequency of $2/T_{span}$, while the highest mode corresponds to the frequency of once per month.} 
\bthis{For example, for PSR J1909$-$3744, which has a data span of 3.38 years, the four values of N, among which the model selection is performed, are 6, 16, 27, and 40. We chose the (integer) truncated values for N in case we got floating point numbers. For PSR J1744$-$1134, where the data span is only 0.44 years, the four values of N chosen are 1, 2, 3, and 4.}
The optimum number of Fourier modes selection is performed in multiple steps, starting from RN, using \textit{WR} model, which contains only WN and RN. We obtain the Bayes factors for the four values of modes described above. We set a Bayes factor threshold of 10 for the selection of the number of modes, similar to ~\cite{ChalumeauBabak+2022}. After the optimum number is obtained for RN, and if the noise model of a pulsar contains DMv, we perform mode selection for DMv using \textit{WRD} model, keeping the RN modes at the optimum mode number, as obtained in the previous step. Similarly, if the model also contains scattering variations, we perform modes selection using \textit{WRDS} model, keeping the RN and DMv modes at their optimum value and performing mode selection for only scattering variations. If the selected model only contains DMv, we use \textit{WD} model to obtain the optimum number. If the model contains DMv and Sv and no RN, then we first use \textit{WD} to obtain the optimum number of DMv and then use \textit{WDS} model to perform the modes selection for Sv, keeping DMv modes at the optimum number. Once again, we calculate the Bayesian evidence and corresponding Bayes factors using the \texttt{DYNESTY} package. The number of modes for each pulsar and noise model can be found in Table~\ref{parameter_values}.

\subsection{Parameter estimation}
After the selection of the \bthis{preferred} noise model, followed by choosing the optimum number of Fourier modes for each pulsar, we perform the parameter estimation using Bayesian regression to obtain the final noise parameter values using \texttt{PTMCMCSAMPLER}~\cite{EllisvanHaasteren2017}. We again use the same Gaussian likelihood and the priors for the selected noise models as those which were used for calculating the evidence. The final values are tabulated in Table~\ref{parameter_values}. All the red noise posterior plots and the full corner plots for all the noise models can be found in Appendix \ref{appendix_b} and \ref{appendix_c}, respectively.

\section{Results and discussion}
\label{sec:discussion}

This section presents the results of custom noise modelling for each pulsar. One thing to note is that as the InPTA DR1 parfiles contain \texttt{DMXs} that absorb the DM variations, we remove them from the parfiles and fit for \texttt{DM1} and \texttt{DM2} using \texttt{TEMPO2}. These final parfiles are used for our noise analysis. We also remove \texttt{T2EFACs} from our parfiles before using them for SPNA work. In the initial step of EQUAD model selection, we find that pulsars J0613$-$1224, J1012+5307, J1744$-$1134 and J2124$-$3358 do not prefer EQUAD, while all the remaining ten pulsars strongly prefer it.
Overall, eight pulsars J1012+5307, J1022+1001, J1643$-$1224, J1713+0747, J1744$-$1134, J1909$-$3744, J1939+2134 and J2145$-$0750 show the presence of DMv in InPTA dataset. For three pulsars J1643$-$1224, J1939+2134, and J2145$-$0750, we can see from Table~\ref{model_select} that \textit{WRDS} is preferred over other models, i.e. all the red noises considered in this paper are present in these pulsars. Except for J0751+1807 and J1600$-$3053, all the other pulsars support the red noises in the InPTA data. Interestingly, we see scattering variations for four pulsars in our sample: J1643$-$1224, J1713+0747, J1939+2134, and J2145$-$0750. Six pulsars J0751+1807, J1012+5307, J1022+1001, J1600$-$3053, J1713+0747 and J1744$-$1134 do not show the presence of achromatic red noise. Pulsars J1909$-$3744, J1939+2134 and J2145$-$0750 prefer the highest number of Fourier modes for RN, denoting the presence of RN at high frequencies. Similarly, Sv is present in high frequencies for PSR J2145$-$0750, while only in low frequencies for J1643$-$1224, J1713+0747 and J1939+2134.

\subsection*{\texorpdfstring{J0437$-$4715}{J0437-4715}}
PSR J0437$-$4715 is one of the brightest pulsars observed by InPTA. In the InPTA DR1, we can see achromatic red noise in the ToA residuals and do not find significant DM variations in the DM time series, whereas both these noises are present in PPTA for this pulsar~\cite{GonchorovReardon+2021}. As the Bayes factor for \textit{WRD} over \textit{WR} was inconclusive, we selected \textit{WR} because it has less number of free parameters. As the baseline for this pulsar is around one year for InPTA, it could be one of the reasons for the differences in models.
To understand this, more work is underway to characterize the jitter across frequencies as we have simultaneous multi-band observations.

\subsection*{\texorpdfstring{J0613$-$0200}{J0613-0200}}
For this pulsar, again, we see slight variations in the ToA residuals, while no such variations are seen in the corresponding DM time series, where the $\Delta\text{DMs}$ have precision up to the fourth decimal place. \textit{WR}, \textit{WRD}, and \textit{WRDS} have nearly the same value of Bayesian Evidence, and \textit{WRD}, \textit{WRDS} have inconclusive Bayes factors over \textit{WR} as seen in Table \ref{model_select}. Hence, we select \textit{WR} based on it having fewer number of free parameters. We also find that the noise model in PPTA~\cite{GonchorovReardon+2021} also contains only achromatic red noise similar to our results, while EPTA~\cite{ChalumeauBabak+2022} contains additional DM variations.

\subsection*{J0751+1807}
For this pulsar, Bayesian analysis suggests no significant Bayes factors of models with any kind of red noise over the white noise only model, i.e. \textit{W} and hence, \textit{W} is chosen based on the simplicity of the model. The DM time series and residuals from the InPTA DR1 also show no significant variation over time, supporting the selected noise model.

\subsection*{J1012+5307}
In this pulsar, the DM variations are evident, especially in the later part of the data, while no such variations are observed in ToA residuals. The Bayes factor for \textit{WRD}, \textit{WDS}, and \textit{WRDS} with respect to \textit{WD} is insignificant as seen in Table \ref{model_select}. Hence we select \textit{WD} based on the model's simplicity. Also, the data before cycle 37 did not have very high precision, causing large error bars on the DM, but the subsequent data has high precision DMs, where a discernible trend is evident.
Hence, one expects the DM variations to be the dominant noise process, and Bayesian analysis favors the same. The prefered noise model for this pulsar in EPTA~\cite{ChalumeauBabak+2022} contains both achromatic red noise and DM variations. The lack of red noise in InPTA DR1 could be due to the short data span and large data gap for this pulsar.

\begin{figure*}[t!]
\includegraphics[keepaspectratio=true,scale=0.40]{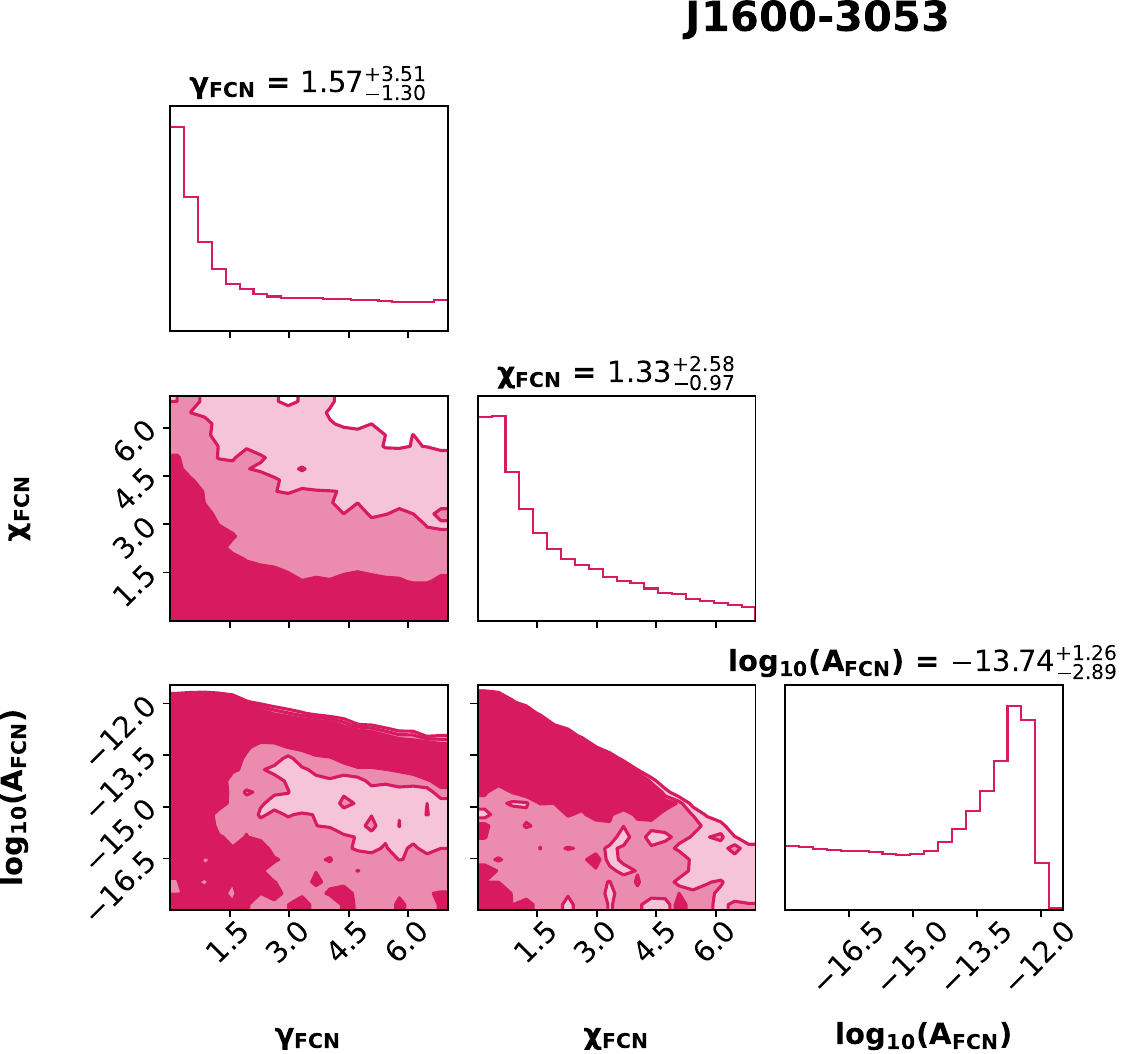}
  \includegraphics[keepaspectratio=true,scale=0.40]{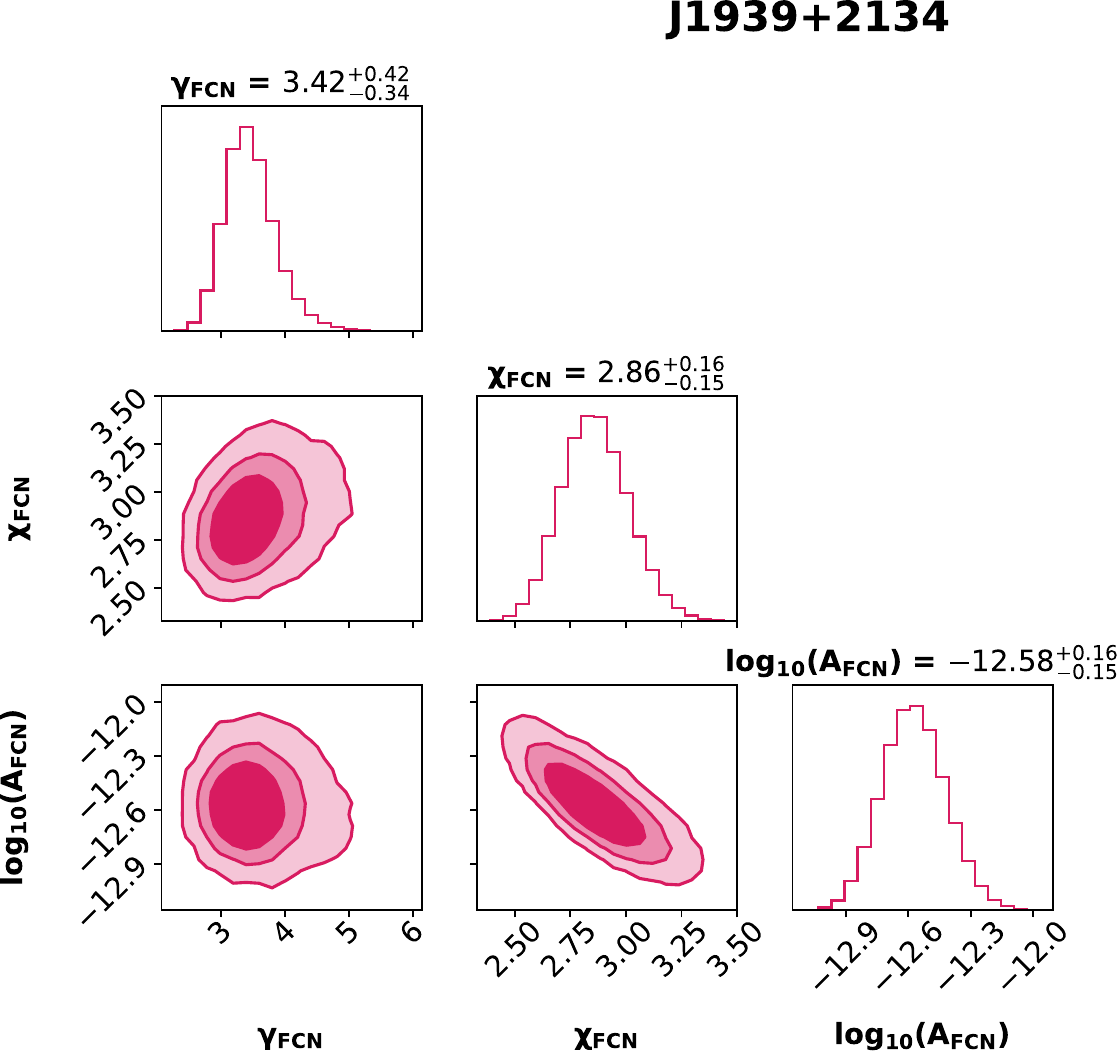}
   \caption{(Left Panel): J1600$-$3053 posterior distributions of the chromatic index $\chi_{FCN}$. (Right Panel): J1939+2134 posterior distributions of the chromatic index $\chi_{FCN}$.}
  \label{fig:16001939crn}
\end{figure*}

\subsection*{J1022+1001}
For this pulsar, we again observe that the DM variations are conspicuous from cycle 37 data onward. Hence, it seems to be the dominant process, with no discernible variations in the ToA residuals. 
The Bayesian analysis gives comparable evidences for \textit{WDS} to \textit{WD} with Bayes factors close to one (see Table \ref{model_select}), which implies there is no preferred model among these. Therefore, we chose \textit{WD} as it had the smallest number of free parameters. 
Our noise model also agrees with the PPTA noise model for this pulsar~\cite{GonchorovReardon+2021}.

\subsection*{\texorpdfstring{J1600$-$3053}{J1600-3053}}
This pulsar does not show signatures of any type of red noises based on Bayesian evidence, where all the models show comparable Bayesian evidence values. This is reaffirmed by the analysis based on the free chromatic index, where the chromatic index is a one-sided distribution with large error bars, as seen in Fig.~\ref{fig:16001939crn}. In contrast, the amplitude and spectral index are unconstrained, which implies that it encapsulates only the white noise. The DM series and ToA residuals show no significant variations, supporting the Bayesian analysis result of \textit{W}. PPTA~\cite{GonchorovReardon+2021} and EPTA~\cite{ChalumeauBabak+2022} report achromatic red noise and DM variations (along with scattering variations in the case of PPTA), and we also see scattering variations in DM time series in NANOGrav 12.5-year dataset \cite{AlamArzoumanian+2020b}. {InPTA dataset for this pulsar begins where the NANOGrav 12.5-year dataset terminates \cite{TarafdarNobleson+2022,AlamArzoumanian+2020b}, and comparing the DMs between their last epoch (52.333508 $pc/cm^3$) and our first epoch (with \texttt{DMX}) (52.3326 $pc/cm^3$), we observe the difference of $9\times10^{-4}~pc/cm^3$, which is consistent within errors. Furthermore, we do not see any significant scatter-broadening in the low-frequency observations of this pulsar. In addition, we also find that in the NANOGrav 12.5-yr dataset, the DMs tend to stabilize towards the end. A complete understanding of this inconsistency shall be explored, but we suspect that the DMs are stable across the 3.5 years InPTA dataset and may have been variable before, as seen in NANOGrav dataset along with EPTA \cite{ChalumeauBabak+2022}.

\subsection*{\texorpdfstring{J1643$-$1224}{J1643-1224}}
Bayesian analysis for this pulsar strongly prefers \textit{WRDS}, which has achromatic red noise, DM, and scattering variations. Low-frequency observations of this pulsar confirm significant scatter-broadening, which varies from epoch to epoch. In addition to that, DM variations are also seen. From the DR1 plots, it is difficult to adjudicate between scatter broadening and DM variations as both lead to similar exponential scatter, but DM variations seem very evident.

 \subsection*{J1713+0747}
PSR J1713+0747 strongly supports the \textit{WDS} based on the estimated Bayes factor, i.e. only DM and scattering variations. This is one of the best-timed PTA pulsars, and there seems to be very less achromatic red noise. There are DM and scattering variations, but their amplitudes are very small. Overall, the pulsar seems to be a good timer, provided these noise sources are included. One thing to note is that the data does not include the DM event \cite{lam+2018} and profile change event \cite{singha+2021}. There is very little scattering variation, as seen by other PTAs \cite{ChalumeauBabak+2022}, while our model contains it. This could be because the InPTA dataset contains low-frequency data, which is absent in the EPTA dataset, hence could be more sensitive to detect this weak scattering, which may have been missed in EPTA.

\subsection*{\texorpdfstring{J1744$-$1134}{J1744-1134}}
The data span for this pulsar is only about six months; hence it is difficult to make definitive conclusions about the RN here. On the other hand, the DM variations are evident in the DM time series. The Bayesian analysis supports this and provides \textit{WD} as the most selected model. The selected model for this pulsar in EPTA~\cite{ChalumeauBabak+2022} has RN along with DMv, which is absent in InPTA data due to a small time span, while 
 \citet{GonchorovReardon+2021} obtain a similar model as ours.

\subsection*{J1857+0943}
In this pulsar, the achromatic red noise is quite visible in the ToA residuals and seems to be the dominant source, while there are no visible DM variations in the DM time series. Bayesian model selection prefers \textit{WR}, which is expected based on our observations and supports our claim. \citet{GonchorovReardon+2021} model also contains DM noise, which is absent in our modelling and can be due to our relatively short data span.

\begin{figure*}
  \includegraphics[keepaspectratio=true,scale=0.40]{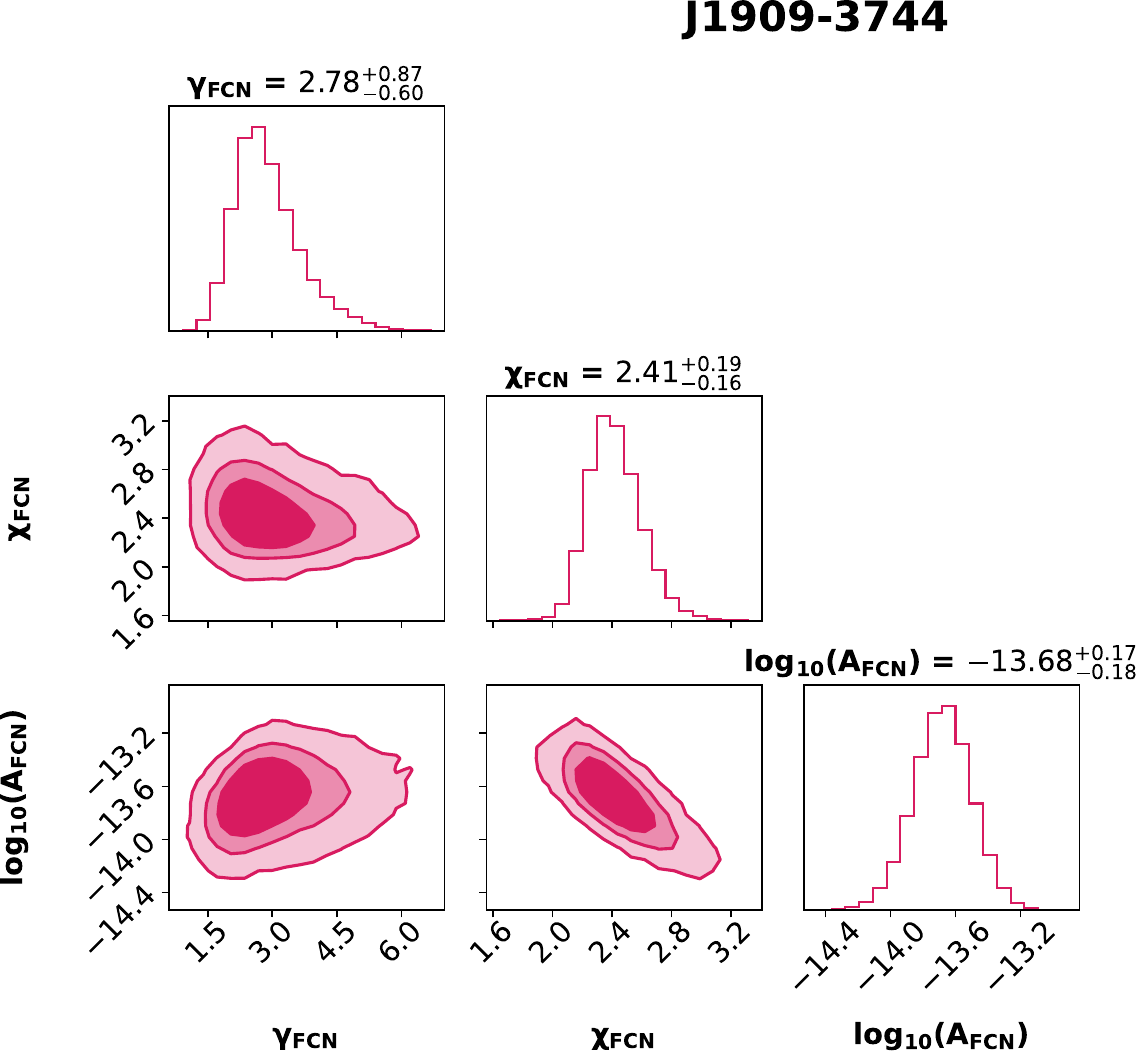}
  \includegraphics[keepaspectratio=true,scale=0.40]{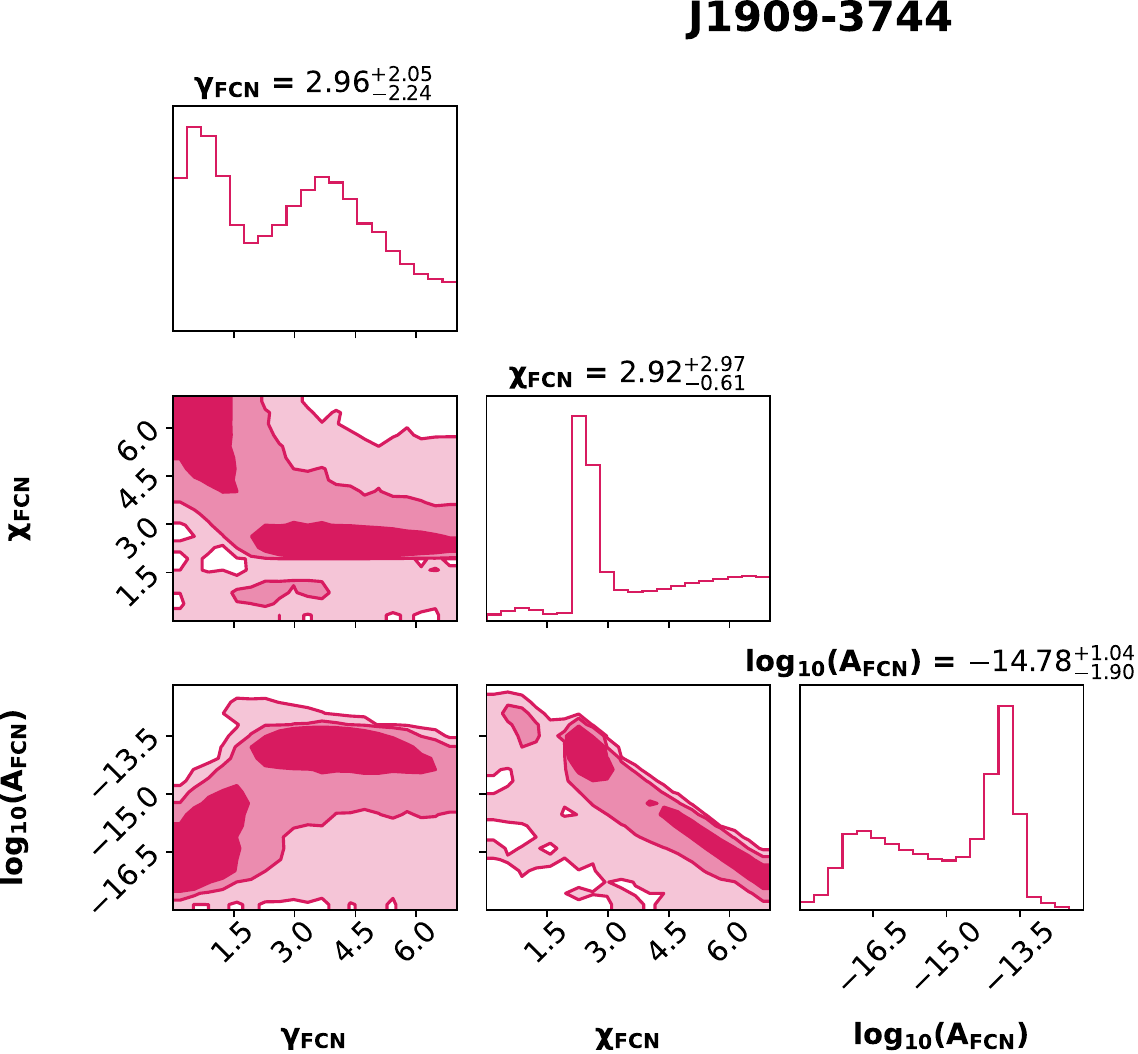}
 \caption{(Left Panel): J1909$-$3744 posterior distributions of the chromatic index $\chi_{FCN}$ \bthis{with the noise model containing WN and RN}. (Right Panel): J1909$-$3744 posterior distributions of the chromatic index $\chi_{FCN}$ \bthis{with the noise model containing WN, RN and DMv}.}
  \label{fig:1909crn}
\end{figure*}

\subsection*{\texorpdfstring{J1909$-$3744}{J1909-3744}}
PSR J1909-3744 exhibits the best stability among the PTA pulsars. Nevertheless, there are significant variations in the DM towards this pulsar in the InPTA DR1, as seen from the DM time series. Our analysis indicates strong evidence for \textit{WRD} and \textit{WRDS}, both of which incorporate achromatic red noise and DM noise. The latter model also includes a GP corresponding to the variations in scatter-broadening and is marginally favored over the former by $\ln(\text{Bayes Factor})$ of 4.4. However, this pulsar has no pulse broadening, even at 300 MHz. To investigate this further, we examined the parameters estimated for the different noise processes in \textit{WR}, \textit{WRD}, and \textit{WRDS}. The amplitude \bthis{(at frequency of $1 yr^{-1}$)}, and high-frequency cut-off for the achromatic noise are consistent for all these models ($\log A_{RN} \sim -12.5$ and $\gamma_{RN} \sim$ 0.68), and so is the case for DM noise ($\log A_{DMv} \sim -13.5$ and $\gamma_{DMv} \sim$ 2) with DM noise process an order of magnitude weaker than the achromatic red-noise with a much larger frequency content than the latter. In contrast, the noise process amplitude (\bthis{at frequency of $1 yr^{-1}$}) representing the scatter broadening is two orders of magnitude smaller than either the achromatic red-noise process or the DM noise process with $\gamma$ essentially consistent with zero (in other words, consistent with no variations). We further investigated this by using a GP with a free chromatic index. A model involving just the achromatic red noise process with a chromatic process having a free index yields hyper-parameters similar to those of chromatic red noise and the DM noise with a chromatic index of $2.4^{+0.19}_{-0.16}$, close to the expected chromatic index of 2 for a DM noise process as seen in Fig.~\ref{fig:1909crn}. Furthermore, parameter estimation with a model consisting of achromatic red noise, DM noise, and chromatic process with a free index again yields a two-order weaker free index process, with significant error bars on the index (see Fig.~\ref{fig:1909crn} again). Thus, there is no strong evidence for a scattering process consistent with the absence of any observed scattering at 300 MHz despite marginally higher evidence for a scattering process. Hence, we have chosen \textit{WRD} for this pulsar which is the simpler model.

\subsection*{J1939+2134}
PSR J1939+2134 is the oldest known MSP. For a long time, it was assumed to be the most stable rotator \citep{krt1992}. Still, high-precision timing campaigns have shown that not only does its rotation rate wander, but it also exhibits significant DM and scatter-broadening variations. Observations near 300 MHz clearly show that the pulse profile is scatter broadened at these frequencies~\citep{joshirama2006}. Thus, we needed to model the  DM and scattering variations apart from the achromatic red noise and white noise for this bright but relatively distant PTA pulsar (\textit{WRDS}). The posterior distributions of the relevant GP are shown in Appendix~\ref{appendix_b}, while those of the full model are presented in Appendix~\ref{appendix_c}. As mentioned before, the GP representing the scatter-broadening variations assumes Kolmogorov turbulence with a chromatic index of 4. Previous studies have shown that this is not true along all lines of sight, and the chromatic index can go as low as -0.7, i.e. shallower than Kolmogorov~\cite{levin2016,kk2019,turner2021}. To investigate this, we used a FCN model apart from the usual white noise, achromatic red noise, DM variations, and scattering variations. While this incorporates an additional free parameter in terms of the chromatic index, we find that the corresponding FCN parameters to be well constrained, as can be seen in Fig.~\ref{fig:16001939crn} with $\chi_{FCN}$ $\sim$ 2.86, and $\log_{10}A_{FCN}$ $\sim$ -12.58, which is much greater than conventional scattering variation model with chromatic index 4, and comparable with other red noise amplitudes (see Table \ref{parameter_values}). This indicates that for this pulsar, the chromatic index does not agree with Kolmogorov turbulence with chromatic index 4. We plan to investigate this further from direct measurements.

\subsection*{\texorpdfstring{J2124$-$3358}{J2124-3358}}
The selected model for this pulsar is \textit{WR}, as \textit{WRD} and \textit{WR} have comparable Bayesian evidence, and hence, inconclusive Bayes factor for \textit{WRD} over \textit{WR}. Therefore, \textit{WR} is chosen based on fewer number of  free parameters. Although, the ToA residuals have large error bars in InPTA DR1, slight achromatic red noise is visible. At the same time, the high precision $\Delta\text{DM}$ estimates vary only in the fourth decimal place, and hence do not showing significant DM variations, which supports the selected model.

\subsection*{\texorpdfstring{J2145$-$0750}{J2145-0750}}
For PSR J2145$-$0750, we obtain \textit{WRDS} as the strongly preferred using Bayesian model selection, suggesting the presence of achromatic red noise, as well as DM and scattering variations. In the DR1 plots, we see in ToA residuals that there is a small amplitude achromatic red noise variation over a large time scale, and we see in the DM series there are short time scale small amplitude DM variations. Scattering noise is not evident as it is difficult to see small scattering in DM series due to smaller amplitude than DM noise in this case.

\section{Conclusions}
\label{sec:conclusion}

We have used the InPTA DR1 data set to carry out noise analysis for individual pulsars. Using Bayesian inference, we have chosen the most optimum noise models for each pulsar in our data set (see Table \ref{model_select}). We have also estimated the optimum number of Fourier modes for the red noise analysis for each pulsar. Even with a relatively modest timing baseline of 3.5 years, 8 out of 14 pulsars show a clear presence of red noise. Finally, given the unique low-frequency coverage of the InPTA data set, we were able to constrain the DM noise for eight pulsars while also detecting scatter-broadening variations in four pulsars (see Table \ref{parameter_values}). While our results are broadly in agreement with the other PTAs, we would like to note the well-constrained DM and scatter-broadening variations, even with a short timing baseline. \bthis{The shallow spectral indices for achromatic red noise for most of the pulsars can be attributed either to the short timing baseline, which is roughly one-fifth of the timespan when compared to other PTAs, or a gap in our observations.} The most noteworthy result from our analysis was obtained for PSR J1939+2134, where we found the residual chromatic noise with a $\chi \sim$ 2.86, with two orders larger amplitude than scattering variation with $\chi_{Sv}$ = 4, providing an indication towards scattering index that doesn't agree with that expected for a Kolmogorov scattering medium.  
Also, for PSR J1909$-$3744, we find no significant scattering variations in profiles that are present in other PTAs noise models and found using the FCN model that residual DMv was masquerading as scattering variations. 
 This exemplifies the pivotal role  of InPTA data towards modelling the noise budget of pulsars in the IPTA data set, especially in the cases of pulsars whose noise budget is dominated by variations in the delays caused by the ISM. 
 \bthis{ In a followup work, we shall also  produce band 3+5 ToAs using the wideband technique~\cite{Paladi2023, Noble22}, and also plan to perform noise analysis on the wideband dataset in the future.}
 
 \bthis{In this work, we have performed single pulsar noise analysis (SPNA), where we implement Bayesian analysis to only fit for the noise models, while keeping the parameters from \tempotwo{} parameter file as best fits. But we also plan to perform single pulsar noise and timing analysis (SPNTA) in the future, where we fit for all the parameters, including noise as well as parameters from \tempotwo{} parameter file. This will facilitate a robust search for gravitational bursts in the InPTA data, and also would answer many questions about the utility of the low-frequency data in modeling and mitigating the ISM contribution in the overall noise budget of the IPTA pulsars.} 

\begin{acknowledgments}

InPTA acknowledges the support of the GMRT staff in resolving technical difficulties and providing technical solutions for high-precision work. We acknowledge the GMRT telescope operators for the observations. The GMRT is run by the National Centre for Radio Astrophysics of the Tata Institute of Fundamental Research, India.
AmS is supported by CSIR fellowship Grant number 09/1001(12656)/2021-EMR-I
and T-641 (DST-ICPS).
BCJ, YG and YM acknowledge the support of the Department of Atomic Energy, Government of India, under Project Identification \# RTI 4002. BCJ acknowledges the support of the Department of Atomic Energy, Government of India, under project No. 12-R\&D-TFR-5.02-0700. 
AS is supported by the NANOGrav NSF Physics Frontiers Center (awards \#1430284 and 2020265).
KT is partially supported by JSPS KAKENHI Grant Numbers 20H00180, 21H01130, and 21H04467, Bilateral Joint Research Projects of JSPS, and the ISM Cooperative Research Program (2021-ISMCRP-2017).
AKP is supported by CSIR fellowship Grant number 09/0079(15784)/2022-EMR-I.
KN is supported by the Birla Institute of Technology \& Science Institute fellowship.
SH is supported by JSPS KAKENHI Grant Number 20J20509.
TK is partially supported by the JSPS Overseas Challenge Program for Young Researchers.
AC acknowledges financial support provided under the European Union’s H2020 ERC Consolidator Grant “Binary Massive Black Hole Astrophysics” (B Massive, Grant Agreement: 818691, PI - A. Sesana).
GS acknowledges financial support provided under the European Union’s H2020 ERC Consolidator Grant “Binary Massive Black Hole Astrophysics” (B Massive, Grant Agreement: 818691, PI - A. Sesana).
We acknowledge the National Supercomputing Mission (NSM) for providing computing resources of ‘PARAM Ganga’ at the Indian Institute of Technology Roorkee as well as `PARAM Seva' at IIT Hyderabad, which is implemented by C-DAC and supported by the Ministry of Electronics and Information Technology (MeitY) and Department of Science and Technology (DST), Government of India.\\

\end{acknowledgments}
\bibliographystyle{apsrev4-1}
\bibliography{spna}

\begin{thebibliography}{67}%
\makeatletter
\providecommand \@ifxundefined [1]{%
 \@ifx{#1\undefined}
}%
\providecommand \@ifnum [1]{%
 \ifnum #1\expandafter \@firstoftwo
 \else \expandafter \@secondoftwo
 \fi
}%
\providecommand \@ifx [1]{%
 \ifx #1\expandafter \@firstoftwo
 \else \expandafter \@secondoftwo
 \fi
}%
\providecommand \natexlab [1]{#1}%
\providecommand \enquote  [1]{``#1''}%
\providecommand \bibnamefont  [1]{#1}%
\providecommand \bibfnamefont [1]{#1}%
\providecommand \citenamefont [1]{#1}%
\providecommand \href@noop [0]{\@secondoftwo}%
\providecommand \href [0]{\begingroup \@sanitize@url \@href}%
\providecommand \@href[1]{\@@startlink{#1}\@@href}%
\providecommand \@@href[1]{\endgroup#1\@@endlink}%
\providecommand \@sanitize@url [0]{\catcode `\\12\catcode `\$12\catcode
  `\&12\catcode `\#12\catcode `\^12\catcode `\_12\catcode `\%12\relax}%
\providecommand \@@startlink[1]{}%
\providecommand \@@endlink[0]{}%
\providecommand \url  [0]{\begingroup\@sanitize@url \@url }%
\providecommand \@url [1]{\endgroup\@href {#1}{\urlprefix }}%
\providecommand \urlprefix  [0]{URL }%
\providecommand \Eprint [0]{\href }%
\providecommand \doibase [0]{http://dx.doi.org/}%
\providecommand \selectlanguage [0]{\@gobble}%
\providecommand \bibinfo  [0]{\@secondoftwo}%
\providecommand \bibfield  [0]{\@secondoftwo}%
\providecommand \translation [1]{[#1]}%
\providecommand \BibitemOpen [0]{}%
\providecommand \bibitemStop [0]{}%
\providecommand \bibitemNoStop [0]{.\EOS\space}%
\providecommand \EOS [0]{\spacefactor3000\relax}%
\providecommand \BibitemShut  [1]{\csname bibitem#1\endcsname}%
\let\auto@bib@innerbib\@empty
\bibitem [{\citenamefont {{Foster}}\ and\ \citenamefont
  {{Backer}}(1990)}]{FosterBacker1990}%
  \BibitemOpen
  \bibfield  {author} {\bibinfo {author} {\bibfnamefont {R.~S.}\ \bibnamefont
  {{Foster}}}\ and\ \bibinfo {author} {\bibfnamefont {D.~C.}\ \bibnamefont
  {{Backer}}},\ }\href {\doibase 10.1086/169195} {\bibfield  {journal}
  {\bibinfo  {journal} {\apj}\ }\textbf {\bibinfo {volume} {361}},\ \bibinfo
  {pages} {300} (\bibinfo {year} {1990})}\BibitemShut {NoStop}%
\bibitem [{\citenamefont {{Burke-Spolaor}}\ \emph {et~al.}(2019)\citenamefont
  {{Burke-Spolaor}}, \citenamefont {{Taylor}}, \citenamefont {{Charisi}},
  \citenamefont {{Dolch}}, \citenamefont {{Hazboun}} \emph
  {et~al.}}]{Burke-SpolaorTaylor+2019}%
  \BibitemOpen
  \bibfield  {author} {\bibinfo {author} {\bibfnamefont {S.}~\bibnamefont
  {{Burke-Spolaor}}}, \bibinfo {author} {\bibfnamefont {S.~R.}\ \bibnamefont
  {{Taylor}}}, \bibinfo {author} {\bibfnamefont {M.}~\bibnamefont {{Charisi}}},
  \bibinfo {author} {\bibfnamefont {T.}~\bibnamefont {{Dolch}}}, \bibinfo
  {author} {\bibfnamefont {J.~S.}\ \bibnamefont {{Hazboun}}},  \emph {et~al.},\
  }\href {\doibase 10.1007/s00159-019-0115-7} {\bibfield  {journal} {\bibinfo
  {journal} {\aapr}\ }\textbf {\bibinfo {volume} {27}},\ \bibinfo {eid} {5}
  (\bibinfo {year} {2019})},\ \Eprint {http://arxiv.org/abs/1811.08826}
  {arXiv:1811.08826 [astro-ph.HE]} \BibitemShut {NoStop}%
\bibitem [{\citenamefont {{Hellings}}\ and\ \citenamefont
  {{Downs}}(1983)}]{HellingsDowns1983}%
  \BibitemOpen
  \bibfield  {author} {\bibinfo {author} {\bibfnamefont {R.~W.}\ \bibnamefont
  {{Hellings}}}\ and\ \bibinfo {author} {\bibfnamefont {G.~S.}\ \bibnamefont
  {{Downs}}},\ }\href {\doibase 10.1086/183954} {\bibfield  {journal} {\bibinfo
   {journal} {\apjl}\ }\textbf {\bibinfo {volume} {265}},\ \bibinfo {pages}
  {L39} (\bibinfo {year} {1983})}\BibitemShut {NoStop}%
\bibitem [{\citenamefont {{Hogan}}(1986)}]{Hogan1986}%
  \BibitemOpen
  \bibfield  {author} {\bibinfo {author} {\bibfnamefont {C.~J.}\ \bibnamefont
  {{Hogan}}},\ }\href {\doibase 10.1093/mnras/218.4.629} {\bibfield  {journal}
  {\bibinfo  {journal} {\mnras}\ }\textbf {\bibinfo {volume} {218}},\ \bibinfo
  {pages} {629} (\bibinfo {year} {1986})}\BibitemShut {NoStop}%
\bibitem [{\citenamefont {Damour}\ and\ \citenamefont
  {Vilenkin}(2000)}]{DamourVilenkin2000}%
  \BibitemOpen
  \bibfield  {author} {\bibinfo {author} {\bibfnamefont {T.}~\bibnamefont
  {Damour}}\ and\ \bibinfo {author} {\bibfnamefont {A.}~\bibnamefont
  {Vilenkin}},\ }\href {\doibase 10.1103/PhysRevLett.85.3761} {\bibfield
  {journal} {\bibinfo  {journal} {Phys. Rev. Lett.}\ }\textbf {\bibinfo
  {volume} {85}},\ \bibinfo {pages} {3761} (\bibinfo {year}
  {2000})}\BibitemShut {NoStop}%
\bibitem [{\citenamefont {Grishchuk}(2005)}]{Grishchuk2005}%
  \BibitemOpen
  \bibfield  {author} {\bibinfo {author} {\bibfnamefont {L.~P.}\ \bibnamefont
  {Grishchuk}},\ }\href {\doibase 10.1070/PU2005v048n12ABEH005795} {\bibfield
  {journal} {\bibinfo  {journal} {Physics-Uspekhi}\ }\textbf {\bibinfo {volume}
  {48}},\ \bibinfo {pages} {1235} (\bibinfo {year} {2005})}\BibitemShut
  {NoStop}%
\bibitem [{\citenamefont {{Kramer}}\ and\ \citenamefont
  {{Champion}}(2013)}]{KramerChampion2013}%
  \BibitemOpen
  \bibfield  {author} {\bibinfo {author} {\bibfnamefont {M.}~\bibnamefont
  {{Kramer}}}\ and\ \bibinfo {author} {\bibfnamefont {D.~J.}\ \bibnamefont
  {{Champion}}},\ }\href {\doibase 10.1088/0264-9381/30/22/224009} {\bibfield
  {journal} {\bibinfo  {journal} {Classical and Quantum Gravity}\ }\textbf
  {\bibinfo {volume} {30}},\ \bibinfo {eid} {224009} (\bibinfo {year}
  {2013})}\BibitemShut {NoStop}%
\bibitem [{\citenamefont {{Hobbs}}(2013)}]{Hobbs2013}%
  \BibitemOpen
  \bibfield  {author} {\bibinfo {author} {\bibfnamefont {G.}~\bibnamefont
  {{Hobbs}}},\ }\href {\doibase 10.1088/0264-9381/30/22/224007} {\bibfield
  {journal} {\bibinfo  {journal} {Classical and Quantum Gravity}\ }\textbf
  {\bibinfo {volume} {30}},\ \bibinfo {eid} {224007} (\bibinfo {year}
  {2013})},\ \Eprint {http://arxiv.org/abs/1307.2629} {arXiv:1307.2629
  [astro-ph.IM]} \BibitemShut {NoStop}%
\bibitem [{\citenamefont {{McLaughlin}}(2013)}]{McLaughlin2013}%
  \BibitemOpen
  \bibfield  {author} {\bibinfo {author} {\bibfnamefont {M.~A.}\ \bibnamefont
  {{McLaughlin}}},\ }\href {\doibase 10.1088/0264-9381/30/22/224008} {\bibfield
   {journal} {\bibinfo  {journal} {Classical and Quantum Gravity}\ }\textbf
  {\bibinfo {volume} {30}},\ \bibinfo {eid} {224008} (\bibinfo {year}
  {2013})},\ \Eprint {http://arxiv.org/abs/1310.0758} {arXiv:1310.0758
  [astro-ph.IM]} \BibitemShut {NoStop}%
\bibitem [{\citenamefont {Joshi}\ \emph {et~al.}(2018)\citenamefont {Joshi},
  \citenamefont {Arumugasamy}, \citenamefont {Bagchi}, \citenamefont
  {Bandyopadhyay}, \citenamefont {Basu} \emph
  {et~al.}}]{JoshiArumugasamy+2018}%
  \BibitemOpen
  \bibfield  {author} {\bibinfo {author} {\bibfnamefont {B.~C.}\ \bibnamefont
  {Joshi}}, \bibinfo {author} {\bibfnamefont {P.}~\bibnamefont {Arumugasamy}},
  \bibinfo {author} {\bibfnamefont {M.}~\bibnamefont {Bagchi}}, \bibinfo
  {author} {\bibfnamefont {D.}~\bibnamefont {Bandyopadhyay}}, \bibinfo {author}
  {\bibfnamefont {A.}~\bibnamefont {Basu}},  \emph {et~al.},\ }\href {\doibase
  10.1007/s12036-018-9549-y} {\bibfield  {journal} {\bibinfo  {journal}
  {Journal of Astrophysics and Astronomy}\ }\textbf {\bibinfo {volume} {39}},\
  \bibinfo {pages} {51} (\bibinfo {year} {2018})}\BibitemShut {NoStop}%
\bibitem [{\citenamefont {{Lee}}(2016)}]{Lee2016}%
  \BibitemOpen
  \bibfield  {author} {\bibinfo {author} {\bibfnamefont {K.~J.}\ \bibnamefont
  {{Lee}}},\ }in\ \href@noop {} {\emph {\bibinfo {booktitle} {Frontiers in
  Radio Astronomy and FAST Early Sciences Symposium 2015}}},\ \bibinfo {series}
  {Astronomical Society of the Pacific Conference Series}, Vol.\ \bibinfo
  {volume} {502},\ \bibinfo {editor} {edited by\ \bibinfo {editor}
  {\bibfnamefont {L.}~\bibnamefont {{Qain}}}\ and\ \bibinfo {editor}
  {\bibfnamefont {D.}~\bibnamefont {{Li}}}}\ (\bibinfo {year} {2016})\
  p.~\bibinfo {pages} {19}\BibitemShut {NoStop}%
\bibitem [{\citenamefont {{Spiewak}}\ \emph {et~al.}(2022)\citenamefont
  {{Spiewak}}, \citenamefont {{Bailes}}, \citenamefont {{Miles}}, \citenamefont
  {{Parthasarathy}}, \citenamefont {{Reardon}} \emph
  {et~al.}}]{SpiewakBailes+2022}%
  \BibitemOpen
  \bibfield  {author} {\bibinfo {author} {\bibfnamefont {R.}~\bibnamefont
  {{Spiewak}}}, \bibinfo {author} {\bibfnamefont {M.}~\bibnamefont {{Bailes}}},
  \bibinfo {author} {\bibfnamefont {M.~T.}\ \bibnamefont {{Miles}}}, \bibinfo
  {author} {\bibfnamefont {A.}~\bibnamefont {{Parthasarathy}}}, \bibinfo
  {author} {\bibfnamefont {D.~J.}\ \bibnamefont {{Reardon}}},  \emph {et~al.},\
  }\href {\doibase 10.1017/pasa.2022.19} {\bibfield  {journal} {\bibinfo
  {journal} {\pasa}\ }\textbf {\bibinfo {volume} {39}},\ \bibinfo {eid} {e027}
  (\bibinfo {year} {2022})},\ \Eprint {http://arxiv.org/abs/2204.04115}
  {arXiv:2204.04115 [astro-ph.HE]} \BibitemShut {NoStop}%
\bibitem [{\citenamefont {Hobbs}\ \emph {et~al.}(2010)\citenamefont {Hobbs},
  \citenamefont {Archibald}, \citenamefont {Arzoumanian}, \citenamefont
  {Backer}, \citenamefont {Bailes} \emph {et~al.}}]{HobbsArchibald+2010}%
  \BibitemOpen
  \bibfield  {author} {\bibinfo {author} {\bibfnamefont {G.}~\bibnamefont
  {Hobbs}}, \bibinfo {author} {\bibfnamefont {A.}~\bibnamefont {Archibald}},
  \bibinfo {author} {\bibfnamefont {Z.}~\bibnamefont {Arzoumanian}}, \bibinfo
  {author} {\bibfnamefont {D.}~\bibnamefont {Backer}}, \bibinfo {author}
  {\bibfnamefont {M.}~\bibnamefont {Bailes}},  \emph {et~al.},\ }\href
  {\doibase 10.1088/0264-9381/27/8/084013} {\bibfield  {journal} {\bibinfo
  {journal} {Classical and Quantum Gravity}\ }\textbf {\bibinfo {volume}
  {27}},\ \bibinfo {pages} {084013} (\bibinfo {year} {2010})}\BibitemShut
  {NoStop}%
\bibitem [{\citenamefont {{Goncharov}}\ \emph
  {et~al.}(2021{\natexlab{a}})\citenamefont {{Goncharov}}, \citenamefont
  {{Reardon}}, \citenamefont {{Shannon}}, \citenamefont {{Zhu}}, \citenamefont
  {{Thrane}}, \citenamefont {{Bailes}} \emph {et~al.}}]{GonchorovReardon+2021}%
  \BibitemOpen
  \bibfield  {author} {\bibinfo {author} {\bibfnamefont {B.}~\bibnamefont
  {{Goncharov}}}, \bibinfo {author} {\bibfnamefont {D.~J.}\ \bibnamefont
  {{Reardon}}}, \bibinfo {author} {\bibfnamefont {R.~M.}\ \bibnamefont
  {{Shannon}}}, \bibinfo {author} {\bibfnamefont {X.-J.}\ \bibnamefont
  {{Zhu}}}, \bibinfo {author} {\bibfnamefont {E.}~\bibnamefont {{Thrane}}},
  \bibinfo {author} {\bibfnamefont {M.}~\bibnamefont {{Bailes}}},  \emph
  {et~al.},\ }\href {\doibase 10.1093/mnras/staa3411} {\bibfield  {journal}
  {\bibinfo  {journal} {\mnras}\ }\textbf {\bibinfo {volume} {502}},\ \bibinfo
  {pages} {478} (\bibinfo {year} {2021}{\natexlab{a}})},\ \Eprint
  {http://arxiv.org/abs/2010.06109} {arXiv:2010.06109 [astro-ph.HE]}
  \BibitemShut {NoStop}%
\bibitem [{\citenamefont {{Chalumeau}}\ \emph {et~al.}(2022)\citenamefont
  {{Chalumeau}}, \citenamefont {{Babak}}, \citenamefont {{Petiteau}},
  \citenamefont {{Chen}}, \citenamefont {{Samajdar}} \emph
  {et~al.}}]{ChalumeauBabak+2022}%
  \BibitemOpen
  \bibfield  {author} {\bibinfo {author} {\bibfnamefont {A.}~\bibnamefont
  {{Chalumeau}}}, \bibinfo {author} {\bibfnamefont {S.}~\bibnamefont
  {{Babak}}}, \bibinfo {author} {\bibfnamefont {A.}~\bibnamefont {{Petiteau}}},
  \bibinfo {author} {\bibfnamefont {S.}~\bibnamefont {{Chen}}}, \bibinfo
  {author} {\bibfnamefont {A.}~\bibnamefont {{Samajdar}}},  \emph {et~al.},\
  }\href {\doibase 10.1093/mnras/stab3283} {\bibfield  {journal} {\bibinfo
  {journal} {\mnras}\ }\textbf {\bibinfo {volume} {509}},\ \bibinfo {pages}
  {5538} (\bibinfo {year} {2022})},\ \Eprint {http://arxiv.org/abs/2111.05186}
  {arXiv:2111.05186 [astro-ph.HE]} \BibitemShut {NoStop}%
\bibitem [{\citenamefont {{Arzoumanian}}\ \emph {et~al.}(2020)\citenamefont
  {{Arzoumanian}}, \citenamefont {{Baker}}, \citenamefont {{Blumer}},
  \citenamefont {{B{\'e}csy}}, \citenamefont {{Brazier}} \emph
  {et~al.}}]{ArzoumanianBaker+2020}%
  \BibitemOpen
  \bibfield  {author} {\bibinfo {author} {\bibfnamefont {Z.}~\bibnamefont
  {{Arzoumanian}}}, \bibinfo {author} {\bibfnamefont {P.~T.}\ \bibnamefont
  {{Baker}}}, \bibinfo {author} {\bibfnamefont {H.}~\bibnamefont {{Blumer}}},
  \bibinfo {author} {\bibfnamefont {B.}~\bibnamefont {{B{\'e}csy}}}, \bibinfo
  {author} {\bibfnamefont {A.}~\bibnamefont {{Brazier}}},  \emph {et~al.},\
  }\href {\doibase 10.3847/2041-8213/abd401} {\bibfield  {journal} {\bibinfo
  {journal} {\apjl}\ }\textbf {\bibinfo {volume} {905}},\ \bibinfo {eid} {L34}
  (\bibinfo {year} {2020})},\ \Eprint {http://arxiv.org/abs/2009.04496}
  {arXiv:2009.04496 [astro-ph.HE]} \BibitemShut {NoStop}%
\bibitem [{\citenamefont {Antoniadis}\ \emph {et~al.}(2022)\citenamefont
  {Antoniadis}, \citenamefont {Arzoumanian}, \citenamefont {Babak},
  \citenamefont {Bailes}, \citenamefont {Bak Nielsen} \emph
  {et~al.}}]{AntoniadisArzoumanian+2022}%
  \BibitemOpen
  \bibfield  {author} {\bibinfo {author} {\bibfnamefont {J.}~\bibnamefont
  {Antoniadis}}, \bibinfo {author} {\bibfnamefont {Z.}~\bibnamefont
  {Arzoumanian}}, \bibinfo {author} {\bibfnamefont {S.}~\bibnamefont {Babak}},
  \bibinfo {author} {\bibfnamefont {M.}~\bibnamefont {Bailes}}, \bibinfo
  {author} {\bibfnamefont {A.-S.}\ \bibnamefont {Bak Nielsen}},  \emph
  {et~al.},\ }\href {\doibase 10.1093/mnras/stab3418} {\bibfield  {journal}
  {\bibinfo  {journal} {Monthly Notices of the Royal Astronomical Society}\
  }\textbf {\bibinfo {volume} {510}},\ \bibinfo {pages} {4873} (\bibinfo {year}
  {2022})}\BibitemShut {NoStop}%
\bibitem [{\citenamefont {{Joshi}}\ \emph {et~al.}(2022)\citenamefont
  {{Joshi}}, \citenamefont {{Gopakumar}}, \citenamefont {{Pandian}},
  \citenamefont {{Prabu}}, \citenamefont {{Dey}}, \citenamefont {{Bagchi}}
  \emph {et~al.}}]{JoshiGopu+2022}%
  \BibitemOpen
  \bibfield  {author} {\bibinfo {author} {\bibfnamefont {B.~C.}\ \bibnamefont
  {{Joshi}}}, \bibinfo {author} {\bibfnamefont {A.}~\bibnamefont
  {{Gopakumar}}}, \bibinfo {author} {\bibfnamefont {A.}~\bibnamefont
  {{Pandian}}}, \bibinfo {author} {\bibfnamefont {T.}~\bibnamefont {{Prabu}}},
  \bibinfo {author} {\bibfnamefont {L.}~\bibnamefont {{Dey}}}, \bibinfo
  {author} {\bibfnamefont {M.}~\bibnamefont {{Bagchi}}},  \emph {et~al.},\
  }\href {\doibase 10.1007/s12036-022-09869-w} {\bibfield  {journal} {\bibinfo
  {journal} {Journal of Astrophysics and Astronomy}\ }\textbf {\bibinfo
  {volume} {43}},\ \bibinfo {eid} {98} (\bibinfo {year} {2022})},\ \Eprint
  {http://arxiv.org/abs/2207.06461} {arXiv:2207.06461 [astro-ph.HE]}
  \BibitemShut {NoStop}%
\bibitem [{\citenamefont {{Gupta}}\ \emph {et~al.}(2017)\citenamefont
  {{Gupta}}, \citenamefont {{Ajithkumar}}, \citenamefont {{Kale}},
  \citenamefont {{Nayak}}, \citenamefont {{Sabhapathy}} \emph
  {et~al.}}]{GuptaAjithkumar+2017}%
  \BibitemOpen
  \bibfield  {author} {\bibinfo {author} {\bibfnamefont {Y.}~\bibnamefont
  {{Gupta}}}, \bibinfo {author} {\bibfnamefont {B.}~\bibnamefont
  {{Ajithkumar}}}, \bibinfo {author} {\bibfnamefont {H.~S.}\ \bibnamefont
  {{Kale}}}, \bibinfo {author} {\bibfnamefont {S.}~\bibnamefont {{Nayak}}},
  \bibinfo {author} {\bibfnamefont {S.}~\bibnamefont {{Sabhapathy}}},  \emph
  {et~al.},\ }\href {\doibase 10.18520/cs/v113/i04/707-714} {\bibfield
  {journal} {\bibinfo  {journal} {Current Science}\ }\textbf {\bibinfo {volume}
  {113}},\ \bibinfo {pages} {707} (\bibinfo {year} {2017})}\BibitemShut
  {NoStop}%
\bibitem [{\citenamefont {{Krishnakumar}}\ \emph {et~al.}(2021)\citenamefont
  {{Krishnakumar}}, \citenamefont {{Manoharan}}, \citenamefont {{Joshi}},
  \citenamefont {{Girgaonkar}}, \citenamefont {{Desai}} \emph
  {et~al.}}]{KrishnakumarManoharan+2021}%
  \BibitemOpen
  \bibfield  {author} {\bibinfo {author} {\bibfnamefont {M.~A.}\ \bibnamefont
  {{Krishnakumar}}}, \bibinfo {author} {\bibfnamefont {P.~K.}\ \bibnamefont
  {{Manoharan}}}, \bibinfo {author} {\bibfnamefont {B.~C.}\ \bibnamefont
  {{Joshi}}}, \bibinfo {author} {\bibfnamefont {R.}~\bibnamefont
  {{Girgaonkar}}}, \bibinfo {author} {\bibfnamefont {S.}~\bibnamefont
  {{Desai}}},  \emph {et~al.},\ }\href {\doibase 10.1051/0004-6361/202140340}
  {\bibfield  {journal} {\bibinfo  {journal} {\aap}\ }\textbf {\bibinfo
  {volume} {651}},\ \bibinfo {eid} {A5} (\bibinfo {year} {2021})},\ \Eprint
  {http://arxiv.org/abs/2101.05334} {arXiv:2101.05334 [astro-ph.HE]}
  \BibitemShut {NoStop}%
\bibitem [{\citenamefont {{Tarafdar}}\ \emph {et~al.}(2022)\citenamefont
  {{Tarafdar}}, \citenamefont {{Nobleson}}, \citenamefont {{Rana}},
  \citenamefont {{Singha}}, \citenamefont {{Krishnakumar}} \emph
  {et~al.}}]{TarafdarNobleson+2022}%
  \BibitemOpen
  \bibfield  {author} {\bibinfo {author} {\bibfnamefont {P.}~\bibnamefont
  {{Tarafdar}}}, \bibinfo {author} {\bibfnamefont {K.}~\bibnamefont
  {{Nobleson}}}, \bibinfo {author} {\bibfnamefont {P.}~\bibnamefont {{Rana}}},
  \bibinfo {author} {\bibfnamefont {J.}~\bibnamefont {{Singha}}}, \bibinfo
  {author} {\bibfnamefont {M.~A.}\ \bibnamefont {{Krishnakumar}}},  \emph
  {et~al.},\ }\href {\doibase 10.1017/pasa.2022.46} {\bibfield  {journal}
  {\bibinfo  {journal} {\pasa}\ }\textbf {\bibinfo {volume} {39}},\ \bibinfo
  {eid} {e053} (\bibinfo {year} {2022})},\ \Eprint
  {http://arxiv.org/abs/2206.09289} {arXiv:2206.09289 [astro-ph.IM]}
  \BibitemShut {NoStop}%
\bibitem [{\citenamefont {{Taylor}}(1992)}]{Taylor1992}%
  \BibitemOpen
  \bibfield  {author} {\bibinfo {author} {\bibfnamefont {J.~H.}\ \bibnamefont
  {{Taylor}}},\ }\href {\doibase 10.1098/rsta.1992.0088} {\bibfield  {journal}
  {\bibinfo  {journal} {Philosophical Transactions of the Royal Society of
  London Series A}\ }\textbf {\bibinfo {volume} {341}},\ \bibinfo {pages} {117}
  (\bibinfo {year} {1992})}\BibitemShut {NoStop}%
\bibitem [{\citenamefont {{Pennucci}}\ \emph {et~al.}(2014)\citenamefont
  {{Pennucci}}, \citenamefont {{Demorest}},\ and\ \citenamefont
  {{Ransom}}}]{PennucciDemorestRansom2014}%
  \BibitemOpen
  \bibfield  {author} {\bibinfo {author} {\bibfnamefont {T.~T.}\ \bibnamefont
  {{Pennucci}}}, \bibinfo {author} {\bibfnamefont {P.~B.}\ \bibnamefont
  {{Demorest}}}, \ and\ \bibinfo {author} {\bibfnamefont {S.~M.}\ \bibnamefont
  {{Ransom}}},\ }\href {\doibase 10.1088/0004-637X/790/2/93} {\bibfield
  {journal} {\bibinfo  {journal} {\apj}\ }\textbf {\bibinfo {volume} {790}},\
  \bibinfo {eid} {93} (\bibinfo {year} {2014})},\ \Eprint
  {http://arxiv.org/abs/1402.1672} {arXiv:1402.1672 [astro-ph.IM]} \BibitemShut
  {NoStop}%
\bibitem [{\citenamefont {{Pennucci}}(2019)}]{Pennucci2019}%
  \BibitemOpen
  \bibfield  {author} {\bibinfo {author} {\bibfnamefont {T.~T.}\ \bibnamefont
  {{Pennucci}}},\ }\href {\doibase 10.3847/1538-4357/aaf6ef} {\bibfield
  {journal} {\bibinfo  {journal} {\apj}\ }\textbf {\bibinfo {volume} {871}},\
  \bibinfo {eid} {34} (\bibinfo {year} {2019})},\ \Eprint
  {http://arxiv.org/abs/1812.02006} {arXiv:1812.02006 [astro-ph.HE]}
  \BibitemShut {NoStop}%
\bibitem [{\citenamefont {{Caballero}}\ \emph {et~al.}(2016)\citenamefont
  {{Caballero}}, \citenamefont {{Lee}}, \citenamefont {{Lentati}},
  \citenamefont {{Desvignes}}, \citenamefont {{Champion}}, \citenamefont
  {{Verbiest}}, \citenamefont {{Janssen}} \emph {et~al.}}]{caballero2016}%
  \BibitemOpen
  \bibfield  {author} {\bibinfo {author} {\bibfnamefont {R.~N.}\ \bibnamefont
  {{Caballero}}}, \bibinfo {author} {\bibfnamefont {K.~J.}\ \bibnamefont
  {{Lee}}}, \bibinfo {author} {\bibfnamefont {L.}~\bibnamefont {{Lentati}}},
  \bibinfo {author} {\bibfnamefont {G.}~\bibnamefont {{Desvignes}}}, \bibinfo
  {author} {\bibfnamefont {D.~J.}\ \bibnamefont {{Champion}}}, \bibinfo
  {author} {\bibfnamefont {J.~P.~W.}\ \bibnamefont {{Verbiest}}}, \bibinfo
  {author} {\bibfnamefont {G.~H.}\ \bibnamefont {{Janssen}}},  \emph {et~al.},\
  }\href {\doibase 10.1093/mnras/stw179} {\bibfield  {journal} {\bibinfo
  {journal} {\mnras}\ }\textbf {\bibinfo {volume} {457}},\ \bibinfo {pages}
  {4421} (\bibinfo {year} {2016})},\ \Eprint {http://arxiv.org/abs/1510.09194}
  {arXiv:1510.09194 [astro-ph.IM]} \BibitemShut {NoStop}%
\bibitem [{\citenamefont {{Lentati}}\ \emph {et~al.}(2016)\citenamefont
  {{Lentati}}, \citenamefont {{Shannon}}, \citenamefont {{Coles}},
  \citenamefont {{Verbiest}}, \citenamefont {{van Haasteren}}, \citenamefont
  {{Ellis}} \emph {et~al.}}]{lentati2016}%
  \BibitemOpen
  \bibfield  {author} {\bibinfo {author} {\bibfnamefont {L.}~\bibnamefont
  {{Lentati}}}, \bibinfo {author} {\bibfnamefont {R.~M.}\ \bibnamefont
  {{Shannon}}}, \bibinfo {author} {\bibfnamefont {W.~A.}\ \bibnamefont
  {{Coles}}}, \bibinfo {author} {\bibfnamefont {J.~P.~W.}\ \bibnamefont
  {{Verbiest}}}, \bibinfo {author} {\bibfnamefont {R.}~\bibnamefont {{van
  Haasteren}}}, \bibinfo {author} {\bibfnamefont {J.~A.}\ \bibnamefont
  {{Ellis}}},  \emph {et~al.},\ }\href {\doibase 10.1093/mnras/stw395}
  {\bibfield  {journal} {\bibinfo  {journal} {\mnras}\ }\textbf {\bibinfo
  {volume} {458}},\ \bibinfo {pages} {2161} (\bibinfo {year} {2016})},\ \Eprint
  {http://arxiv.org/abs/1602.05570} {arXiv:1602.05570 [astro-ph.IM]}
  \BibitemShut {NoStop}%
\bibitem [{\citenamefont {{Chen}}\ \emph {et~al.}(2021)\citenamefont {{Chen}},
  \citenamefont {{Caballero}}, \citenamefont {{Guo}}, \citenamefont
  {{Chalumeau}}, \citenamefont {{Liu}} \emph {et~al.}}]{chen2021}%
  \BibitemOpen
  \bibfield  {author} {\bibinfo {author} {\bibfnamefont {S.}~\bibnamefont
  {{Chen}}}, \bibinfo {author} {\bibfnamefont {R.~N.}\ \bibnamefont
  {{Caballero}}}, \bibinfo {author} {\bibfnamefont {Y.~J.}\ \bibnamefont
  {{Guo}}}, \bibinfo {author} {\bibfnamefont {A.}~\bibnamefont {{Chalumeau}}},
  \bibinfo {author} {\bibfnamefont {K.}~\bibnamefont {{Liu}}},  \emph
  {et~al.},\ }\href {\doibase 10.1093/mnras/stab2833} {\bibfield  {journal}
  {\bibinfo  {journal} {\mnras}\ }\textbf {\bibinfo {volume} {508}},\ \bibinfo
  {pages} {4970} (\bibinfo {year} {2021})},\ \Eprint
  {http://arxiv.org/abs/2110.13184} {arXiv:2110.13184 [astro-ph.HE]}
  \BibitemShut {NoStop}%
\bibitem [{\citenamefont {{Ellis}}\ \emph {et~al.}(2019)\citenamefont
  {{Ellis}}, \citenamefont {{Vallisneri}}, \citenamefont {{Taylor}},\ and\
  \citenamefont {{Baker}}}]{EllisVallisneri+2019}%
  \BibitemOpen
  \bibfield  {author} {\bibinfo {author} {\bibfnamefont {J.~A.}\ \bibnamefont
  {{Ellis}}}, \bibinfo {author} {\bibfnamefont {M.}~\bibnamefont
  {{Vallisneri}}}, \bibinfo {author} {\bibfnamefont {S.~R.}\ \bibnamefont
  {{Taylor}}}, \ and\ \bibinfo {author} {\bibfnamefont {P.~T.}\ \bibnamefont
  {{Baker}}},\ }\href@noop {} {\enquote {\bibinfo {title} {{ENTERPRISE:
  Enhanced Numerical Toolbox Enabling a Robust PulsaR Inference SuitE}},}\
  }\bibinfo {howpublished} {Astrophysics Source Code Library, record
  ascl:1912.015} (\bibinfo {year} {2019}),\ \Eprint
  {http://arxiv.org/abs/1912.015} {ascl:1912.015} \BibitemShut {NoStop}%
\bibitem [{\citenamefont {{Speagle}}(2020)}]{Speagle2020}%
  \BibitemOpen
  \bibfield  {author} {\bibinfo {author} {\bibfnamefont {J.~S.}\ \bibnamefont
  {{Speagle}}},\ }\href {\doibase 10.1093/mnras/staa278} {\bibfield  {journal}
  {\bibinfo  {journal} {\mnras}\ }\textbf {\bibinfo {volume} {493}},\ \bibinfo
  {pages} {3132} (\bibinfo {year} {2020})},\ \Eprint
  {http://arxiv.org/abs/1904.02180} {arXiv:1904.02180 [astro-ph.IM]}
  \BibitemShut {NoStop}%
\bibitem [{\citenamefont {{Skilling}}(2004)}]{nestedsampling}%
  \BibitemOpen
  \bibfield  {author} {\bibinfo {author} {\bibfnamefont {J.}~\bibnamefont
  {{Skilling}}},\ }in\ \href {\doibase 10.1063/1.1835238} {\emph {\bibinfo
  {booktitle} {Bayesian Inference and Maximum Entropy Methods in Science and
  Engineering: 24th International Workshop on Bayesian Inference and Maximum
  Entropy Methods in Science and Engineering}}},\ \bibinfo {series} {American
  Institute of Physics Conference Series}, Vol.\ \bibinfo {volume} {735},\
  \bibinfo {editor} {edited by\ \bibinfo {editor} {\bibfnamefont
  {R.}~\bibnamefont {{Fischer}}}, \bibinfo {editor} {\bibfnamefont
  {R.}~\bibnamefont {{Preuss}}}, \ and\ \bibinfo {editor} {\bibfnamefont
  {U.~V.}\ \bibnamefont {{Toussaint}}}}\ (\bibinfo {year} {2004})\ pp.\
  \bibinfo {pages} {395--405}\BibitemShut {NoStop}%
\bibitem [{\citenamefont {Ellis}\ and\ \citenamefont {van
  Haasteren}(2017)}]{EllisvanHaasteren2017}%
  \BibitemOpen
  \bibfield  {author} {\bibinfo {author} {\bibfnamefont {J.}~\bibnamefont
  {Ellis}}\ and\ \bibinfo {author} {\bibfnamefont {R.}~\bibnamefont {van
  Haasteren}},\ }\href {\doibase 10.5281/zenodo.1037579} {\enquote {\bibinfo
  {title} {jellis18/ptmcmcsampler: Official release},}\ } (\bibinfo {year}
  {2017})\BibitemShut {NoStop}%
\bibitem [{\citenamefont {Reddy}\ \emph {et~al.}(2017)\citenamefont {Reddy},
  \citenamefont {Kudale}, \citenamefont {Gokhale}, \citenamefont {Halagalli},
  \citenamefont {Raskar}, \citenamefont {De}, \citenamefont {Gnanaraj},
  \citenamefont {Ajith~Kumar},\ and\ \citenamefont {Gupta}}]{ReddyKudale+2017}%
  \BibitemOpen
  \bibfield  {author} {\bibinfo {author} {\bibfnamefont {S.~H.}\ \bibnamefont
  {Reddy}}, \bibinfo {author} {\bibfnamefont {S.}~\bibnamefont {Kudale}},
  \bibinfo {author} {\bibfnamefont {U.}~\bibnamefont {Gokhale}}, \bibinfo
  {author} {\bibfnamefont {I.}~\bibnamefont {Halagalli}}, \bibinfo {author}
  {\bibfnamefont {N.}~\bibnamefont {Raskar}}, \bibinfo {author} {\bibfnamefont
  {K.}~\bibnamefont {De}}, \bibinfo {author} {\bibfnamefont {S.}~\bibnamefont
  {Gnanaraj}}, \bibinfo {author} {\bibfnamefont {B.}~\bibnamefont
  {Ajith~Kumar}}, \ and\ \bibinfo {author} {\bibfnamefont {Y.}~\bibnamefont
  {Gupta}},\ }\href {\doibase 10.1142/S2251171716410117} {\bibfield  {journal}
  {\bibinfo  {journal} {Journal of Astronomical Instrumentation}\ }\textbf
  {\bibinfo {volume} {6}},\ \bibinfo {pages} {1641011} (\bibinfo {year}
  {2017})}\BibitemShut {NoStop}%
\bibitem [{\citenamefont {Susobhanan}\ \emph {et~al.}(2021)\citenamefont
  {Susobhanan}, \citenamefont {Maan}, \citenamefont {Joshi}, \citenamefont
  {Prabu}, \citenamefont {Desai} \emph {et~al.}}]{SusobhananMaan+2021}%
  \BibitemOpen
  \bibfield  {author} {\bibinfo {author} {\bibfnamefont {A.}~\bibnamefont
  {Susobhanan}}, \bibinfo {author} {\bibfnamefont {Y.}~\bibnamefont {Maan}},
  \bibinfo {author} {\bibfnamefont {B.~C.}\ \bibnamefont {Joshi}}, \bibinfo
  {author} {\bibfnamefont {T.}~\bibnamefont {Prabu}}, \bibinfo {author}
  {\bibfnamefont {S.}~\bibnamefont {Desai}},  \emph {et~al.},\ }\href {\doibase
  10.1017/pasa.2021.12} {\bibfield  {journal} {\bibinfo  {journal}
  {Publications of the Astronomical Society of Australia}\ }\textbf {\bibinfo
  {volume} {38}},\ \bibinfo {pages} {e017} (\bibinfo {year}
  {2021})}\BibitemShut {NoStop}%
\bibitem [{\citenamefont {Hobbs}\ \emph {et~al.}(2006)\citenamefont {Hobbs},
  \citenamefont {Edwards},\ and\ \citenamefont
  {Manchester}}]{HobbsEdwardsManchester2006}%
  \BibitemOpen
  \bibfield  {author} {\bibinfo {author} {\bibfnamefont {G.~B.}\ \bibnamefont
  {Hobbs}}, \bibinfo {author} {\bibfnamefont {R.~T.}\ \bibnamefont {Edwards}},
  \ and\ \bibinfo {author} {\bibfnamefont {R.~N.}\ \bibnamefont {Manchester}},\
  }\href {\doibase 10.1111/j.1365-2966.2006.10302.x} {\bibfield  {journal}
  {\bibinfo  {journal} {Monthly Notices of the Royal Astronomical Society}\
  }\textbf {\bibinfo {volume} {369}},\ \bibinfo {pages} {655} (\bibinfo {year}
  {2006})}\BibitemShut {NoStop}%
\bibitem [{\citenamefont {{Alam}}\ \emph
  {et~al.}(2021{\natexlab{a}})\citenamefont {{Alam}}, \citenamefont
  {{Arzoumanian}}, \citenamefont {{Baker}}, \citenamefont {{Blumer}},
  \citenamefont {{Bohler}}, \citenamefont {{Brazier}} \emph
  {et~al.}}]{AlamArzoumanian+2020b}%
  \BibitemOpen
  \bibfield  {author} {\bibinfo {author} {\bibfnamefont {M.~F.}\ \bibnamefont
  {{Alam}}}, \bibinfo {author} {\bibfnamefont {Z.}~\bibnamefont
  {{Arzoumanian}}}, \bibinfo {author} {\bibfnamefont {P.~T.}\ \bibnamefont
  {{Baker}}}, \bibinfo {author} {\bibfnamefont {H.}~\bibnamefont {{Blumer}}},
  \bibinfo {author} {\bibfnamefont {K.~E.}\ \bibnamefont {{Bohler}}}, \bibinfo
  {author} {\bibfnamefont {A.}~\bibnamefont {{Brazier}}},  \emph {et~al.},\
  }\href {\doibase 10.3847/1538-4365/abc6a1} {\bibfield  {journal} {\bibinfo
  {journal} {\apjs}\ }\textbf {\bibinfo {volume} {252}},\ \bibinfo {eid} {5}
  (\bibinfo {year} {2021}{\natexlab{a}})},\ \Eprint
  {http://arxiv.org/abs/2005.06495} {arXiv:2005.06495 [astro-ph.HE]}
  \BibitemShut {NoStop}%
\bibitem [{\citenamefont {Nice}\ \emph {et~al.}(2015)\citenamefont {Nice},
  \citenamefont {Demorest}, \citenamefont {Stairs}, \citenamefont {Manchester},
  \citenamefont {Taylor}, \citenamefont {Peters}, \citenamefont {Weisberg},
  \citenamefont {Irwin}, \citenamefont {Wex},\ and\ \citenamefont
  {Huang}}]{NiceDemorest+2015}%
  \BibitemOpen
  \bibfield  {author} {\bibinfo {author} {\bibfnamefont {D.}~\bibnamefont
  {Nice}}, \bibinfo {author} {\bibfnamefont {P.}~\bibnamefont {Demorest}},
  \bibinfo {author} {\bibfnamefont {I.}~\bibnamefont {Stairs}}, \bibinfo
  {author} {\bibfnamefont {R.}~\bibnamefont {Manchester}}, \bibinfo {author}
  {\bibfnamefont {J.}~\bibnamefont {Taylor}}, \bibinfo {author} {\bibfnamefont
  {W.}~\bibnamefont {Peters}}, \bibinfo {author} {\bibfnamefont
  {J.}~\bibnamefont {Weisberg}}, \bibinfo {author} {\bibfnamefont
  {A.}~\bibnamefont {Irwin}}, \bibinfo {author} {\bibfnamefont
  {N.}~\bibnamefont {Wex}}, \ and\ \bibinfo {author} {\bibfnamefont
  {Y.}~\bibnamefont {Huang}},\ }\href {http://tempo.sourceforge.net/} {\enquote
  {\bibinfo {title} {{Tempo: Pulsar timing data analysis}},}\ } (\bibinfo
  {year} {2015})\BibitemShut {NoStop}%
\bibitem [{\citenamefont {{Hazboun}}\ \emph {et~al.}(2020)\citenamefont
  {{Hazboun}}, \citenamefont {{Simon}}, \citenamefont {{Siemens}},\ and\
  \citenamefont {{Romano}}}]{hazboun2020}%
  \BibitemOpen
  \bibfield  {author} {\bibinfo {author} {\bibfnamefont {J.~S.}\ \bibnamefont
  {{Hazboun}}}, \bibinfo {author} {\bibfnamefont {J.}~\bibnamefont {{Simon}}},
  \bibinfo {author} {\bibfnamefont {X.}~\bibnamefont {{Siemens}}}, \ and\
  \bibinfo {author} {\bibfnamefont {J.~D.}\ \bibnamefont {{Romano}}},\ }\href
  {\doibase 10.3847/2041-8213/abca92} {\bibfield  {journal} {\bibinfo
  {journal} {\apjl}\ }\textbf {\bibinfo {volume} {905}},\ \bibinfo {eid} {L6}
  (\bibinfo {year} {2020})},\ \Eprint {http://arxiv.org/abs/2009.05143}
  {arXiv:2009.05143 [astro-ph.IM]} \BibitemShut {NoStop}%
\bibitem [{\citenamefont {{van Haasteren}}\ and\ \citenamefont
  {{Vallisneri}}(2014)}]{vhaasteren+2014}%
  \BibitemOpen
  \bibfield  {author} {\bibinfo {author} {\bibfnamefont {R.}~\bibnamefont {{van
  Haasteren}}}\ and\ \bibinfo {author} {\bibfnamefont {M.}~\bibnamefont
  {{Vallisneri}}},\ }\href {\doibase 10.1103/PhysRevD.90.104012} {\bibfield
  {journal} {\bibinfo  {journal} {\prd}\ }\textbf {\bibinfo {volume} {90}},\
  \bibinfo {eid} {104012} (\bibinfo {year} {2014})},\ \Eprint
  {http://arxiv.org/abs/1407.1838} {arXiv:1407.1838 [gr-qc]} \BibitemShut
  {NoStop}%
\bibitem [{\citenamefont {{Liu}}\ \emph {et~al.}(2012)\citenamefont {{Liu}},
  \citenamefont {{Keane}}, \citenamefont {{Lee}}, \citenamefont {{Kramer}},
  \citenamefont {{Cordes}},\ and\ \citenamefont {{Purver}}}]{Liu.et.al}%
  \BibitemOpen
  \bibfield  {author} {\bibinfo {author} {\bibfnamefont {K.}~\bibnamefont
  {{Liu}}}, \bibinfo {author} {\bibfnamefont {E.~F.}\ \bibnamefont {{Keane}}},
  \bibinfo {author} {\bibfnamefont {K.~J.}\ \bibnamefont {{Lee}}}, \bibinfo
  {author} {\bibfnamefont {M.}~\bibnamefont {{Kramer}}}, \bibinfo {author}
  {\bibfnamefont {J.~M.}\ \bibnamefont {{Cordes}}}, \ and\ \bibinfo {author}
  {\bibfnamefont {M.~B.}\ \bibnamefont {{Purver}}},\ }\href {\doibase
  10.1111/j.1365-2966.2011.20041.x} {\bibfield  {journal} {\bibinfo  {journal}
  {\mnras}\ }\textbf {\bibinfo {volume} {420}},\ \bibinfo {pages} {361}
  (\bibinfo {year} {2012})},\ \Eprint {http://arxiv.org/abs/1110.4759}
  {arXiv:1110.4759 [astro-ph.HE]} \BibitemShut {NoStop}%
\bibitem [{\citenamefont {{Shannon}}\ \emph {et~al.}(2014)\citenamefont
  {{Shannon}}, \citenamefont {{Os{\l}owski}}, \citenamefont {{Dai}},
  \citenamefont {{Bailes}}, \citenamefont {{Hobbs}}, \citenamefont
  {{Manchester}} \emph {et~al.}}]{shannon2014}%
  \BibitemOpen
  \bibfield  {author} {\bibinfo {author} {\bibfnamefont {R.~M.}\ \bibnamefont
  {{Shannon}}}, \bibinfo {author} {\bibfnamefont {S.}~\bibnamefont
  {{Os{\l}owski}}}, \bibinfo {author} {\bibfnamefont {S.}~\bibnamefont
  {{Dai}}}, \bibinfo {author} {\bibfnamefont {M.}~\bibnamefont {{Bailes}}},
  \bibinfo {author} {\bibfnamefont {G.}~\bibnamefont {{Hobbs}}}, \bibinfo
  {author} {\bibfnamefont {R.~N.}\ \bibnamefont {{Manchester}}},  \emph
  {et~al.},\ }\href {\doibase 10.1093/mnras/stu1213} {\bibfield  {journal}
  {\bibinfo  {journal} {\mnras}\ }\textbf {\bibinfo {volume} {443}},\ \bibinfo
  {pages} {1463} (\bibinfo {year} {2014})},\ \Eprint
  {http://arxiv.org/abs/1406.4716} {arXiv:1406.4716 [astro-ph.SR]} \BibitemShut
  {NoStop}%
\bibitem [{\citenamefont {{Lam}}\ \emph {et~al.}(2016)\citenamefont {{Lam}},
  \citenamefont {{Cordes}}, \citenamefont {{Chatterjee}}, \citenamefont
  {{Arzoumanian}}, \citenamefont {{Crowter}}, \citenamefont {{Demorest}} \emph
  {et~al.}}]{lam2016}%
  \BibitemOpen
  \bibfield  {author} {\bibinfo {author} {\bibfnamefont {M.~T.}\ \bibnamefont
  {{Lam}}}, \bibinfo {author} {\bibfnamefont {J.~M.}\ \bibnamefont {{Cordes}}},
  \bibinfo {author} {\bibfnamefont {S.}~\bibnamefont {{Chatterjee}}}, \bibinfo
  {author} {\bibfnamefont {Z.}~\bibnamefont {{Arzoumanian}}}, \bibinfo {author}
  {\bibfnamefont {K.}~\bibnamefont {{Crowter}}}, \bibinfo {author}
  {\bibfnamefont {P.~B.}\ \bibnamefont {{Demorest}}},  \emph {et~al.},\ }\href
  {\doibase 10.3847/0004-637X/819/2/155} {\bibfield  {journal} {\bibinfo
  {journal} {\apj}\ }\textbf {\bibinfo {volume} {819}},\ \bibinfo {eid} {155}
  (\bibinfo {year} {2016})},\ \Eprint {http://arxiv.org/abs/1512.08326}
  {arXiv:1512.08326 [astro-ph.IM]} \BibitemShut {NoStop}%
\bibitem [{\citenamefont {{Lentati}}\ \emph
  {et~al.}(2014{\natexlab{a}})\citenamefont {{Lentati}}, \citenamefont
  {{Hobson}},\ and\ \citenamefont {{Alexander}}}]{lentati2014}%
  \BibitemOpen
  \bibfield  {author} {\bibinfo {author} {\bibfnamefont {L.}~\bibnamefont
  {{Lentati}}}, \bibinfo {author} {\bibfnamefont {M.~P.}\ \bibnamefont
  {{Hobson}}}, \ and\ \bibinfo {author} {\bibfnamefont {P.}~\bibnamefont
  {{Alexander}}},\ }\href {\doibase 10.1093/mnras/stu1721} {\bibfield
  {journal} {\bibinfo  {journal} {\mnras}\ }\textbf {\bibinfo {volume} {444}},\
  \bibinfo {pages} {3863} (\bibinfo {year} {2014}{\natexlab{a}})},\ \Eprint
  {http://arxiv.org/abs/1405.2460} {arXiv:1405.2460 [astro-ph.IM]} \BibitemShut
  {NoStop}%
\bibitem [{\citenamefont {{Vallisneri}}\ and\ \citenamefont {{van
  Haasteren}}(2017)}]{vallisneri2017}%
  \BibitemOpen
  \bibfield  {author} {\bibinfo {author} {\bibfnamefont {M.}~\bibnamefont
  {{Vallisneri}}}\ and\ \bibinfo {author} {\bibfnamefont {R.}~\bibnamefont
  {{van Haasteren}}},\ }\href {\doibase 10.1093/mnras/stx069} {\bibfield
  {journal} {\bibinfo  {journal} {\mnras}\ }\textbf {\bibinfo {volume} {466}},\
  \bibinfo {pages} {4954} (\bibinfo {year} {2017})},\ \Eprint
  {http://arxiv.org/abs/1609.02144} {arXiv:1609.02144 [astro-ph.IM]}
  \BibitemShut {NoStop}%
\bibitem [{\citenamefont {{Shannon}}\ and\ \citenamefont
  {{Cordes}}(2010)}]{shannon2010}%
  \BibitemOpen
  \bibfield  {author} {\bibinfo {author} {\bibfnamefont {R.~M.}\ \bibnamefont
  {{Shannon}}}\ and\ \bibinfo {author} {\bibfnamefont {J.~M.}\ \bibnamefont
  {{Cordes}}},\ }\href {\doibase 10.1088/0004-637X/725/2/1607} {\bibfield
  {journal} {\bibinfo  {journal} {\apj}\ }\textbf {\bibinfo {volume} {725}},\
  \bibinfo {pages} {1607} (\bibinfo {year} {2010})},\ \Eprint
  {http://arxiv.org/abs/1010.4794} {arXiv:1010.4794 [astro-ph.SR]} \BibitemShut
  {NoStop}%
\bibitem [{\citenamefont {{Lentati}}\ \emph
  {et~al.}(2014{\natexlab{b}})\citenamefont {{Lentati}}, \citenamefont
  {{Alexander}}, \citenamefont {{Hobson}}, \citenamefont {{Feroz}},
  \citenamefont {{van Haasteren}}, \citenamefont {{Lee}},\ and\ \citenamefont
  {{Shannon}}}]{temponest2014}%
  \BibitemOpen
  \bibfield  {author} {\bibinfo {author} {\bibfnamefont {L.}~\bibnamefont
  {{Lentati}}}, \bibinfo {author} {\bibfnamefont {P.}~\bibnamefont
  {{Alexander}}}, \bibinfo {author} {\bibfnamefont {M.~P.}\ \bibnamefont
  {{Hobson}}}, \bibinfo {author} {\bibfnamefont {F.}~\bibnamefont {{Feroz}}},
  \bibinfo {author} {\bibfnamefont {R.}~\bibnamefont {{van Haasteren}}},
  \bibinfo {author} {\bibfnamefont {K.~J.}\ \bibnamefont {{Lee}}}, \ and\
  \bibinfo {author} {\bibfnamefont {R.~M.}\ \bibnamefont {{Shannon}}},\ }\href
  {\doibase 10.1093/mnras/stt2122} {\bibfield  {journal} {\bibinfo  {journal}
  {\mnras}\ }\textbf {\bibinfo {volume} {437}},\ \bibinfo {pages} {3004}
  (\bibinfo {year} {2014}{\natexlab{b}})},\ \Eprint
  {http://arxiv.org/abs/1310.2120} {arXiv:1310.2120 [astro-ph.IM]} \BibitemShut
  {NoStop}%
\bibitem [{\citenamefont {{Cordes}}\ and\ \citenamefont
  {{Downs}}(1985)}]{cordes1985}%
  \BibitemOpen
  \bibfield  {author} {\bibinfo {author} {\bibfnamefont {J.~M.}\ \bibnamefont
  {{Cordes}}}\ and\ \bibinfo {author} {\bibfnamefont {G.~S.}\ \bibnamefont
  {{Downs}}},\ }\href {\doibase 10.1086/191076} {\bibfield  {journal} {\bibinfo
   {journal} {\apjs}\ }\textbf {\bibinfo {volume} {59}},\ \bibinfo {pages}
  {343} (\bibinfo {year} {1985})}\BibitemShut {NoStop}%
\bibitem [{\citenamefont {{D'Alessandro}}\ \emph {et~al.}(1995)\citenamefont
  {{D'Alessandro}}, \citenamefont {{McCulloch}}, \citenamefont {{Hamilton}},\
  and\ \citenamefont {{Deshpande}}}]{allessandro1995}%
  \BibitemOpen
  \bibfield  {author} {\bibinfo {author} {\bibfnamefont {F.}~\bibnamefont
  {{D'Alessandro}}}, \bibinfo {author} {\bibfnamefont {P.~M.}\ \bibnamefont
  {{McCulloch}}}, \bibinfo {author} {\bibfnamefont {P.~A.}\ \bibnamefont
  {{Hamilton}}}, \ and\ \bibinfo {author} {\bibfnamefont {A.~A.}\ \bibnamefont
  {{Deshpande}}},\ }\href {\doibase 10.1093/mnras/277.3.1033} {\bibfield
  {journal} {\bibinfo  {journal} {\mnras}\ }\textbf {\bibinfo {volume} {277}},\
  \bibinfo {pages} {1033} (\bibinfo {year} {1995})}\BibitemShut {NoStop}%
\bibitem [{\citenamefont {{Alam}}\ \emph
  {et~al.}(2021{\natexlab{b}})\citenamefont {{Alam}}, \citenamefont
  {{Arzoumanian}}, \citenamefont {{Baker}}, \citenamefont {{Blumer}},
  \citenamefont {{Bohler}}, \citenamefont {{Brazier}} \emph
  {et~al.}}]{alam2021}%
  \BibitemOpen
  \bibfield  {author} {\bibinfo {author} {\bibfnamefont {M.~F.}\ \bibnamefont
  {{Alam}}}, \bibinfo {author} {\bibfnamefont {Z.}~\bibnamefont
  {{Arzoumanian}}}, \bibinfo {author} {\bibfnamefont {P.~T.}\ \bibnamefont
  {{Baker}}}, \bibinfo {author} {\bibfnamefont {H.}~\bibnamefont {{Blumer}}},
  \bibinfo {author} {\bibfnamefont {K.~E.}\ \bibnamefont {{Bohler}}}, \bibinfo
  {author} {\bibfnamefont {A.}~\bibnamefont {{Brazier}}},  \emph {et~al.},\
  }\href {\doibase 10.3847/1538-4365/abc6a0} {\bibfield  {journal} {\bibinfo
  {journal} {\apjs}\ }\textbf {\bibinfo {volume} {252}},\ \bibinfo {eid} {4}
  (\bibinfo {year} {2021}{\natexlab{b}})},\ \Eprint
  {http://arxiv.org/abs/2005.06490} {arXiv:2005.06490 [astro-ph.HE]}
  \BibitemShut {NoStop}%
\bibitem [{\citenamefont {Lorimer}\ and\ \citenamefont
  {Kramer}(2004)}]{LorimerKramer2004}%
  \BibitemOpen
  \bibfield  {author} {\bibinfo {author} {\bibfnamefont {D.}~\bibnamefont
  {Lorimer}}\ and\ \bibinfo {author} {\bibfnamefont {M.}~\bibnamefont
  {Kramer}},\ }\href@noop {} {\emph {\bibinfo {title} {Handbook of pulsar
  astronomy}}}\ (\bibinfo  {publisher} {Cambridge University Press},\ \bibinfo
  {year} {2004})\BibitemShut {NoStop}%
\bibitem [{\citenamefont {{You}}\ \emph {et~al.}(2007)\citenamefont {{You}},
  \citenamefont {{Hobbs}}, \citenamefont {{Coles}}, \citenamefont
  {{Manchester}}, \citenamefont {{Edwards}}, \citenamefont {{Bailes}} \emph
  {et~al.}}]{you2007}%
  \BibitemOpen
  \bibfield  {author} {\bibinfo {author} {\bibfnamefont {X.~P.}\ \bibnamefont
  {{You}}}, \bibinfo {author} {\bibfnamefont {G.}~\bibnamefont {{Hobbs}}},
  \bibinfo {author} {\bibfnamefont {W.~A.}\ \bibnamefont {{Coles}}}, \bibinfo
  {author} {\bibfnamefont {R.~N.}\ \bibnamefont {{Manchester}}}, \bibinfo
  {author} {\bibfnamefont {R.}~\bibnamefont {{Edwards}}}, \bibinfo {author}
  {\bibfnamefont {M.}~\bibnamefont {{Bailes}}},  \emph {et~al.},\ }\href
  {\doibase 10.1111/j.1365-2966.2007.11617.x} {\bibfield  {journal} {\bibinfo
  {journal} {\mnras}\ }\textbf {\bibinfo {volume} {378}},\ \bibinfo {pages}
  {493} (\bibinfo {year} {2007})},\ \Eprint
  {http://arxiv.org/abs/astro-ph/0702366} {arXiv:astro-ph/0702366 [astro-ph]}
  \BibitemShut {NoStop}%
\bibitem [{\citenamefont {{Keith}}\ \emph {et~al.}(2013)\citenamefont
  {{Keith}}, \citenamefont {{Coles}}, \citenamefont {{Shannon}}, \citenamefont
  {{Hobbs}}, \citenamefont {{Manchester}}, \citenamefont {{Bailes}} \emph
  {et~al.}}]{keith2013}%
  \BibitemOpen
  \bibfield  {author} {\bibinfo {author} {\bibfnamefont {M.~J.}\ \bibnamefont
  {{Keith}}}, \bibinfo {author} {\bibfnamefont {W.}~\bibnamefont {{Coles}}},
  \bibinfo {author} {\bibfnamefont {R.~M.}\ \bibnamefont {{Shannon}}}, \bibinfo
  {author} {\bibfnamefont {G.~B.}\ \bibnamefont {{Hobbs}}}, \bibinfo {author}
  {\bibfnamefont {R.~N.}\ \bibnamefont {{Manchester}}}, \bibinfo {author}
  {\bibfnamefont {M.}~\bibnamefont {{Bailes}}},  \emph {et~al.},\ }\href
  {\doibase 10.1093/mnras/sts486} {\bibfield  {journal} {\bibinfo  {journal}
  {\mnras}\ }\textbf {\bibinfo {volume} {429}},\ \bibinfo {pages} {2161}
  (\bibinfo {year} {2013})},\ \Eprint {http://arxiv.org/abs/1211.5887}
  {arXiv:1211.5887 [astro-ph.GA]} \BibitemShut {NoStop}%
\bibitem [{\citenamefont {{Goncharov}}\ \emph
  {et~al.}(2021{\natexlab{b}})\citenamefont {{Goncharov}}, \citenamefont
  {{Shannon}}, \citenamefont {{Reardon}}, \citenamefont {{Hobbs}},
  \citenamefont {{Zic}} \emph {et~al.}}]{GoncharovShannon+2021}%
  \BibitemOpen
  \bibfield  {author} {\bibinfo {author} {\bibfnamefont {B.}~\bibnamefont
  {{Goncharov}}}, \bibinfo {author} {\bibfnamefont {R.~M.}\ \bibnamefont
  {{Shannon}}}, \bibinfo {author} {\bibfnamefont {D.~J.}\ \bibnamefont
  {{Reardon}}}, \bibinfo {author} {\bibfnamefont {G.}~\bibnamefont {{Hobbs}}},
  \bibinfo {author} {\bibfnamefont {A.}~\bibnamefont {{Zic}}},  \emph
  {et~al.},\ }\href {\doibase 10.3847/2041-8213/ac17f4} {\bibfield  {journal}
  {\bibinfo  {journal} {\apjl}\ }\textbf {\bibinfo {volume} {917}},\ \bibinfo
  {eid} {L19} (\bibinfo {year} {2021}{\natexlab{b}})},\ \Eprint
  {http://arxiv.org/abs/2107.12112} {arXiv:2107.12112 [astro-ph.HE]}
  \BibitemShut {NoStop}%
\bibitem [{\citenamefont {{Sharma}}(2017)}]{Sanjib}%
  \BibitemOpen
  \bibfield  {author} {\bibinfo {author} {\bibfnamefont {S.}~\bibnamefont
  {{Sharma}}},\ }\href {\doibase 10.1146/annurev-astro-082214-122339}
  {\bibfield  {journal} {\bibinfo  {journal} {\araa}\ }\textbf {\bibinfo
  {volume} {55}},\ \bibinfo {pages} {213} (\bibinfo {year} {2017})},\ \Eprint
  {http://arxiv.org/abs/1706.01629} {arXiv:1706.01629 [astro-ph.IM]}
  \BibitemShut {NoStop}%
\bibitem [{\citenamefont {{Kerscher}}\ and\ \citenamefont
  {{Weller}}(2019)}]{Weller}%
  \BibitemOpen
  \bibfield  {author} {\bibinfo {author} {\bibfnamefont {M.}~\bibnamefont
  {{Kerscher}}}\ and\ \bibinfo {author} {\bibfnamefont {J.}~\bibnamefont
  {{Weller}}},\ }\href {\doibase 10.21468/SciPostPhysLectNotes.9} {\bibfield
  {journal} {\bibinfo  {journal} {SciPost Physics Lecture Notes}\ }\textbf
  {\bibinfo {volume} {9}} (\bibinfo {year} {2019}),\
  10.21468/SciPostPhysLectNotes.9},\ \Eprint {http://arxiv.org/abs/1901.07726}
  {arXiv:1901.07726 [astro-ph.CO]} \BibitemShut {NoStop}%
\bibitem [{\citenamefont {{Krishak}}\ and\ \citenamefont
  {{Desai}}(2020)}]{Krishak}%
  \BibitemOpen
  \bibfield  {author} {\bibinfo {author} {\bibfnamefont {A.}~\bibnamefont
  {{Krishak}}}\ and\ \bibinfo {author} {\bibfnamefont {S.}~\bibnamefont
  {{Desai}}},\ }\href {\doibase 10.1088/1475-7516/2020/07/006} {\bibfield
  {journal} {\bibinfo  {journal} {\jcap}\ }\textbf {\bibinfo {volume} {2020}},\
  \bibinfo {eid} {006} (\bibinfo {year} {2020})},\ \Eprint
  {http://arxiv.org/abs/2003.10127} {arXiv:2003.10127 [gr-qc]} \BibitemShut
  {NoStop}%
\bibitem [{\citenamefont {Van~Haasteren}\ \emph {et~al.}(2009)\citenamefont
  {Van~Haasteren}, \citenamefont {Levin}, \citenamefont {McDonald},\ and\
  \citenamefont {Lu}}]{haasteren2009}%
  \BibitemOpen
  \bibfield  {author} {\bibinfo {author} {\bibfnamefont {R.}~\bibnamefont
  {Van~Haasteren}}, \bibinfo {author} {\bibfnamefont {Y.}~\bibnamefont
  {Levin}}, \bibinfo {author} {\bibfnamefont {P.}~\bibnamefont {McDonald}}, \
  and\ \bibinfo {author} {\bibfnamefont {T.}~\bibnamefont {Lu}},\ }\href
  {\doibase 10.1111/j.1365-2966.2009.14590.x} {\bibfield  {journal} {\bibinfo
  {journal} {Monthly Notices of the Royal Astronomical Society}\ }\textbf
  {\bibinfo {volume} {395}},\ \bibinfo {pages} {1005} (\bibinfo {year}
  {2009})},\ \Eprint
  {http://arxiv.org/abs/https://academic.oup.com/mnras/article-pdf/395/2/1005/4900586/mnras0395-1005.pdf}
  {https://academic.oup.com/mnras/article-pdf/395/2/1005/4900586/mnras0395-1005.pdf}
  \BibitemShut {NoStop}%
\bibitem [{\citenamefont {van Haasteren}\ and\ \citenamefont
  {Levin}(2012)}]{haasteren2012}%
  \BibitemOpen
  \bibfield  {author} {\bibinfo {author} {\bibfnamefont {R.}~\bibnamefont {van
  Haasteren}}\ and\ \bibinfo {author} {\bibfnamefont {Y.}~\bibnamefont
  {Levin}},\ }\href {\doibase 10.1093/mnras/sts097} {\bibfield  {journal}
  {\bibinfo  {journal} {Monthly Notices of the Royal Astronomical Society}\
  }\textbf {\bibinfo {volume} {428}},\ \bibinfo {pages} {1147} (\bibinfo {year}
  {2012})},\ \Eprint
  {http://arxiv.org/abs/https://academic.oup.com/mnras/article-pdf/428/2/1147/3206977/sts097.pdf}
  {https://academic.oup.com/mnras/article-pdf/428/2/1147/3206977/sts097.pdf}
  \BibitemShut {NoStop}%
\bibitem [{\citenamefont {{Taylor}}\ \emph {et~al.}(2017)\citenamefont
  {{Taylor}}, \citenamefont {{Lentati}}, \citenamefont {{Babak}}, \citenamefont
  {{Brem}}, \citenamefont {{Gair}}, \citenamefont {{Sesana}},\ and\
  \citenamefont {{Vecchio}}}]{taylor2017}%
  \BibitemOpen
  \bibfield  {author} {\bibinfo {author} {\bibfnamefont {S.~R.}\ \bibnamefont
  {{Taylor}}}, \bibinfo {author} {\bibfnamefont {L.}~\bibnamefont {{Lentati}}},
  \bibinfo {author} {\bibfnamefont {S.}~\bibnamefont {{Babak}}}, \bibinfo
  {author} {\bibfnamefont {P.}~\bibnamefont {{Brem}}}, \bibinfo {author}
  {\bibfnamefont {J.~R.}\ \bibnamefont {{Gair}}}, \bibinfo {author}
  {\bibfnamefont {A.}~\bibnamefont {{Sesana}}}, \ and\ \bibinfo {author}
  {\bibfnamefont {A.}~\bibnamefont {{Vecchio}}},\ }\href {\doibase
  10.1103/PhysRevD.95.042002} {\bibfield  {journal} {\bibinfo  {journal}
  {\prd}\ }\textbf {\bibinfo {volume} {95}},\ \bibinfo {eid} {042002} (\bibinfo
  {year} {2017})},\ \Eprint {http://arxiv.org/abs/1606.09180} {arXiv:1606.09180
  [astro-ph.IM]} \BibitemShut {NoStop}%
\bibitem [{\citenamefont {{Lam}}\ \emph {et~al.}(2018)\citenamefont {{Lam}},
  \citenamefont {{Ellis}}, \citenamefont {{Grillo}}, \citenamefont {{Jones}},
  \citenamefont {{Hazboun}}, \citenamefont {{Brook}} \emph
  {et~al.}}]{lam+2018}%
  \BibitemOpen
  \bibfield  {author} {\bibinfo {author} {\bibfnamefont {M.~T.}\ \bibnamefont
  {{Lam}}}, \bibinfo {author} {\bibfnamefont {J.~A.}\ \bibnamefont {{Ellis}}},
  \bibinfo {author} {\bibfnamefont {G.}~\bibnamefont {{Grillo}}}, \bibinfo
  {author} {\bibfnamefont {M.~L.}\ \bibnamefont {{Jones}}}, \bibinfo {author}
  {\bibfnamefont {J.~S.}\ \bibnamefont {{Hazboun}}}, \bibinfo {author}
  {\bibfnamefont {P.~R.}\ \bibnamefont {{Brook}}},  \emph {et~al.},\ }\href
  {\doibase 10.3847/1538-4357/aac770} {\bibfield  {journal} {\bibinfo
  {journal} {\apj}\ }\textbf {\bibinfo {volume} {861}},\ \bibinfo {eid} {132}
  (\bibinfo {year} {2018})},\ \Eprint {http://arxiv.org/abs/1712.03651}
  {arXiv:1712.03651 [astro-ph.HE]} \BibitemShut {NoStop}%
\bibitem [{\citenamefont {{Singha}}\ \emph {et~al.}(2021)\citenamefont
  {{Singha}}, \citenamefont {{Surnis}}, \citenamefont {{Joshi}}, \citenamefont
  {{Tarafdar}}, \citenamefont {{Rana}}, \citenamefont {{Susobhanan}} \emph
  {et~al.}}]{singha+2021}%
  \BibitemOpen
  \bibfield  {author} {\bibinfo {author} {\bibfnamefont {J.}~\bibnamefont
  {{Singha}}}, \bibinfo {author} {\bibfnamefont {M.~P.}\ \bibnamefont
  {{Surnis}}}, \bibinfo {author} {\bibfnamefont {B.~C.}\ \bibnamefont
  {{Joshi}}}, \bibinfo {author} {\bibfnamefont {P.}~\bibnamefont {{Tarafdar}}},
  \bibinfo {author} {\bibfnamefont {P.}~\bibnamefont {{Rana}}}, \bibinfo
  {author} {\bibfnamefont {A.}~\bibnamefont {{Susobhanan}}},  \emph {et~al.},\
  }\href {\doibase 10.1093/mnrasl/slab098} {\bibfield  {journal} {\bibinfo
  {journal} {\mnras}\ }\textbf {\bibinfo {volume} {507}},\ \bibinfo {pages}
  {L57} (\bibinfo {year} {2021})},\ \Eprint {http://arxiv.org/abs/2107.04607}
  {arXiv:2107.04607 [astro-ph.HE]} \BibitemShut {NoStop}%
\bibitem [{\citenamefont {{Stinebring}}\ \emph {et~al.}(1992)\citenamefont
  {{Stinebring}}, \citenamefont {{Kaspi}}, \citenamefont {{Nice}},
  \citenamefont {{Ryba}}, \citenamefont {{Taylor}}, \citenamefont
  {{Thorsett}},\ and\ \citenamefont {{Hankins}}}]{krt1992}%
  \BibitemOpen
  \bibfield  {author} {\bibinfo {author} {\bibfnamefont {D.~R.}\ \bibnamefont
  {{Stinebring}}}, \bibinfo {author} {\bibfnamefont {V.~M.}\ \bibnamefont
  {{Kaspi}}}, \bibinfo {author} {\bibfnamefont {D.~J.}\ \bibnamefont {{Nice}}},
  \bibinfo {author} {\bibfnamefont {M.~F.}\ \bibnamefont {{Ryba}}}, \bibinfo
  {author} {\bibfnamefont {J.~H.}\ \bibnamefont {{Taylor}}}, \bibinfo {author}
  {\bibfnamefont {S.~E.}\ \bibnamefont {{Thorsett}}}, \ and\ \bibinfo {author}
  {\bibfnamefont {T.~H.}\ \bibnamefont {{Hankins}}},\ }\href {\doibase
  10.1063/1.1143763} {\bibfield  {journal} {\bibinfo  {journal} {Review of
  Scientific Instruments}\ }\textbf {\bibinfo {volume} {63}},\ \bibinfo {pages}
  {3551} (\bibinfo {year} {1992})}\BibitemShut {NoStop}%
\bibitem [{\citenamefont {Joshi}\ and\ \citenamefont
  {Ramakrishna}(2006)}]{joshirama2006}%
  \BibitemOpen
  \bibfield  {author} {\bibinfo {author} {\bibfnamefont {B.~C.}\ \bibnamefont
  {Joshi}}\ and\ \bibinfo {author} {\bibfnamefont {S.}~\bibnamefont
  {Ramakrishna}},\ }\href@noop {} {\bibfield  {journal} {\bibinfo  {journal}
  {Bull. Astron. Soc. India}\ }\textbf {\bibinfo {volume} {34}},\ \bibinfo
  {pages} {401} (\bibinfo {year} {2006})},\ \Eprint
  {http://arxiv.org/abs/astro-ph/0611331} {arXiv:astro-ph/0611331} \BibitemShut
  {NoStop}%
\bibitem [{\citenamefont {{Levin}}\ \emph {et~al.}(2016)\citenamefont
  {{Levin}}, \citenamefont {{McLaughlin}}, \citenamefont {{Jones}},
  \citenamefont {{Cordes}}, \citenamefont {{Stinebring}}, \citenamefont
  {{Chatterjee}} \emph {et~al.}}]{levin2016}%
  \BibitemOpen
  \bibfield  {author} {\bibinfo {author} {\bibfnamefont {L.}~\bibnamefont
  {{Levin}}}, \bibinfo {author} {\bibfnamefont {M.~A.}\ \bibnamefont
  {{McLaughlin}}}, \bibinfo {author} {\bibfnamefont {G.}~\bibnamefont
  {{Jones}}}, \bibinfo {author} {\bibfnamefont {J.~M.}\ \bibnamefont
  {{Cordes}}}, \bibinfo {author} {\bibfnamefont {D.~R.}\ \bibnamefont
  {{Stinebring}}}, \bibinfo {author} {\bibfnamefont {S.}~\bibnamefont
  {{Chatterjee}}},  \emph {et~al.},\ }\href {\doibase
  10.3847/0004-637X/818/2/166} {\bibfield  {journal} {\bibinfo  {journal}
  {\apj}\ }\textbf {\bibinfo {volume} {818}},\ \bibinfo {eid} {166} (\bibinfo
  {year} {2016})},\ \Eprint {http://arxiv.org/abs/1601.04490} {arXiv:1601.04490
  [astro-ph.HE]} \BibitemShut {NoStop}%
\bibitem [{\citenamefont {Krishnakumar}\ \emph {et~al.}(2019)\citenamefont
  {Krishnakumar}, \citenamefont {Maan}, \citenamefont {Joshi},\ and\
  \citenamefont {Manoharan}}]{kk2019}%
  \BibitemOpen
  \bibfield  {author} {\bibinfo {author} {\bibfnamefont {M.~A.}\ \bibnamefont
  {Krishnakumar}}, \bibinfo {author} {\bibfnamefont {Y.}~\bibnamefont {Maan}},
  \bibinfo {author} {\bibfnamefont {B.~C.}\ \bibnamefont {Joshi}}, \ and\
  \bibinfo {author} {\bibfnamefont {P.~K.}\ \bibnamefont {Manoharan}},\ }\href
  {\doibase 10.3847/1538-4357/ab20c5} {\bibfield  {journal} {\bibinfo
  {journal} {The Astrophysical Journal}\ }\textbf {\bibinfo {volume} {878}},\
  \bibinfo {pages} {130} (\bibinfo {year} {2019})}\BibitemShut {NoStop}%
\bibitem [{\citenamefont {{Turner}}\ \emph {et~al.}(2021)\citenamefont
  {{Turner}}, \citenamefont {{McLaughlin}}, \citenamefont {{Cordes}},
  \citenamefont {{Lam}}, \citenamefont {{Shapiro-Albert}}, \citenamefont
  {{Stinebring}} \emph {et~al.}}]{turner2021}%
  \BibitemOpen
  \bibfield  {author} {\bibinfo {author} {\bibfnamefont {J.~E.}\ \bibnamefont
  {{Turner}}}, \bibinfo {author} {\bibfnamefont {M.~A.}\ \bibnamefont
  {{McLaughlin}}}, \bibinfo {author} {\bibfnamefont {J.~M.}\ \bibnamefont
  {{Cordes}}}, \bibinfo {author} {\bibfnamefont {M.~T.}\ \bibnamefont {{Lam}}},
  \bibinfo {author} {\bibfnamefont {B.~J.}\ \bibnamefont {{Shapiro-Albert}}},
  \bibinfo {author} {\bibfnamefont {D.~R.}\ \bibnamefont {{Stinebring}}},
  \emph {et~al.},\ }\href {\doibase 10.3847/1538-4357/abfafe} {\bibfield
  {journal} {\bibinfo  {journal} {\apj}\ }\textbf {\bibinfo {volume} {917}},\
  \bibinfo {eid} {10} (\bibinfo {year} {2021})},\ \Eprint
  {http://arxiv.org/abs/2012.09884} {arXiv:2012.09884 [astro-ph.HE]}
  \BibitemShut {NoStop}%
\bibitem [{\citenamefont {{Paladi}}\ \emph {et~al.}(2023)\citenamefont
  {{Paladi}}, \citenamefont {{Dwivedi}}, \citenamefont {{Rana}}, \citenamefont
  {{K}}, \citenamefont {{Susobhanan}}, \citenamefont {{Joshi}} \emph
  {et~al.}}]{Paladi2023}%
  \BibitemOpen
  \bibfield  {author} {\bibinfo {author} {\bibfnamefont {A.~K.}\ \bibnamefont
  {{Paladi}}}, \bibinfo {author} {\bibfnamefont {C.}~\bibnamefont {{Dwivedi}}},
  \bibinfo {author} {\bibfnamefont {P.}~\bibnamefont {{Rana}}}, \bibinfo
  {author} {\bibfnamefont {N.}~\bibnamefont {{K}}}, \bibinfo {author}
  {\bibfnamefont {A.}~\bibnamefont {{Susobhanan}}}, \bibinfo {author}
  {\bibfnamefont {B.~C.}\ \bibnamefont {{Joshi}}},  \emph {et~al.},\ }\href
  {\doibase 10.48550/arXiv.2304.13072} {\bibfield  {journal} {\bibinfo
  {journal} {arXiv e-prints}\ ,\ \bibinfo {eid} {arXiv:2304.13072}} (\bibinfo
  {year} {2023})},\ \Eprint {http://arxiv.org/abs/2304.13072} {arXiv:2304.13072
  [astro-ph.IM]} \BibitemShut {NoStop}%
\bibitem [{\citenamefont {{Nobleson}}\ \emph {et~al.}(2022)\citenamefont
  {{Nobleson}}, \citenamefont {{Agarwal}}, \citenamefont {{Girgaonkar}},
  \citenamefont {{Pandian}}, \citenamefont {{Joshi}}, \citenamefont
  {{Krishnakumar}} \emph {et~al.}}]{Noble22}%
  \BibitemOpen
  \bibfield  {author} {\bibinfo {author} {\bibfnamefont {K.}~\bibnamefont
  {{Nobleson}}}, \bibinfo {author} {\bibfnamefont {N.}~\bibnamefont
  {{Agarwal}}}, \bibinfo {author} {\bibfnamefont {R.}~\bibnamefont
  {{Girgaonkar}}}, \bibinfo {author} {\bibfnamefont {A.}~\bibnamefont
  {{Pandian}}}, \bibinfo {author} {\bibfnamefont {B.~C.}\ \bibnamefont
  {{Joshi}}}, \bibinfo {author} {\bibfnamefont {M.~A.}\ \bibnamefont
  {{Krishnakumar}}},  \emph {et~al.},\ }\href {\doibase 10.1093/mnras/stac532}
  {\bibfield  {journal} {\bibinfo  {journal} {\mnras}\ }\textbf {\bibinfo
  {volume} {512}},\ \bibinfo {pages} {1234} (\bibinfo {year} {2022})},\ \Eprint
  {http://arxiv.org/abs/2112.06908} {arXiv:2112.06908 [astro-ph.IM]}
  \BibitemShut {NoStop}%
\end{thebibliography}%

\appendix

\section{Parameters obtained from noise analysis}
\label{appendix_a}
Here, we provide the final models and the parameters in Table \ref{parameter_values} for all 14 pulsars for each noise component with a 68 \% confidence interval.

\begin{sidewaystable*}[h!]
	\centering
 \vspace{9cm}
	\caption{Median and 16 $-$ 84 \% credible intervals of the posterior distributions of each single-pulsar noise model parameters for all 14 DR1 pulsars.}
	\setlength{\tabcolsep}{4.5pt}
 {\renewcommand\arraystretch{2.0}
\begin{tabular}{|c|c|c|c|c|c|c|c|c|c|c|c|}
\hline
\multirow{2}{*}{Pulsar} & \multirow{2}{*}{Final Models} &  \multirow{2}{*}{\bthis{Timespan }} & \multicolumn{3}{c|}{Red noise} & 
 \multicolumn{3}{c|}{DM variations} & \multicolumn{3}{c|}{Scattering variations} \\

\cline{4-12}
  & & \bthis{(yr)} & \multicolumn{1}{c|}{$\log_{10}\rm{A}_{RN}$} & \multicolumn{1}{c|}{$\gamma_{RN}$} & \multicolumn{1}{c|}{Fourier modes} & \multicolumn{1}{c|}{$\log_{10}\rm{A}_{DM}$} & \multicolumn{1}{c|}{$\gamma_{DM}$} & \multicolumn{1}{c|}{Fourier modes} & \multicolumn{1}{c|}{$\log_{10}\rm{A}_{Sv}$} & \multicolumn{1}{c|}{$\gamma_{Sv}$} & \multicolumn{1}{c|}{Fourier modes}\\
\hline
J0437$-$4715 & Model2 & 0.85 & $-12.17^{+0.17}_{-0.12}$ & $0.28^{+0.36}_{-0.20}$ & 10 & ------ & ------ & ------ & ------ & ------ & ------ \\  

J0613$-$0200 & Model2 & 3.39 & $-11.85^{+0.16}_{-0.14}$  & $1.10^{+0.59}_{-0.57}$ & 27 & ------ & ------ & ------& ------ & ------ &------ \\

J0751+1807 &  Model1 & 3.39 & ------ & ------ & & ------ & ------ & ------ & ------ & ------ &  ------\\

J1012+5307 &  Model6 & 3.36 & ------ & ------ & ------ & $-13.17^{+0.19}_{-0.20}$ & $0.53^{+0.55}_{-0.37}$ & 16 & ------ & ------ & ------ \\

J1022+1001 &  Model6 & 3.36 & ------ & ------ &------ & $-12.64^{+0.15}_{-0.13}$ & $0.59^{+0.44}_{-0.36}$ & 40 & ------ & ------ &------ \\

 J1600$-$3053 &  Model1 & 3.36 & ------ & ------ & & ------ & ------ &------ & ------ & ------ & ------\\

J1643$-$1224 &  Model5 & 3.39 & $-12.44^{+0.25}_{-0.33}$ & $4.11^{+1.62}_{-1.37}$ & 27 & $-12.42^{+0.09}_{-0.08}$ & $2.46^{+0.29}_{-0.24}$ & 40 & $-13.46^{+0.17}_{-0.16}$ & $3.44^{+1.17}_{-0.99}$ & 6 \\

J1713+0747 &  Model4 & 2.87 & ------ & ------ & ------ & $-13.85^{+0.25}_{-0.17}$ & $0.48^{+0.62}_{-0.34}$ & 34 & $-14.48^{+0.22}_{-0.23}$ & $3.70^{+1.61}_{-1.39}$ & 5\\

J1744$-$1134 &  Model6 & 0.44 & ------ & ------ & ------ & $-11.97^{+0.43}_{-0.36}$ & $3.65^{+1.09}_{-1.02}$ & 4 & ------ & ------ & ------ \\

J1857+0943 & Model2 & 3.39 & $-12.09^{+0.22}_{-0.52}$ & $1.43^{+2.40}_{-0.97}$ & 6 & ------ & ------ & ------ & ------ &  ------ & ------ \\

J1909$-$3744 &  Model5  & 3.38 & $-12.45^{+0.13}_{-0.13}$ & $0.67^{+0.32}_{-0.33}$ & 40 & $-13.43^{+0.12}_{-0.10}$ & $2.06^{+0.69}_{-0.46}$ & 40 & ------ & ------ & ------ \\

J1939+2134 &  Model5 & 3.38 & $-12.52^{+0.10}_{-0.10}$ & $1.09^{+0.26}_{-0.27}$ & 40 & $-12.75^{+0.08}_{-0.07}$ & $1.84^{+0.24}_{-0.23}$ & 40 & $-14.12^{+0.10}_{-0.09}$ & $3.13^{+0.54}_{-0.48}$ & 16 \\

J2124$-$3358 &  Model2 & 3.38 & $-12.23^{+0.15}_{-0.14}$ & $1.04^{+0.96}_{-0.70}$ & 6 & ------ & ------ & ------ & ------ & ------ &------ \\

J2145$-$0750 &  Model5 & 3.38 & $-12.23^{+0.12}_{-0.11}$ & $1.98^{+0.71}_{-0.52}$ & 40 & $-13.43^{+0.30}_{-0.37}$ & $4.50^{+1.55}_{-1.49}$ & 27 & $-14.07^{+0.11}_{-0.10}$ & $1.63^{+0.29}_{-0.27}$ & 40 \\ \hline

\end{tabular}
}
\label{parameter_values}
\end{sidewaystable*}

\section{Red noise posterior plots for all pulsars}
\label{appendix_b}
Here, we provide the posteriors plots for each noise component present for all the pulsars, which can be found in Fig \ref{posterior_plots}.

\begin{figure*}[h!]
\caption{1D marginalized posterior distributions with 68\%,90\%,99\% credible intervals for red noise components present in respective pulsars.}
 \begin{subfigure}
		\centering
   \label{posterior_plots}
 \includegraphics[keepaspectratio=true,scale=0.4]{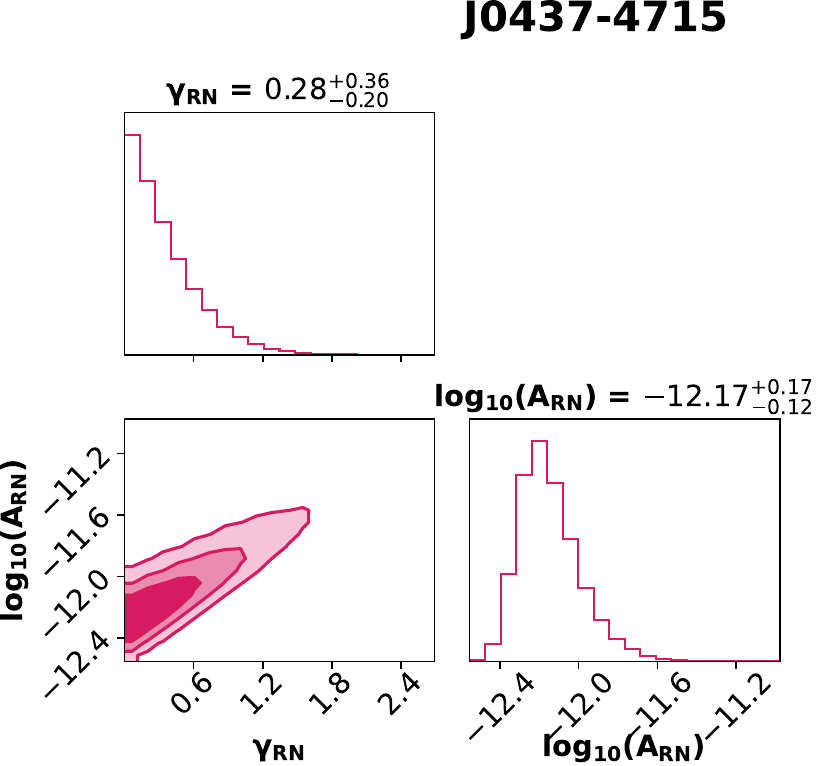}
	\end{subfigure}
 \begin{subfigure}
		\centering
 \includegraphics[keepaspectratio=true,scale=0.4]{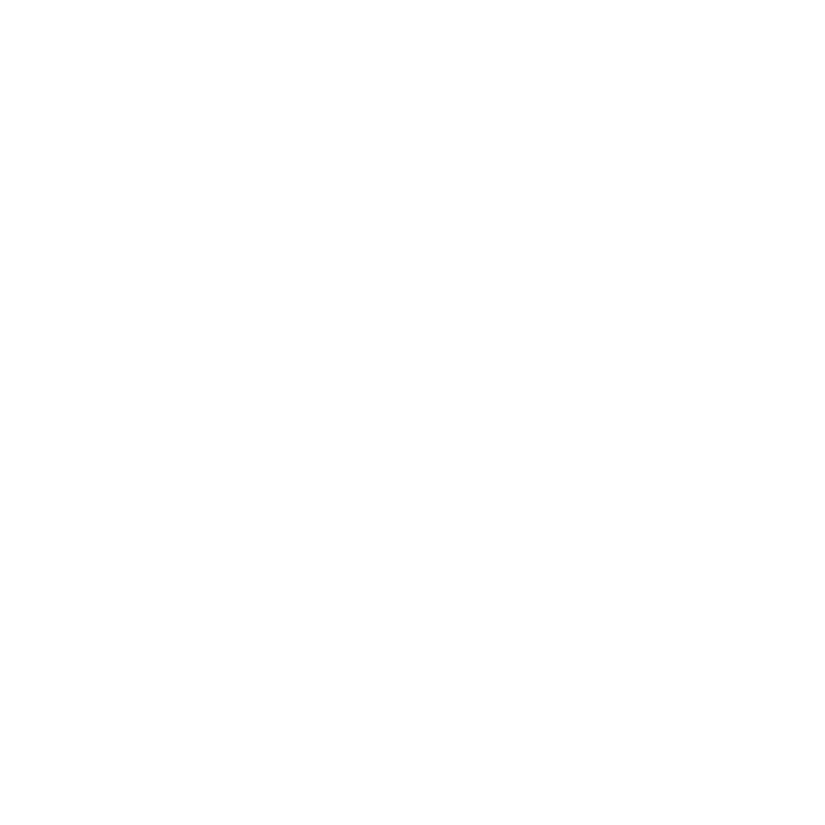}
	\end{subfigure}
 \begin{subfigure}
		\centering
 \includegraphics[keepaspectratio=true,scale=0.4]{figures/blank.pdf}
 \captionsetup{labelformat=empty}
  \caption{FIG. 3(A): J0437$-$4715 posterior distributions with 68\%,90\%,99\% credible intervals for achromatic red noise for \textit{WR} model.}
	\end{subfigure}
 
 \vspace{1cm}

 \begin{subfigure}
		\centering
 \includegraphics[keepaspectratio=true,scale=0.4]{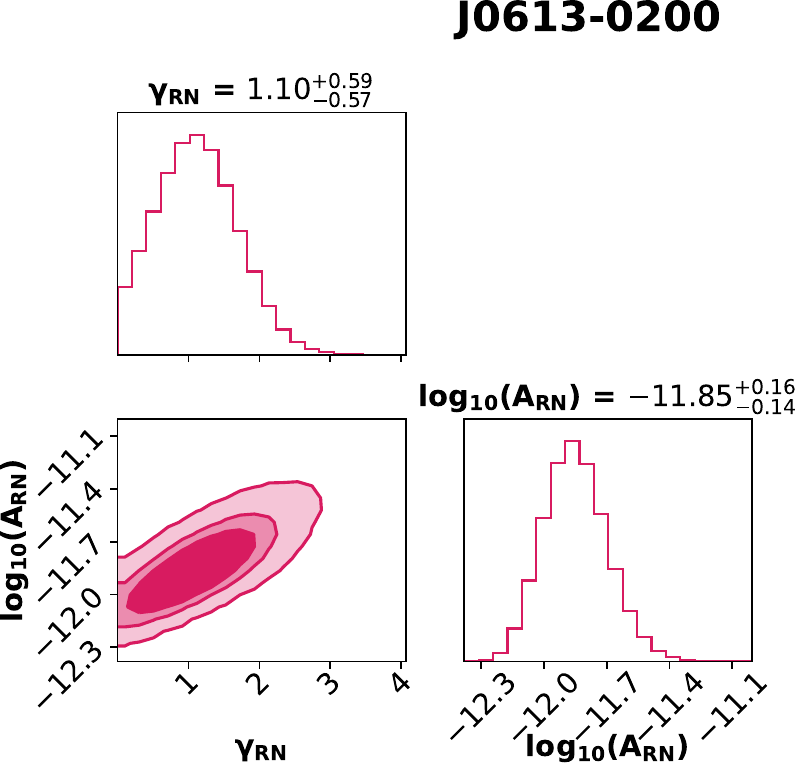}
	\end{subfigure}
 \begin{subfigure}
		\centering
 \includegraphics[keepaspectratio=true,scale=0.4]{figures/blank.pdf}
	\end{subfigure}
  \begin{subfigure}
		\centering
 \includegraphics[keepaspectratio=true,scale=0.4]{figures/blank.pdf}
  \captionsetup{labelformat=empty}
  \caption{FIG. 3(B): J10613$-$0200 posterior distributions with 68\%,90\%,99\% credible intervals for achromatic red noise for \textit{WR} model.}
	\end{subfigure}
 
 \vspace{1cm}

 	\begin{subfigure}
		\centering	
 \includegraphics[keepaspectratio=true,scale=0.4]{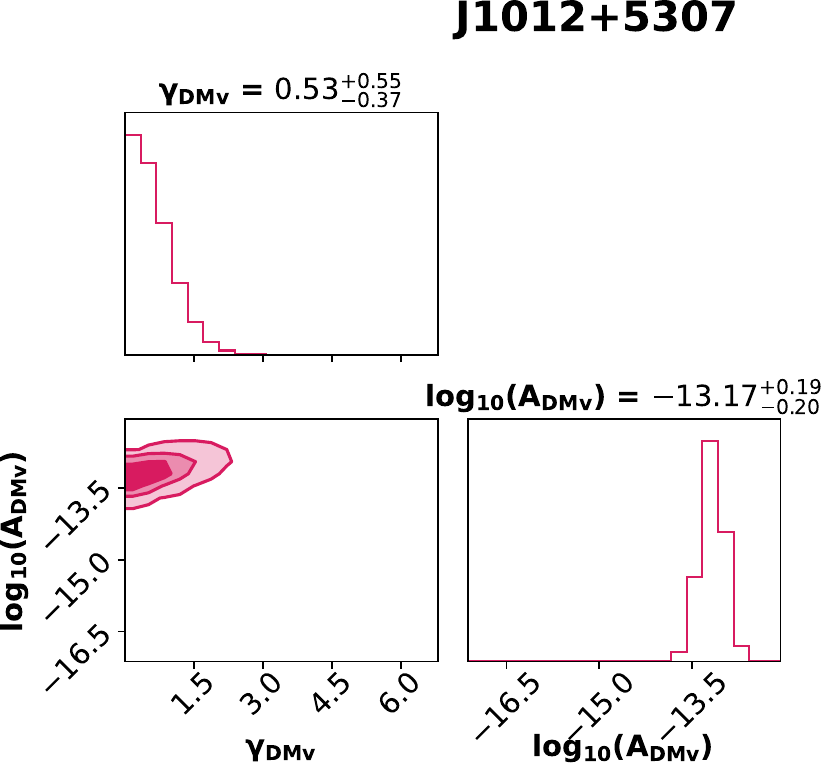}
	\end{subfigure}
    \begin{subfigure}
		\centering
 \includegraphics[keepaspectratio=true,scale=0.4]{figures/blank.pdf}
	\end{subfigure}
  \begin{subfigure}
		\centering
 \includegraphics[keepaspectratio=true,scale=0.4]{figures/blank.pdf}
  \captionsetup{labelformat=empty}
  \caption{FIG. 3(C): J1012+5307 posterior distributions with 68\%,90\%,99\% credible intervals for DMv for \textit{WD} model.}
	\end{subfigure}
 \end{figure*}

\begin{figure*}[h!]
 	\begin{subfigure}
	\centering	
 \includegraphics[keepaspectratio=true,scale=0.4]{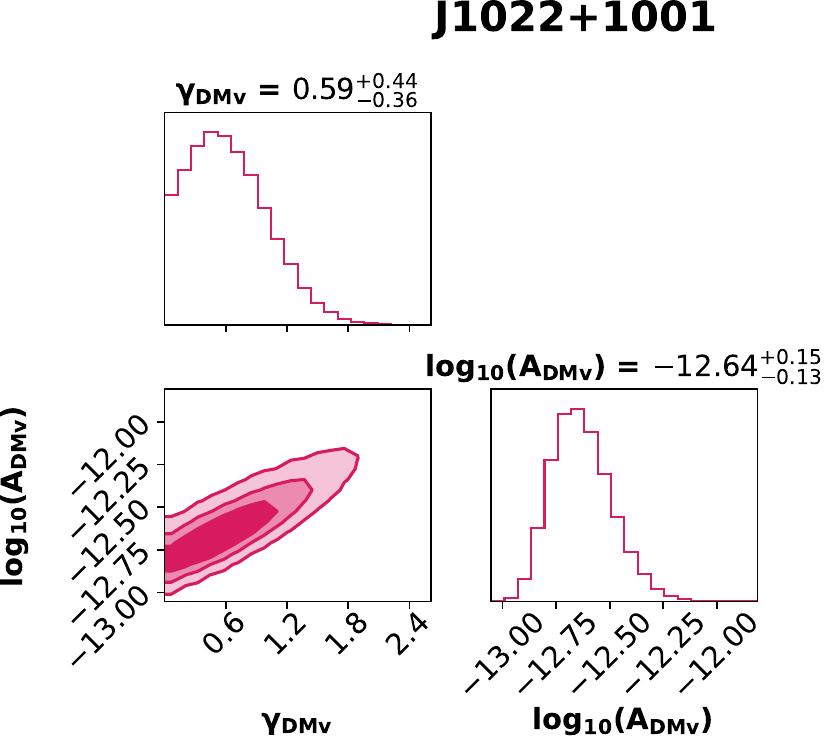}
 	\end{subfigure}
   \begin{subfigure}
		\centering
 \includegraphics[keepaspectratio=true,scale=0.4]{figures/blank.pdf}
	\end{subfigure}
    \begin{subfigure}
		\centering
 \includegraphics[keepaspectratio=true,scale=0.4]{figures/blank.pdf}
  \captionsetup{labelformat=empty}
  \caption{FIG. 3(D): J1022+1001 posterior distributions with 68\%,90\%,99\% credible intervals for DMv for \textit{WD} model.}
	\end{subfigure}
 
\vspace{1.5cm}

\begin{subfigure}
 \centering
 \includegraphics[keepaspectratio=true,scale=0.4]{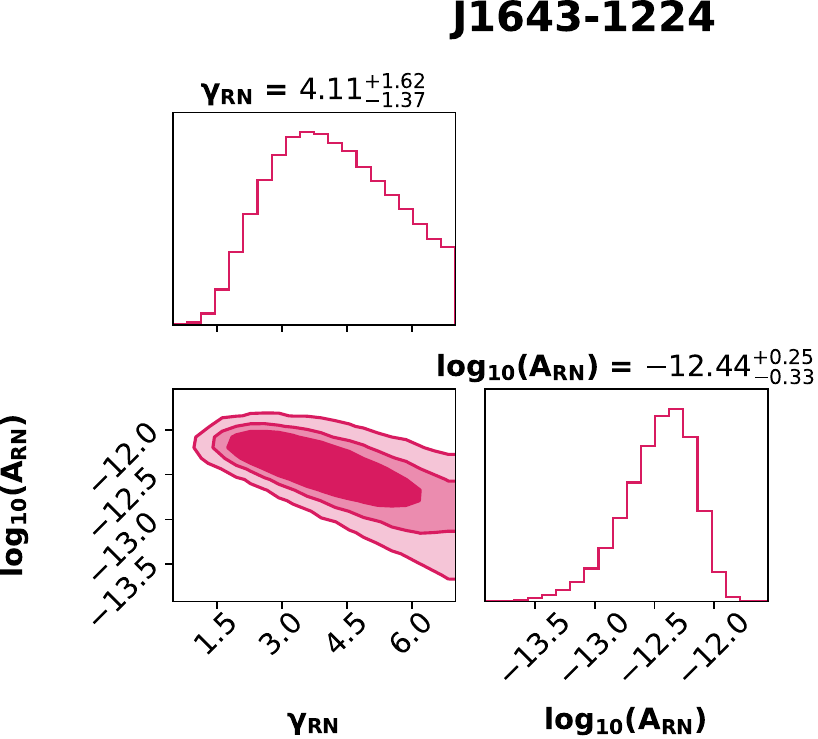}
\end{subfigure} 
 \begin{subfigure}
 \centering
 \includegraphics[keepaspectratio=true,scale=0.4]{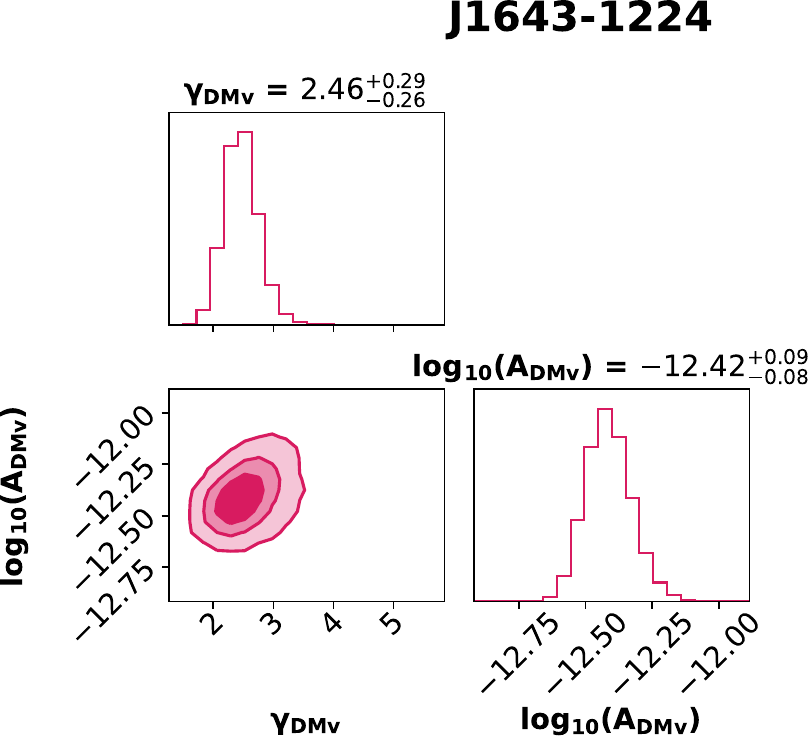}
\end{subfigure} \begin{subfigure}
 \centering
 \includegraphics[keepaspectratio=true,scale=0.4]{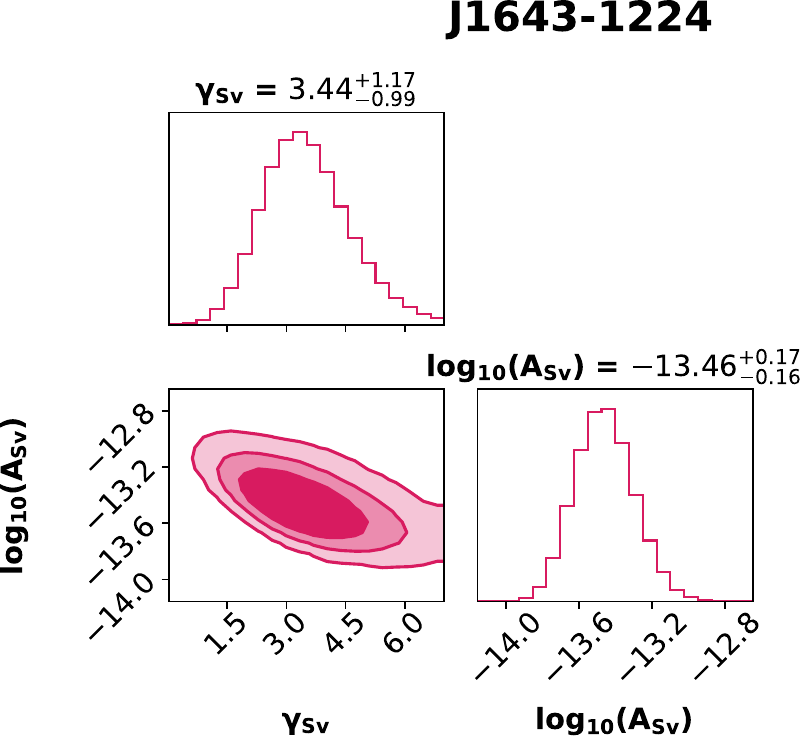}
  \captionsetup{labelformat=empty}
  \caption{FIG. 3(E): J1643$-$1224 posterior distributions with 68\%,90\%,99\% credible intervals for achromatic red noise, DMv and Sv for \textit{WRDS} model.}
\end{subfigure} 

\vspace{1.5cm} 

\begin{subfigure}
 \centering
 \includegraphics[keepaspectratio=true,scale=0.4]{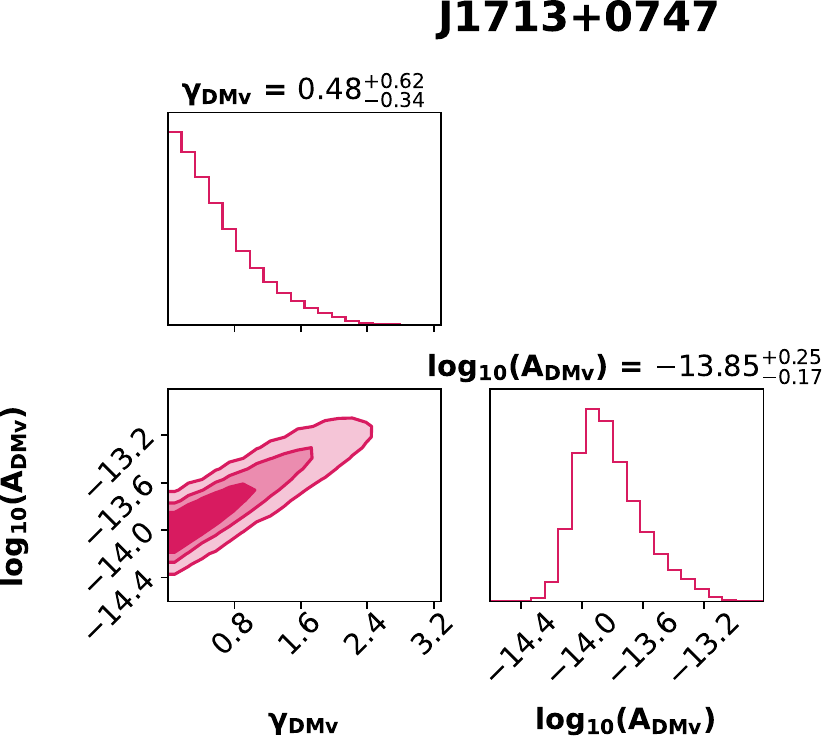}
\end{subfigure} 
\begin{subfigure}
 \centering
 \includegraphics[keepaspectratio=true,scale=0.4]{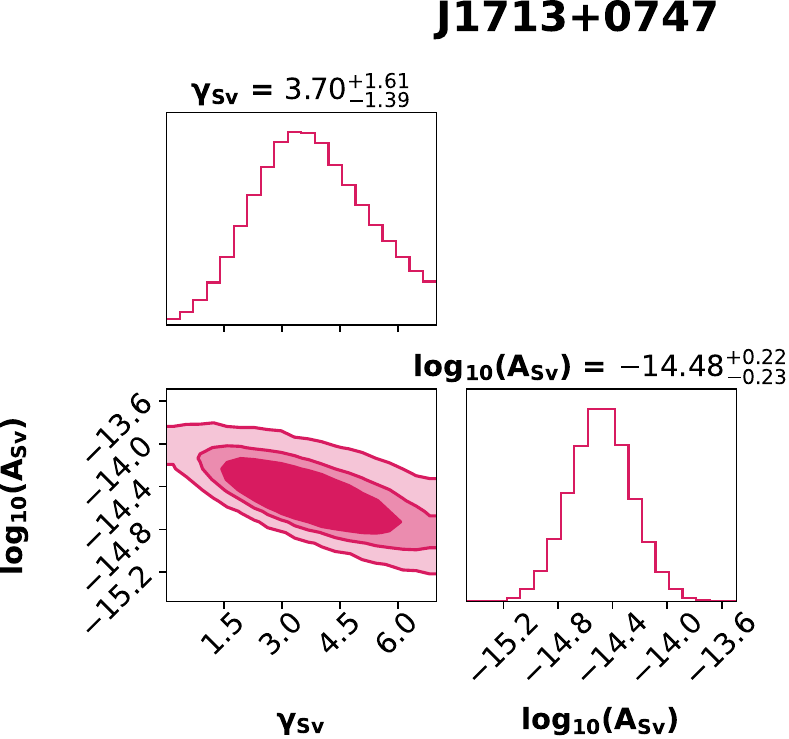}
\end{subfigure}
    \begin{subfigure}
		\centering
 \includegraphics[keepaspectratio=true,scale=0.4]{figures/blank.pdf}
  \captionsetup{labelformat=empty}
  \caption{FIG. 3(F): J1713+0747 posterior distributions with 68\%,90\%,99\% credible intervals for DMv and Sv for \textit{WDS} model.}
	\end{subfigure}

\end{figure*}

\begin{figure*}[h!]

 \begin{subfigure}
		\centering
 \includegraphics[keepaspectratio=true,scale=0.4]{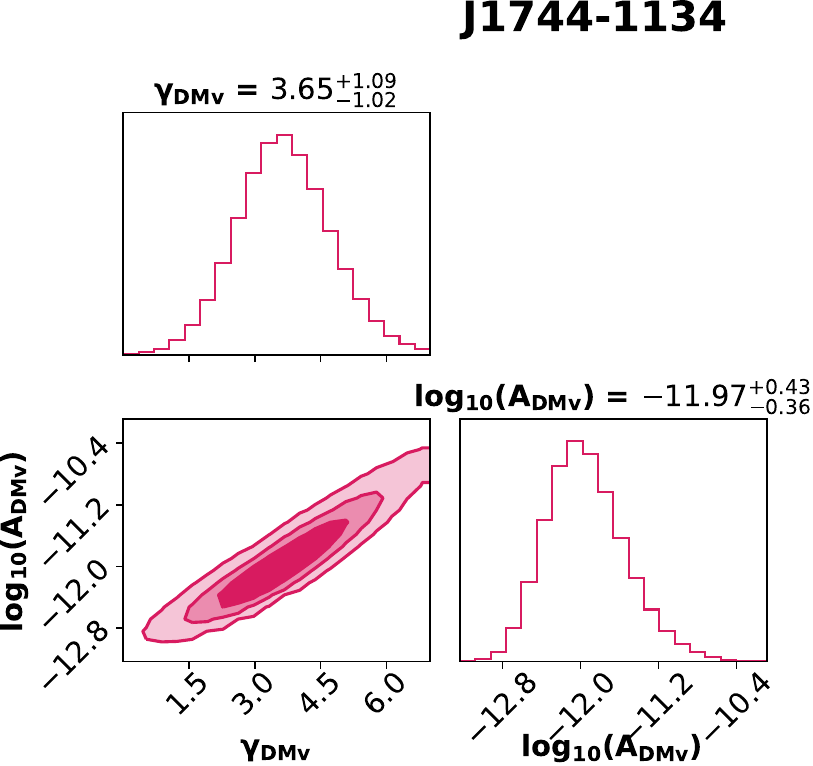}
	\end{subfigure}
  \begin{subfigure}
		\centering
 \includegraphics[keepaspectratio=true,scale=0.4]{figures/blank.pdf}
	\end{subfigure}
  \begin{subfigure}
		\centering
 \includegraphics[keepaspectratio=true,scale=0.4]{figures/blank.pdf}
   \captionsetup{labelformat=empty}
  \caption{FIG. 3(G): J1744$-$1134 posterior distributions with 68\%,90\%,99\% credible intervals for DMv for \textit{WD} model.}
	\end{subfigure}

 \vspace{1.5cm}

	\begin{subfigure}
		\centering
 \includegraphics[keepaspectratio=true,scale=0.4]{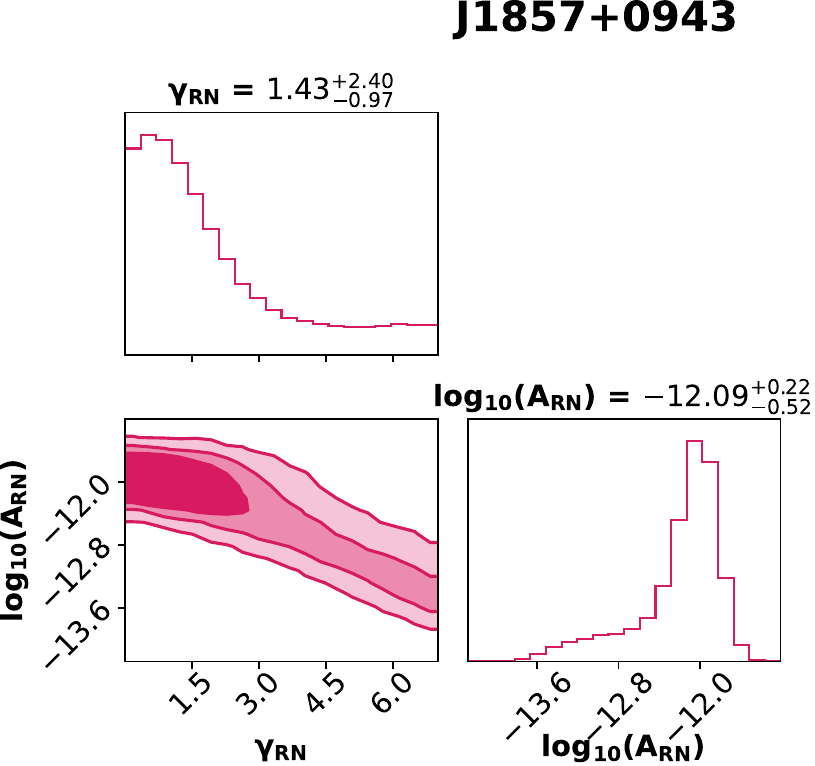}
	\end{subfigure}
  \begin{subfigure}
		\centering
 \includegraphics[keepaspectratio=true,scale=0.4]{figures/blank.pdf}
	\end{subfigure}
  \begin{subfigure}
		\centering
 \includegraphics[keepaspectratio=true,scale=0.4]{figures/blank.pdf}
   \captionsetup{labelformat=empty}
  \caption{FIG. 3(H): J1857+0943 68\%,90\%,99\% credible intervals for achromatic red noise for \textit{WR} model.}
	\end{subfigure}
 
\vspace{1.5cm}

\begin{subfigure}
		\centering
 \includegraphics[keepaspectratio=true,scale=0.4]{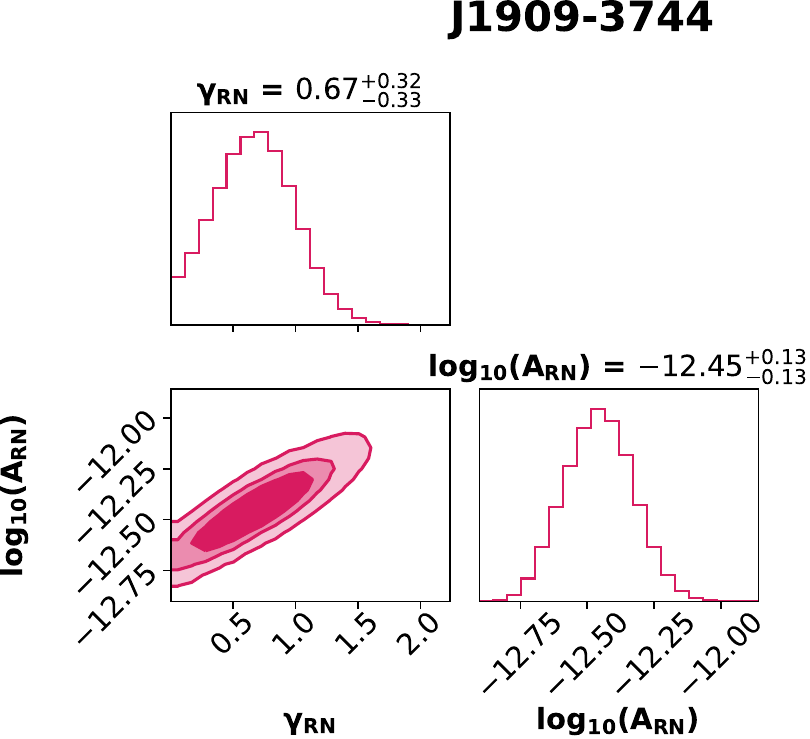}
	\end{subfigure}
 	\begin{subfigure}
		\centering	
 \includegraphics[keepaspectratio=true,scale=0.4]{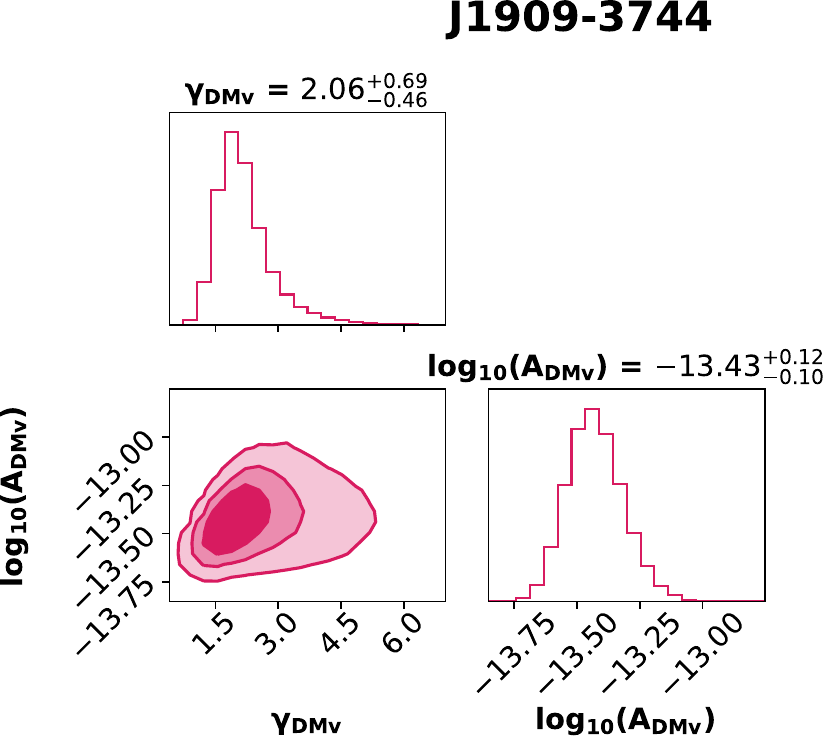}
	\end{subfigure}
  \begin{subfigure}
		\centering
 \includegraphics[keepaspectratio=true,scale=0.4]{figures/blank.pdf}
    \captionsetup{labelformat=empty}
  \caption{FIG. 3(I): J1909$-$3744 posterior distributions with 68\%,90\%,99\% credible intervals for achromatic red noise and DMv for \textit{WRD} model.}
	\end{subfigure}

 \end{figure*}

\begin{figure*}[h!]

 	\begin{subfigure}
		\centering	
 \includegraphics[keepaspectratio=true,scale=0.4]{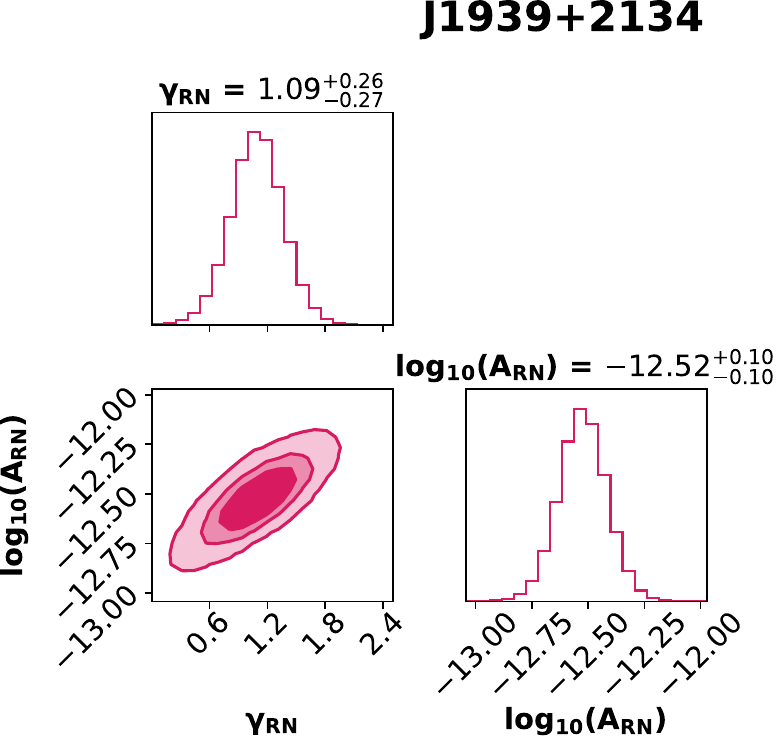}
	\end{subfigure}	
 \begin{subfigure}
		\centering	
  \includegraphics[keepaspectratio=true,scale=0.4]{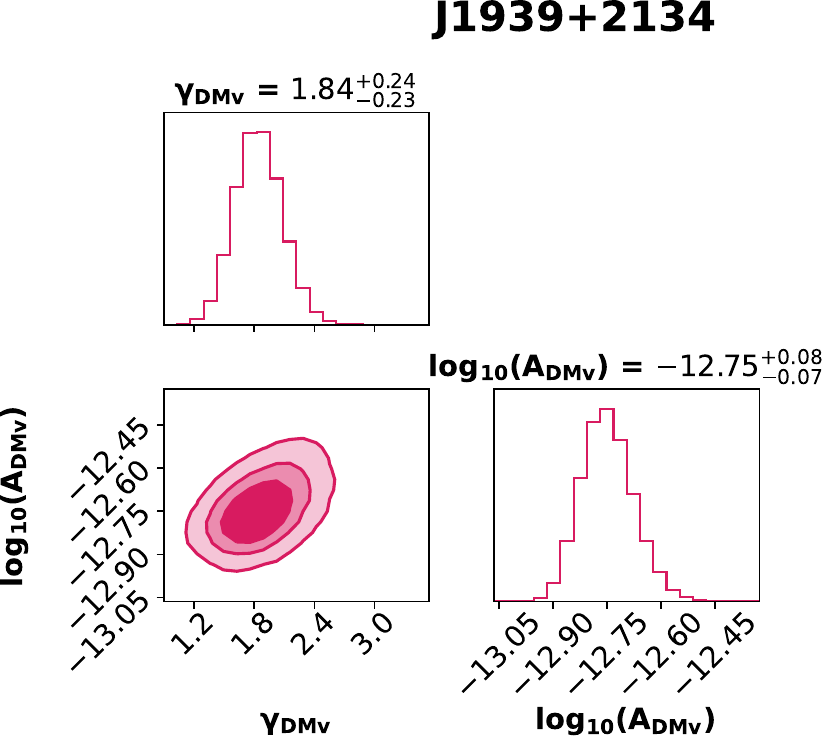}
	\end{subfigure}
 	\begin{subfigure}
		\centering	
 \includegraphics[keepaspectratio=true,scale=0.4]{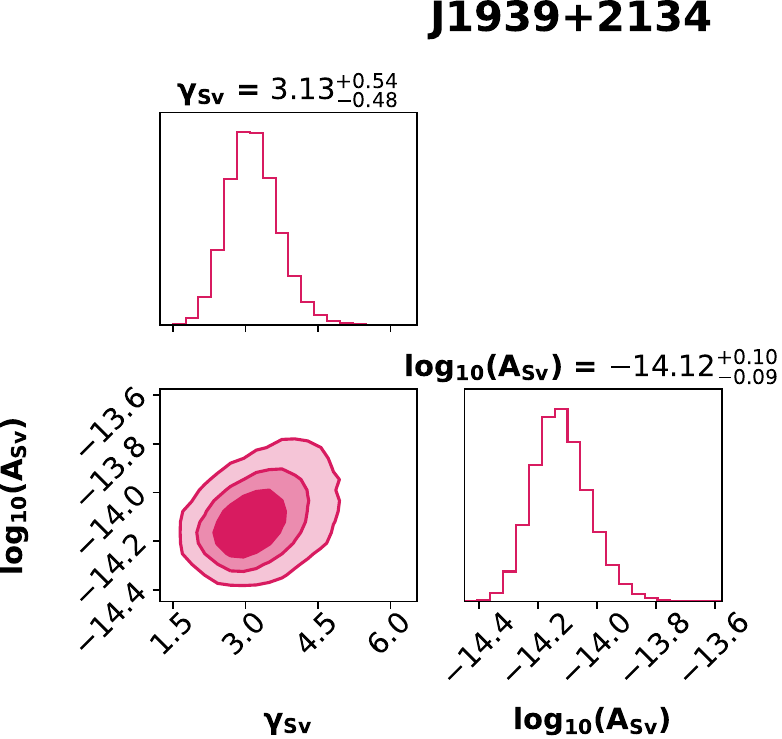}
    \captionsetup{labelformat=empty}
  \caption{FIG. 3(J): J1939+2134 posterior distributions with 68\%,90\%,99\% credible intervals for achromatic red noise, DMv and Sv for \textit{WRDS} model.}
	\end{subfigure}
 
\vspace{1.5cm}

	\begin{subfigure}
		\centering	 
  \includegraphics[keepaspectratio=true,scale=0.4]{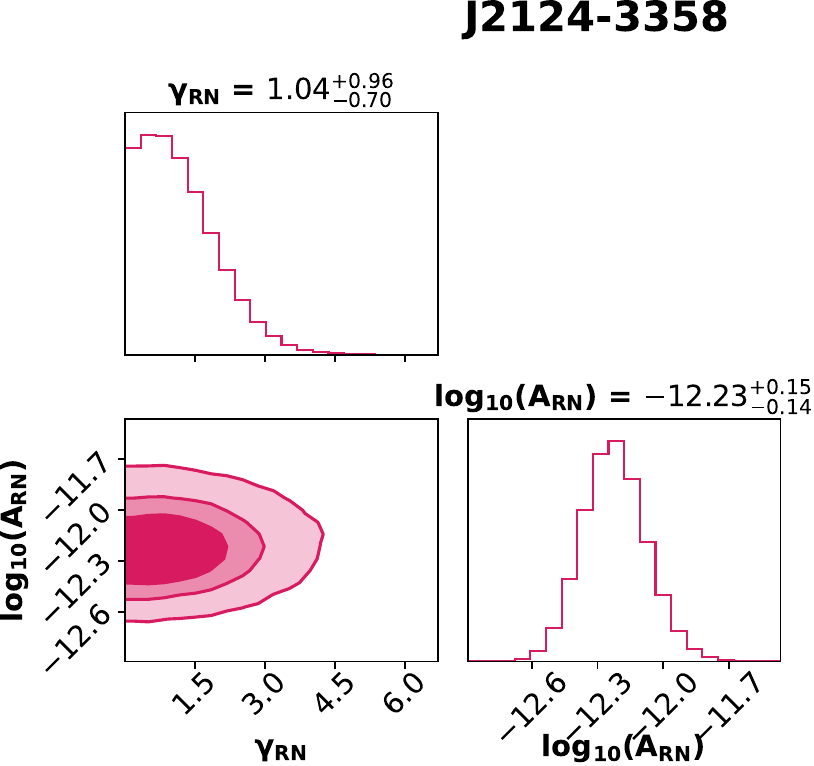}
	\end{subfigure}
  \begin{subfigure}
		\centering
 \includegraphics[keepaspectratio=true,scale=0.4]{figures/blank.pdf}
	\end{subfigure}
  \begin{subfigure}
		\centering
 \includegraphics[keepaspectratio=true,scale=0.4]{figures/blank.pdf}
    \captionsetup{labelformat=empty}
  \caption{FIG. 3(K): J2124$-$3358 posterior distributions with 68\%,90\%,99\% credible intervals for achromatic red noise for \textit{WR} model.}
	\end{subfigure}

 \vspace{1.5cm}

	\begin{subfigure}
		\centering	
 \includegraphics[keepaspectratio=true,scale=0.4]{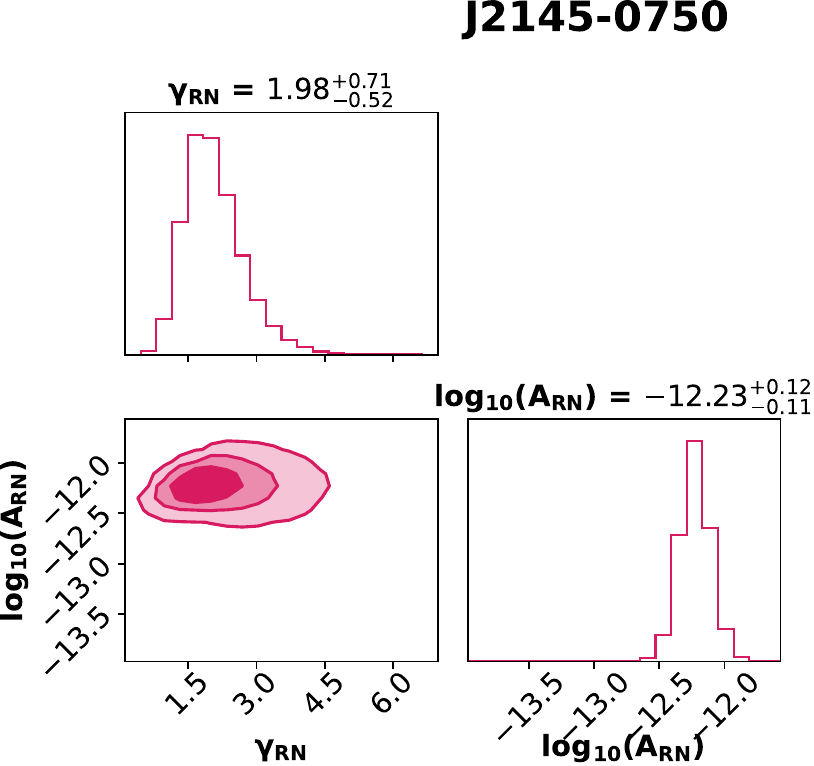}
	\end{subfigure}
 	\begin{subfigure}
		\centering	
 \includegraphics[keepaspectratio=true,scale=0.4]{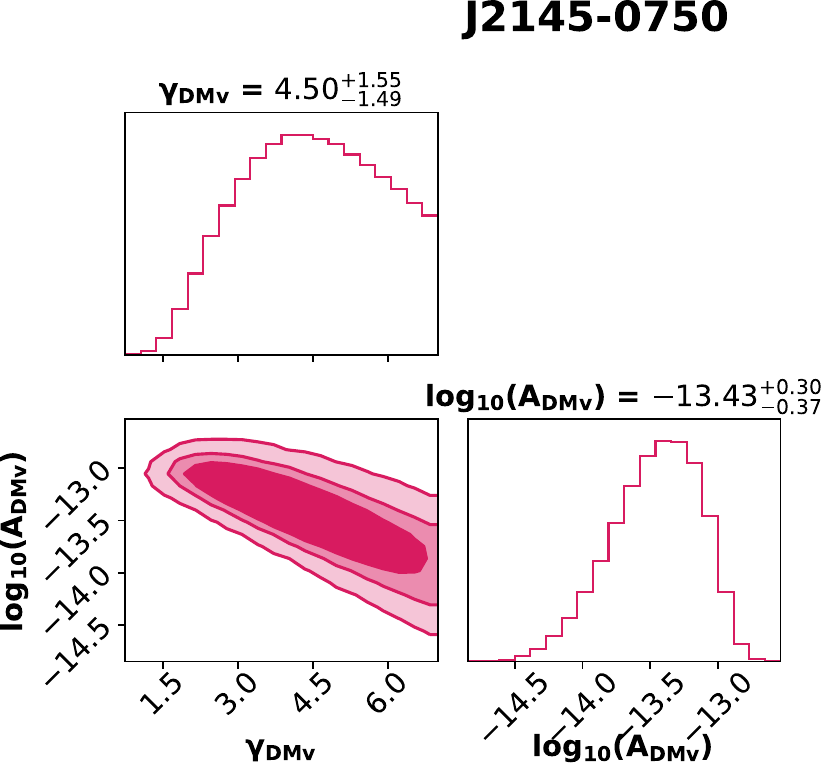}
 	\end{subfigure}
 \begin{subfigure}
		\centering
 \includegraphics[keepaspectratio=true,scale=0.4]{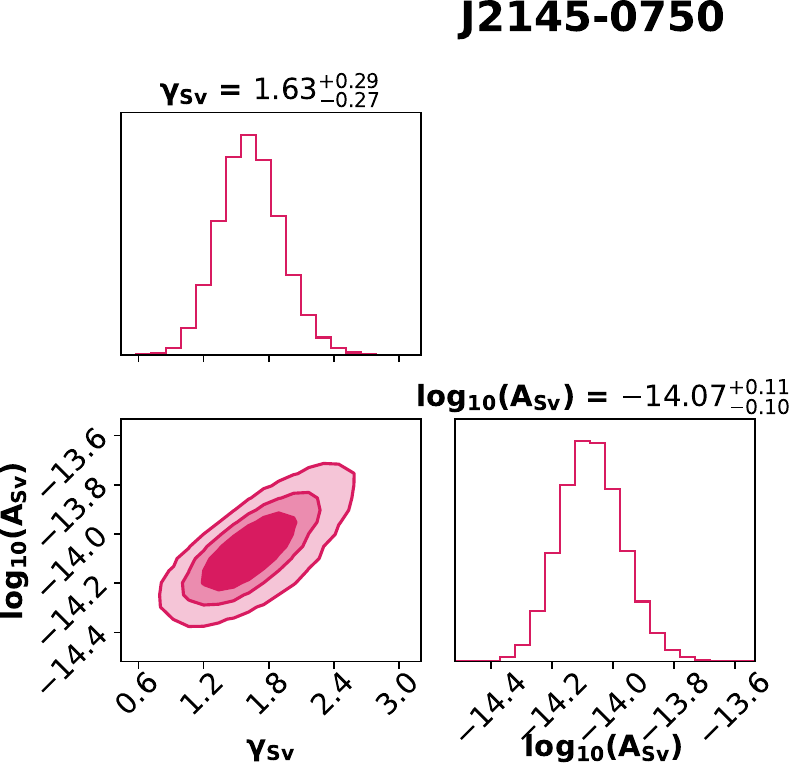}
     \captionsetup{labelformat=empty}
  \caption{FIG. 3(L): J2145$-$0750 posterior distributions with 68\%,90\%,99\% credible intervals for achromatic red noise, DMv and Sv for \textit{WRDS} model.}
	\end{subfigure}
\end{figure*}

\section{Full corner plots for all pulsars}
\label{appendix_c}
The marginalized posterior credible intervals  with white and red noises for each pulsar are shown here in FIG.4 on page \pageref{fullcorner}.

\begin{figure*}[h!]

 \captionsetup{labelformat=empty}
\caption{FIG. 4: Posterior distributions with 68\%,90\%,99\% credible intervals for all noise components present in respective pulsars. For white noises, we used abbreviations such that \textbf{B3} and \textbf{B5} stand for band3 and band5 data, followed by \textbf{A} or \textbf{B}, which denotes pre-cycle36 or post-cycle36 data respectively. \textbf{EFAC} is \texttt{efac} while \textbf{EQ} is \texttt{log10\_t2equad} (for eg: \textbf{B5BEQ} is \texttt{log10\_t2equad} for band5 post-cycle36 data).}
\label{fullcorner}
 \vspace*{1in}
 \centering
 \includegraphics[keepaspectratio=true,scale=0.40]{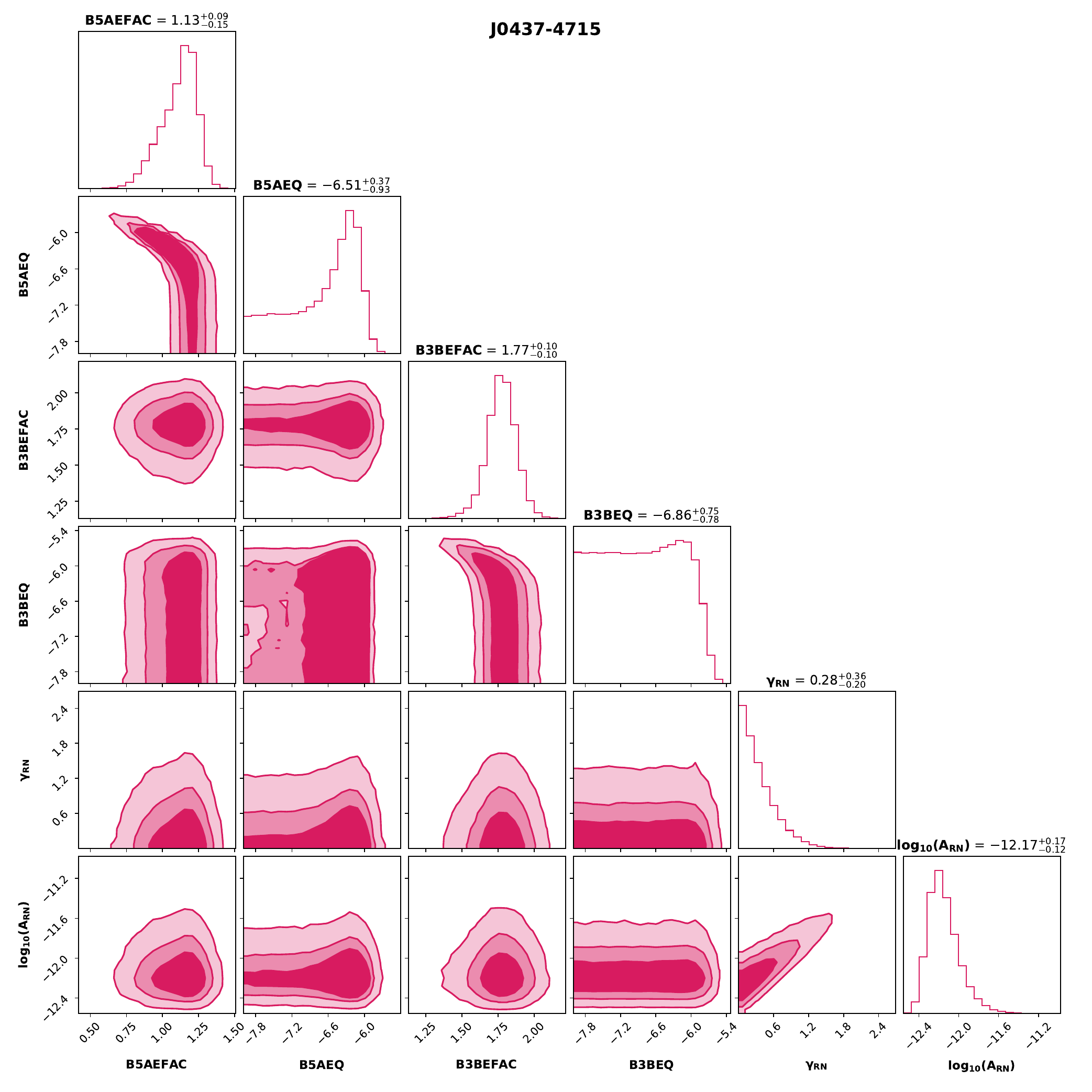}
 \captionsetup{labelformat=empty}
  \caption{FIG. 4(A): J0437$-$4715 posterior distributions with 68\%,90\%,99\% credible intervals for white noise and achromatic red noise for \textit{WR} model.}
\end{figure*}

\begin{figure*}[h!]
 \centering
 \vspace*{1.75in}
 \includegraphics[keepaspectratio=true,scale=0.40]{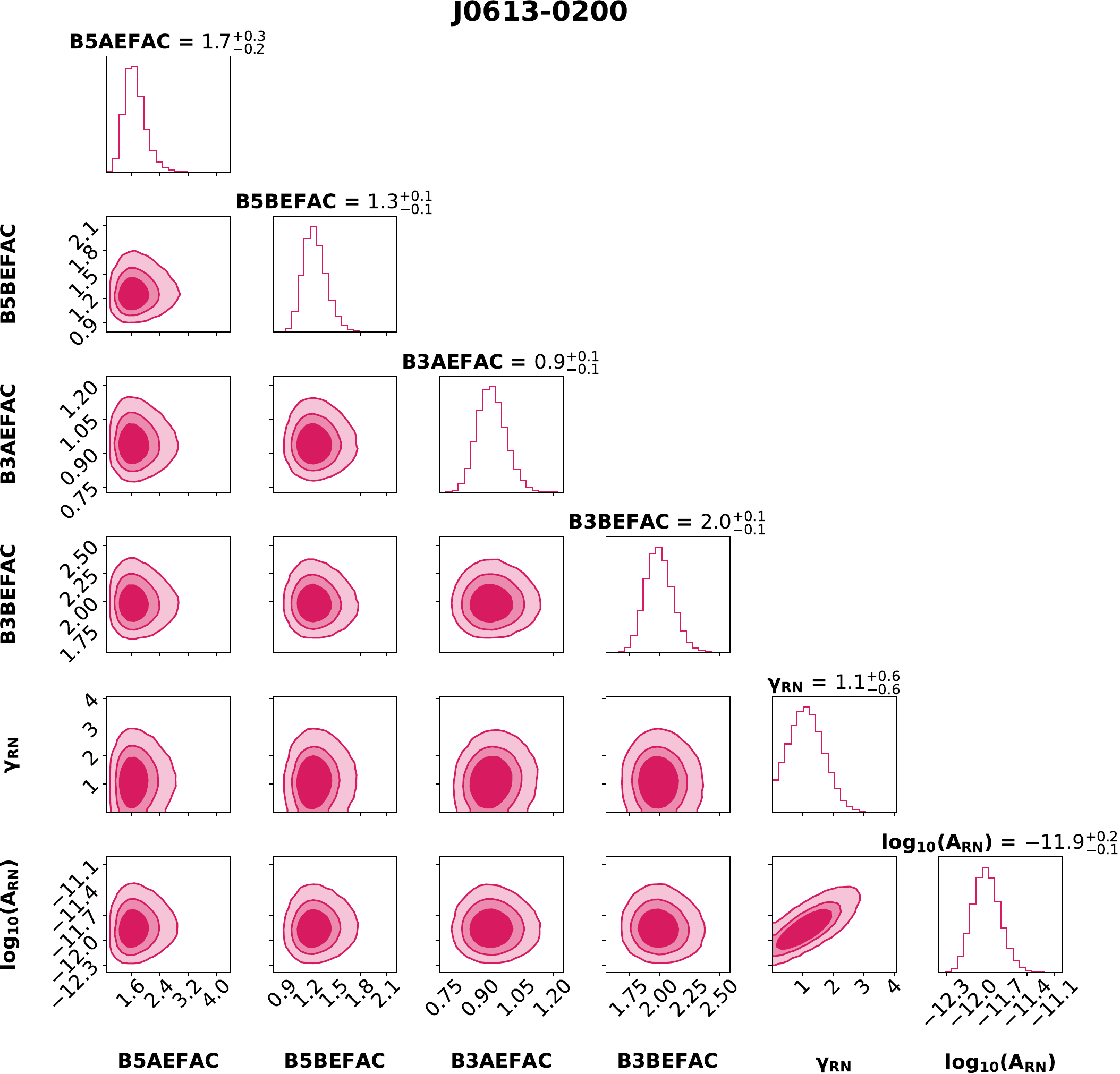}
 \captionsetup{labelformat=empty}
 \caption{FIG. 4(B): J0613$-$0200 posterior distributions with 68\%,90\%,99\% credible intervals for white noise and achromatic red noise for \textit{WR} model.}
\end{figure*}

\begin{figure*}
 \centering
 \vspace*{2in}
 \includegraphics[keepaspectratio=true,scale=0.35]{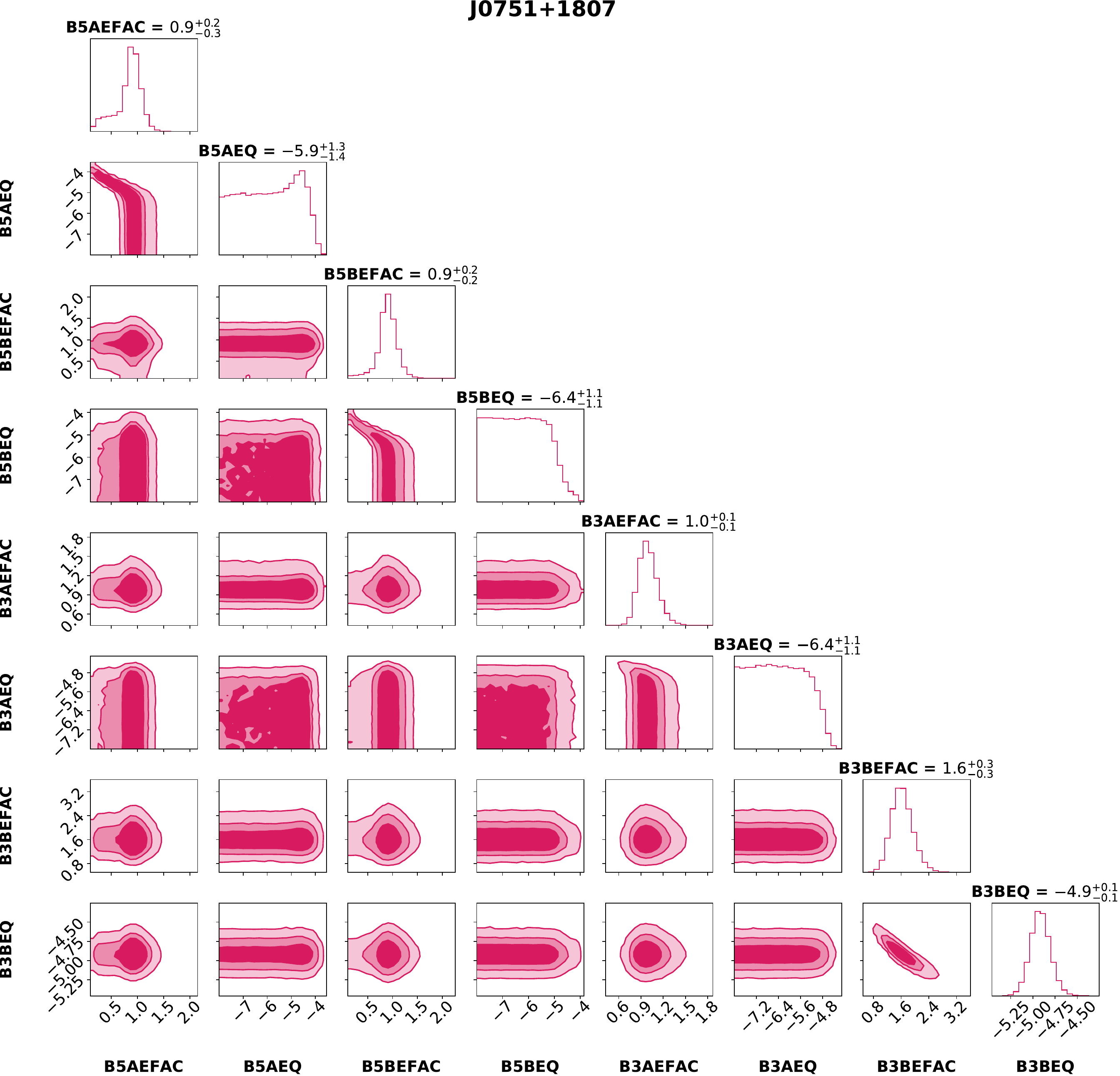}
 \captionsetup{labelformat=empty}
  \caption{FIG. 4(C): J0751+1807 posterior distributions with 68\%,90\%,99\% credible intervals for white noise for \textit{W} model.}

\end{figure*}

\begin{figure*}
\begin{subfigure}
 \centering
 \includegraphics[keepaspectratio=true,scale=0.35]{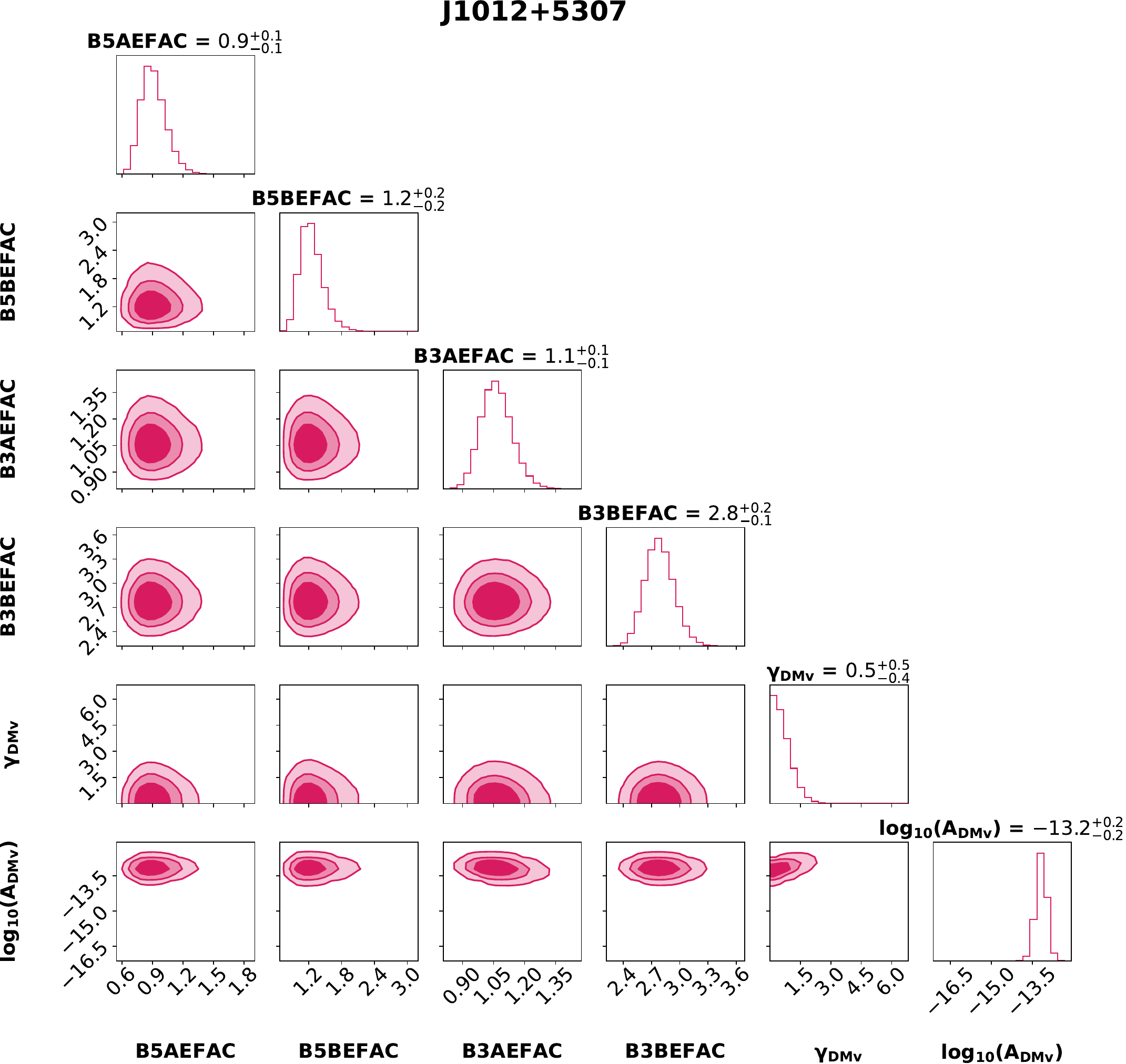}
   \captionsetup{labelformat=empty}
  \caption{FIG. 4(D): J1012+5307 posterior distributions with 68\%,90\%,99\% credible intervals for white noise and DMv for \textit{WD} model.}
\end{subfigure}
 \vspace{0.3in}
\begin{subfigure}
 \centering
 \includegraphics[keepaspectratio=true,scale=0.35]{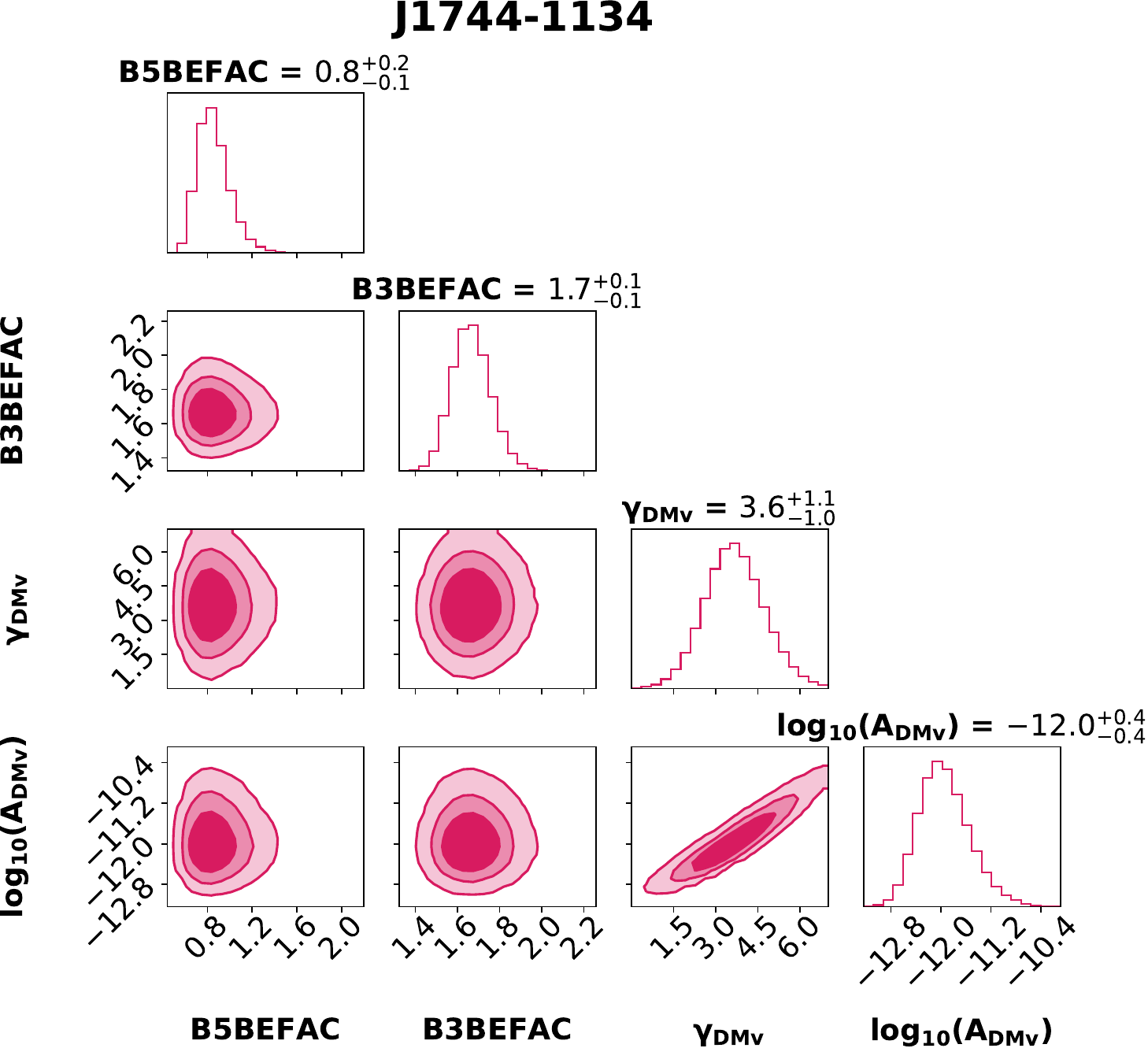}
  \captionsetup{labelformat=empty}
 \caption{FIG. 4(E): J1744$-$1134 posterior distributions with 68\%,90\%,99\% credible intervals for white noise and DMv for \textit{WD} model.}
\end{subfigure}
\end{figure*}

\begin{figure*}
 \centering
 \vspace*{1.25in}
 \includegraphics[keepaspectratio=true,scale=0.33]{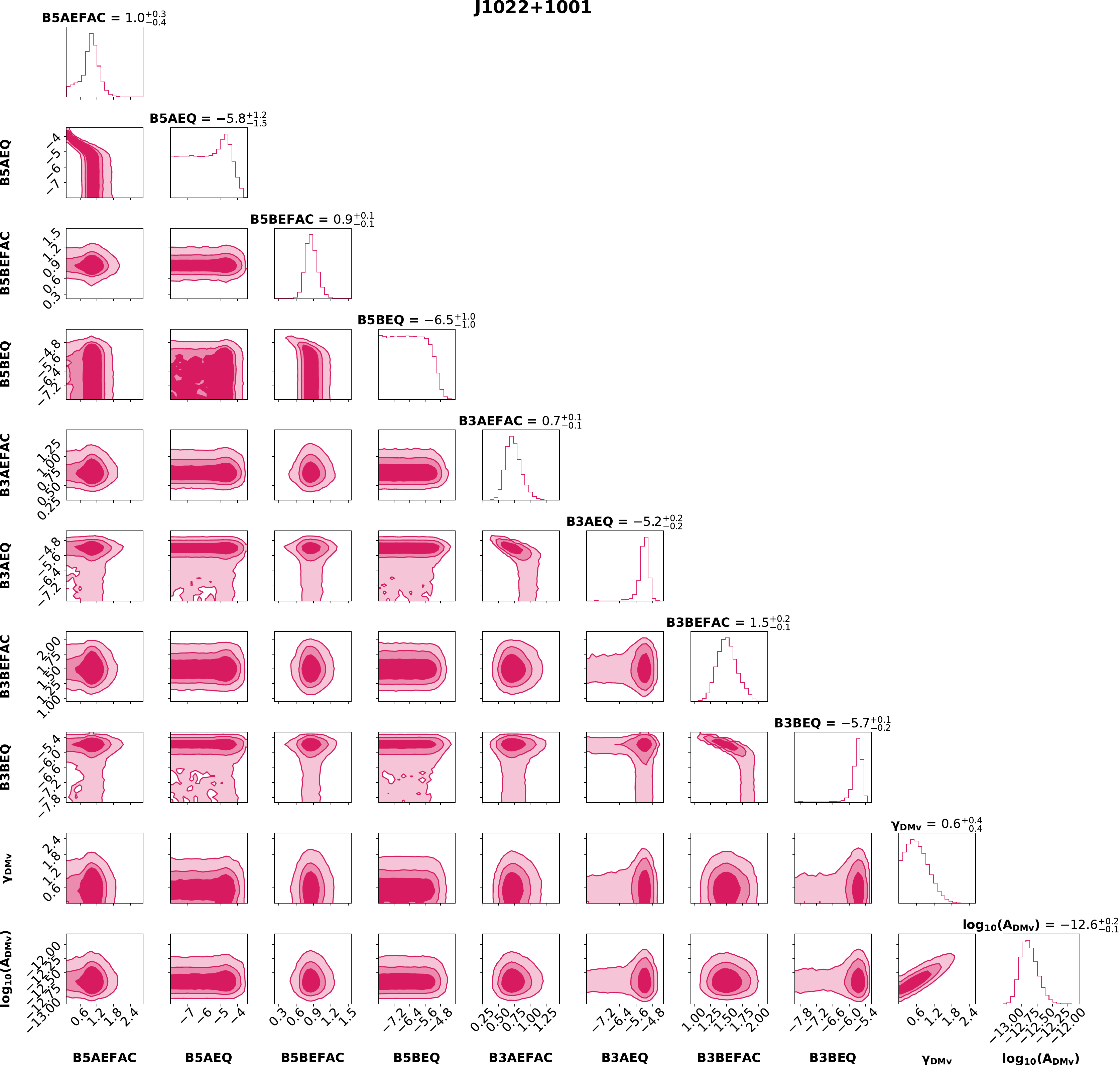}
   \captionsetup{labelformat=empty}
  \caption{FIG. 4(F): J1022+1001 posterior distributions with 68\%,90\%,99\% credible intervals for white noise and DMv for \textit{WD} model.}
\end{figure*}

\begin{figure*}
 \centering
 \vspace*{1.25in}
 \includegraphics[keepaspectratio=true,scale=0.40]{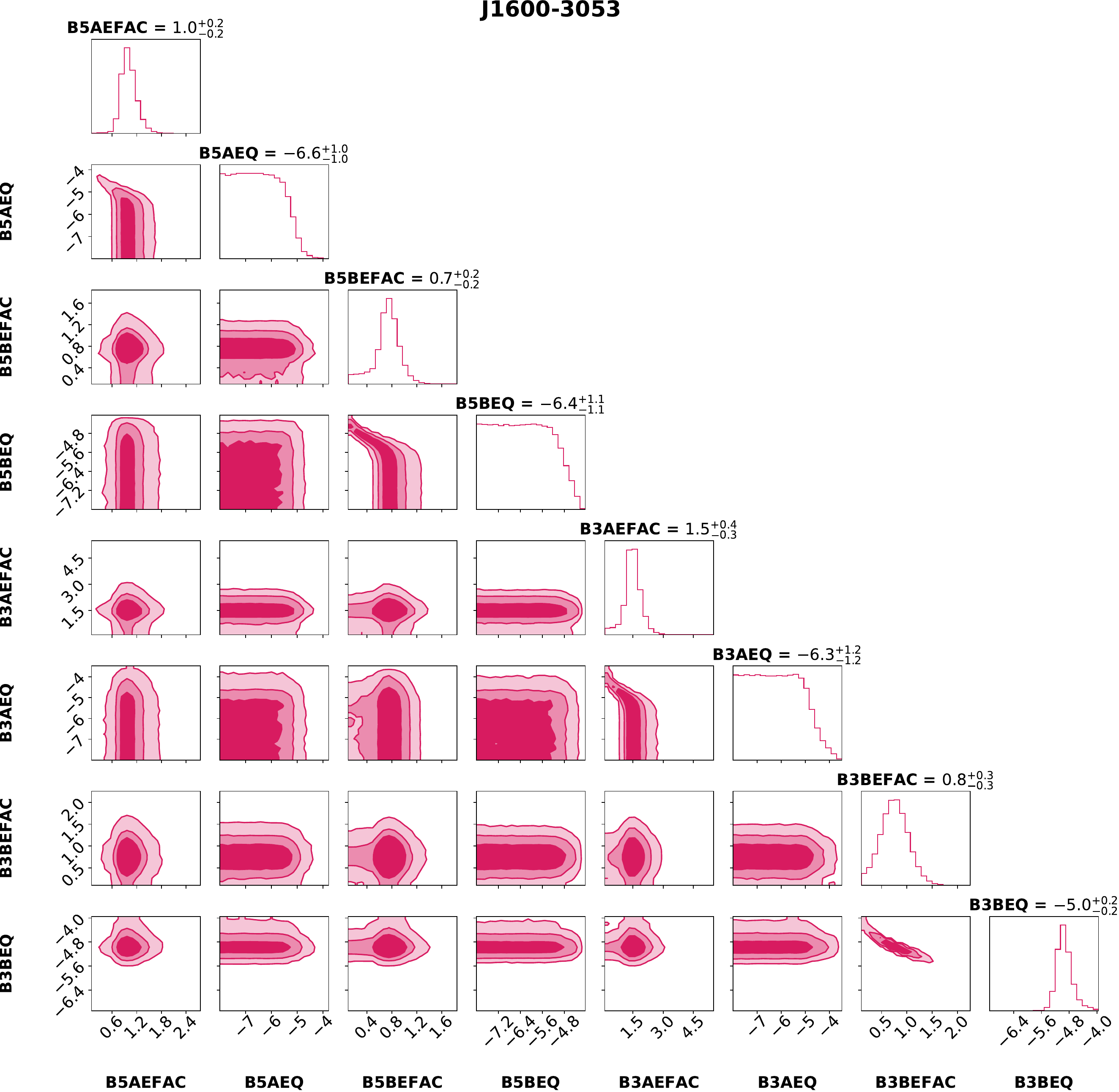}
  \captionsetup{labelformat=empty}
  \caption{FIG. 4(G): J1600$-$3053 posterior distributions with 68\%,90\%,99\% credible intervals for white noise for \textit{W} model.}
\end{figure*}

\begin{figure*}
 \centering
 \vspace*{1.25in}
 \includegraphics[keepaspectratio=true,scale=0.23]{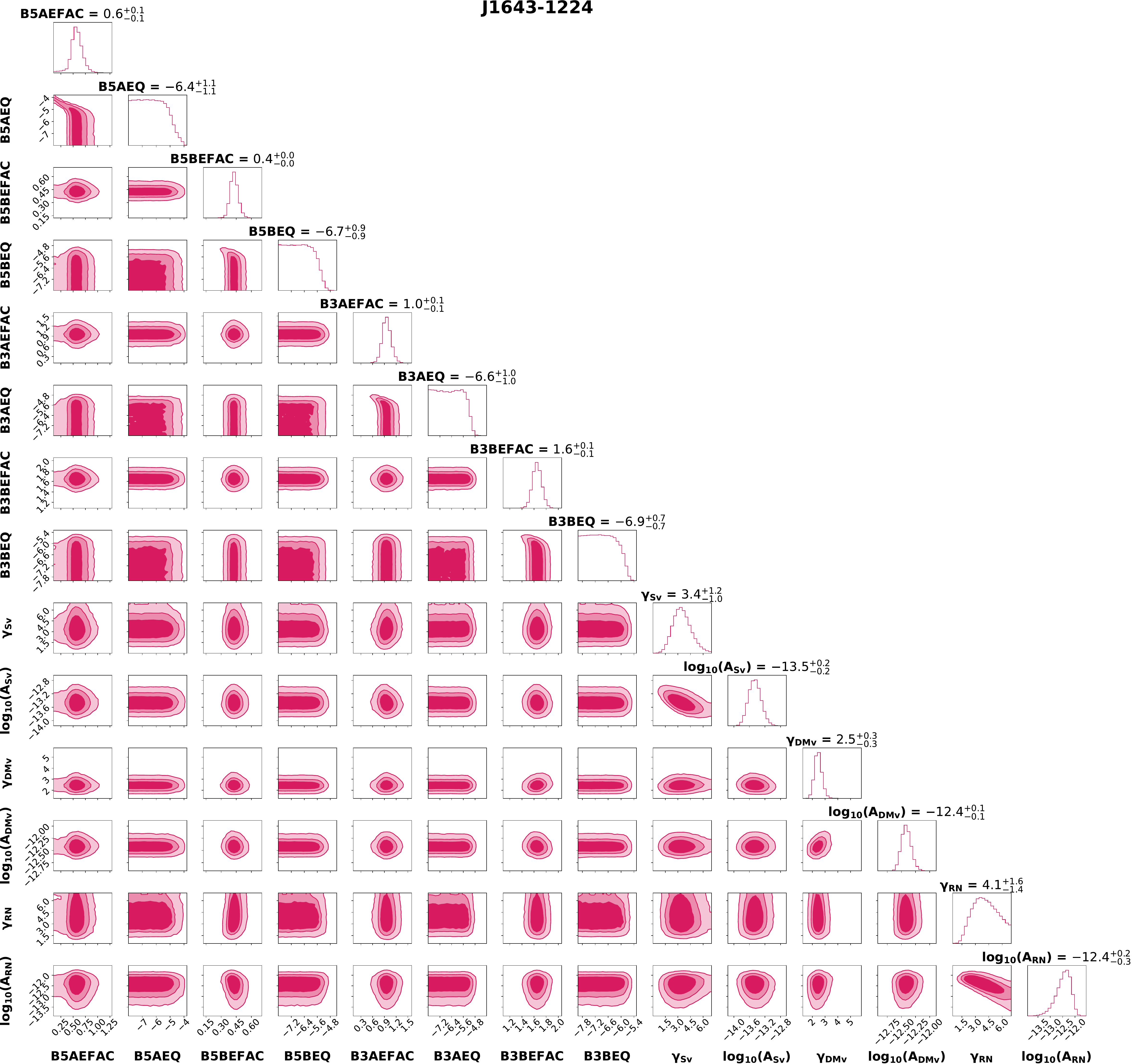}
  \captionsetup{labelformat=empty}
  \caption{FIG. 4(H): J1643$-$1224 posterior distributions with 68\%,90\%,99\% credible intervals for white noise, achromatic red noise, DMv and Sv for \textit{WRDS} model.}
\end{figure*}

\begin{figure*}
 \centering
 \vspace*{1.25in}
 \includegraphics[keepaspectratio=true,scale=0.27]{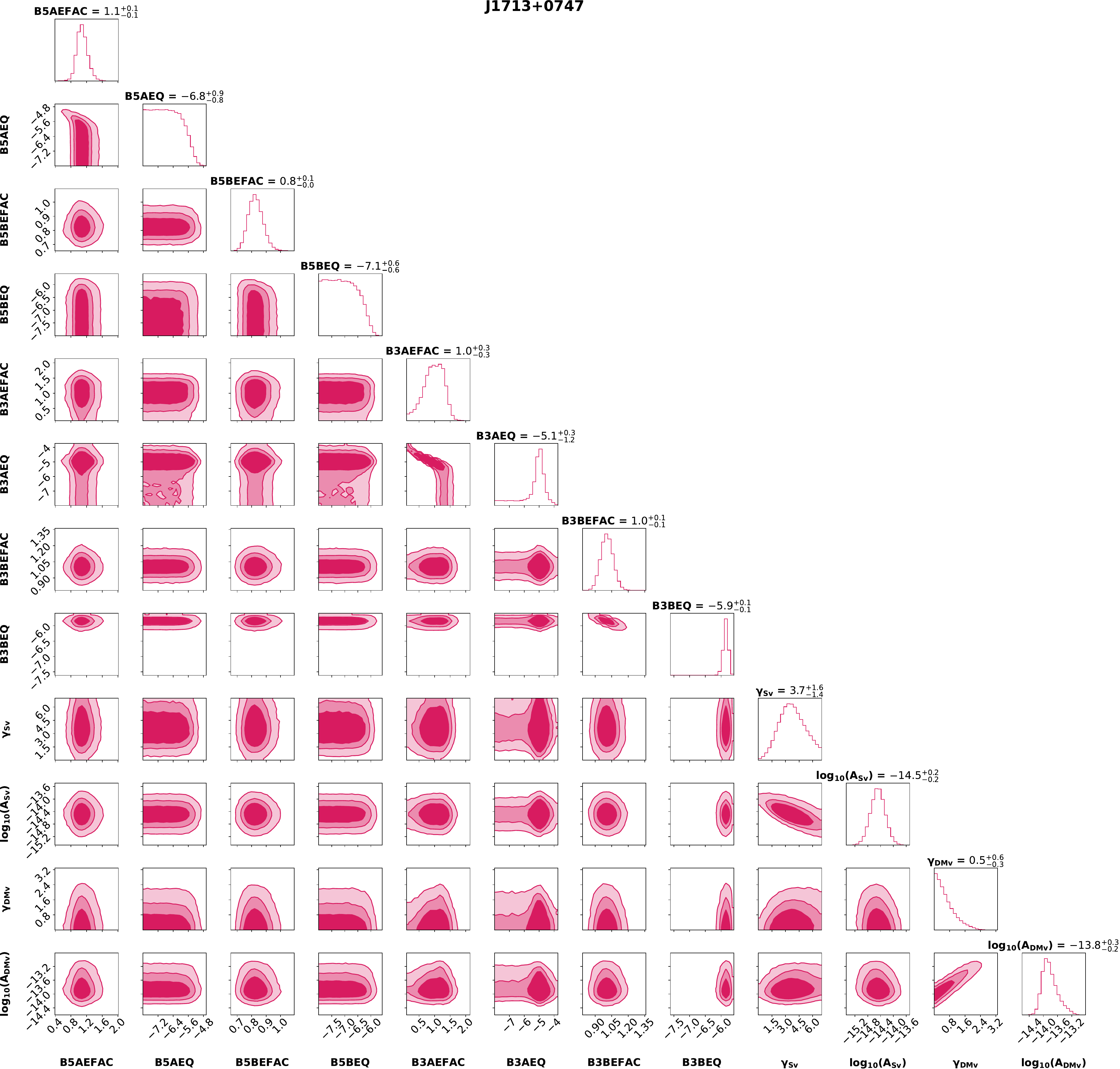}
   \captionsetup{labelformat=empty}
  \caption{FIG. 4(I): J1713+0747 posterior distributions with 68\%,90\%,99\% credible intervals for white noise, DMv and Sv for \textit{WDS} model.}
\end{figure*}

\begin{figure*}
 \centering
 \vspace*{1.25in}
 \includegraphics[keepaspectratio=true,scale=0.33]{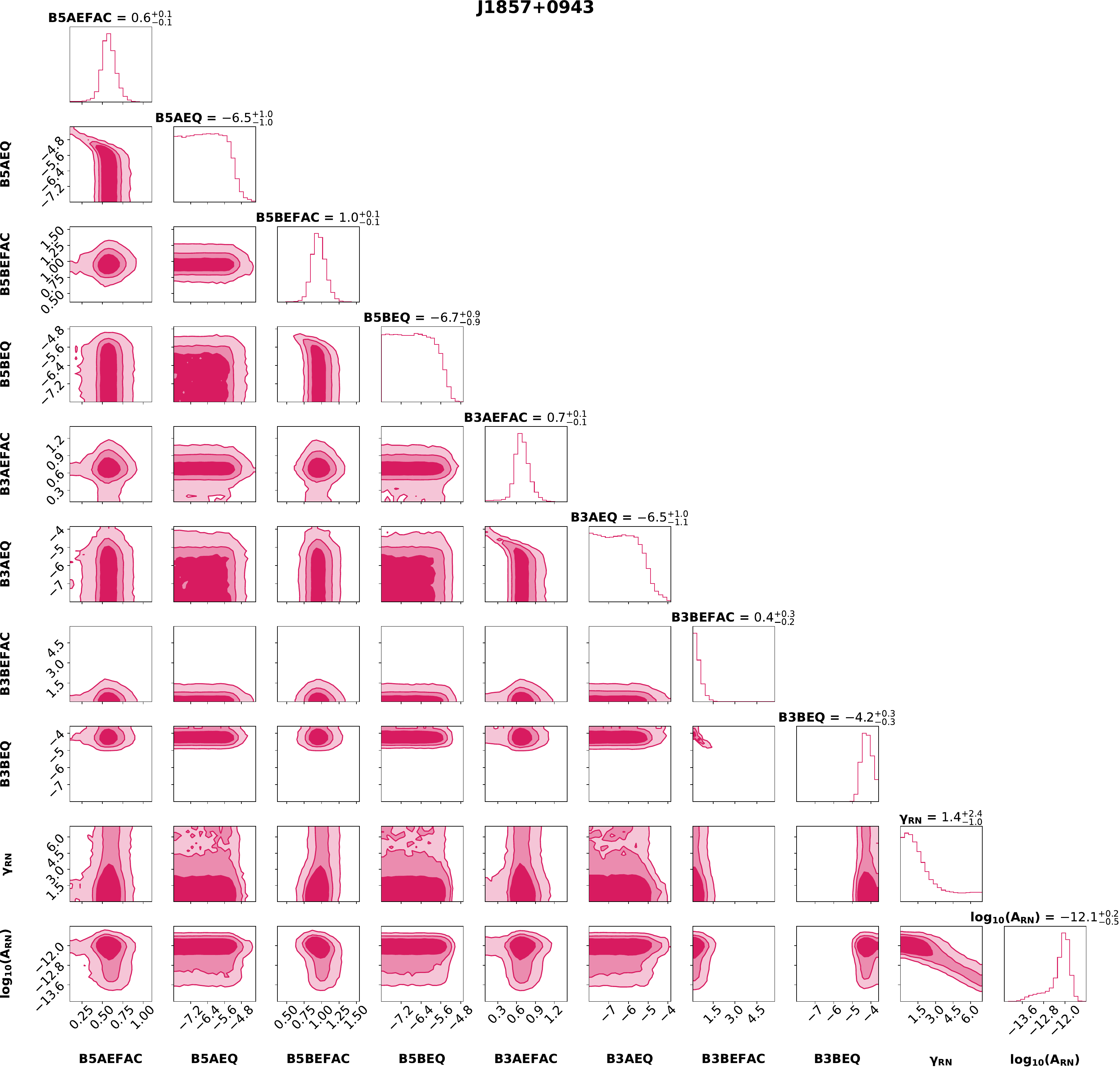}
   \captionsetup{labelformat=empty}
  \caption{FIG. 4(J): J1857+0943 posterior distributions with 68\%,90\%,99\% credible intervals for white noise and achromatic red noise for \textit{WR} model.}
\end{figure*}

\begin{figure*}
 \centering
 \vspace*{1.25in}
   \includegraphics[keepaspectratio=true,scale=0.28]{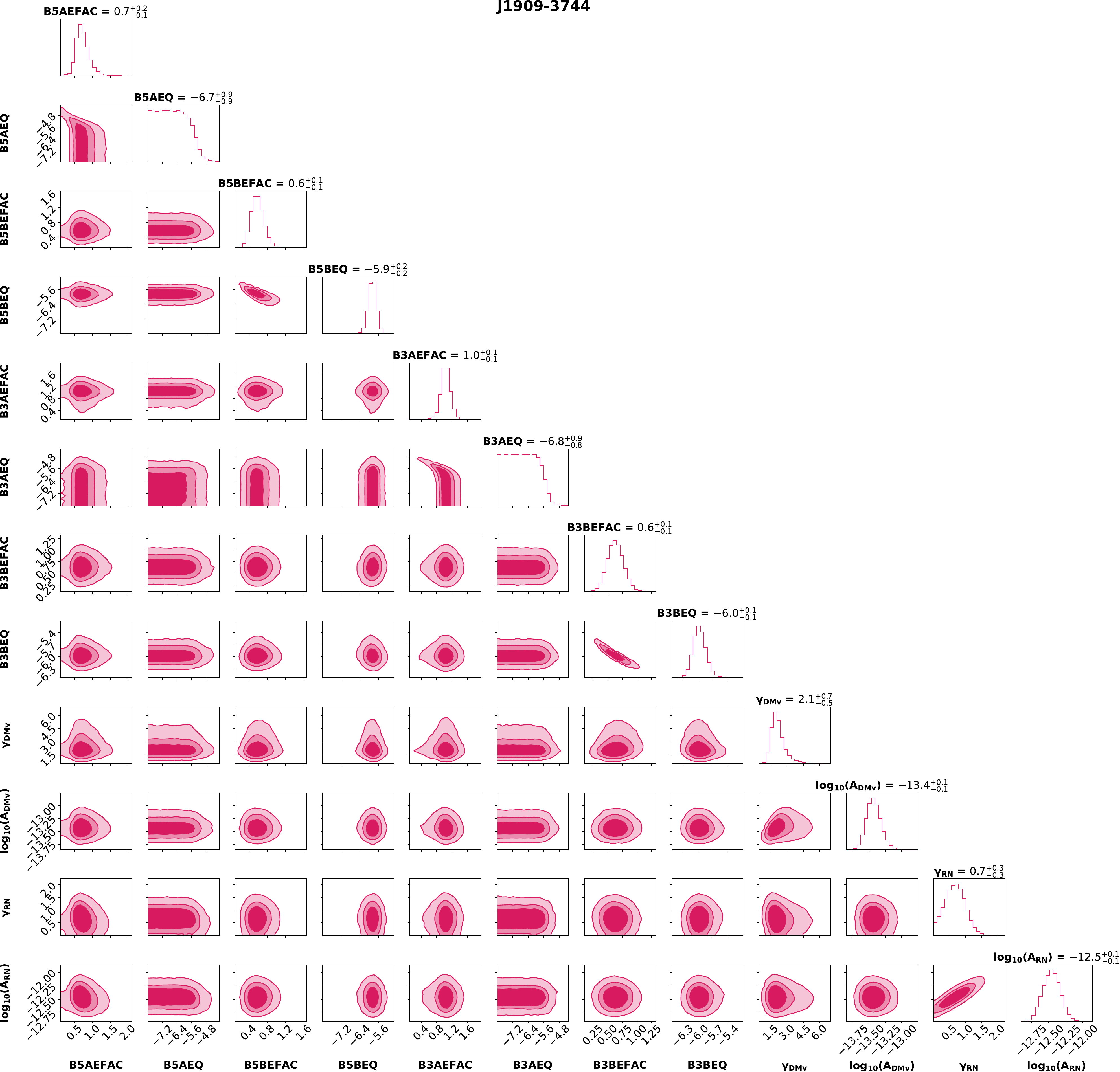}
  \captionsetup{labelformat=empty}
  \caption{FIG. 4(K): J1909$-$3744 posterior distributions with 68\%,90\%,99\% credible intervals for white noise, achromatic red noise and DMv for \textit{WRD} model.}
\end{figure*}

\begin{figure*}

 \centering
 \vspace*{1.25in}
 \includegraphics[keepaspectratio=true,scale=0.23]{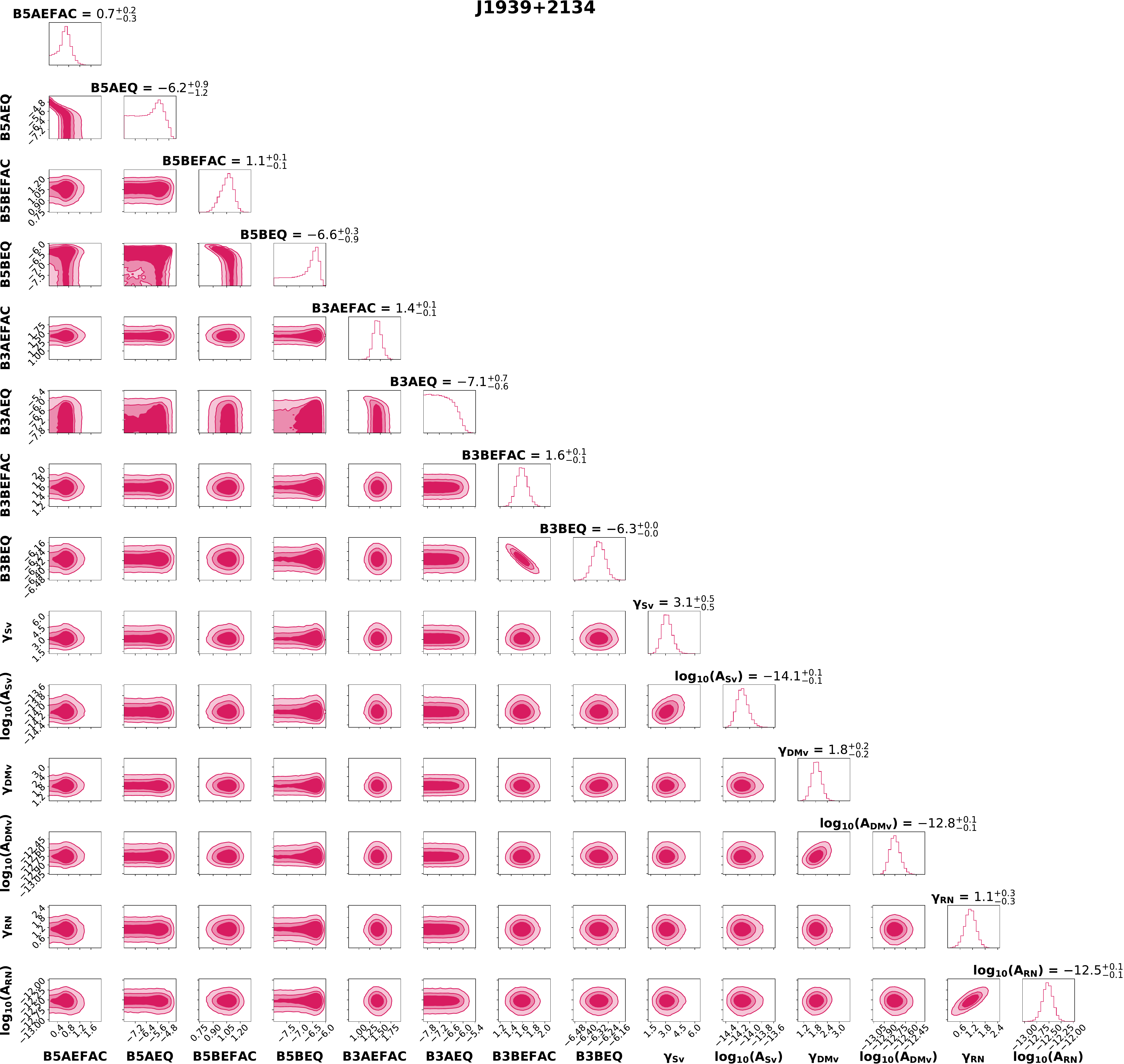}
   \captionsetup{labelformat=empty}
  \caption{FIG. 4(L): J1939+2134 posterior distributions with 68\%,90\%,99\% credible intervals for white noise, achromatic red noise, DMv and Sv for \textit{WRDS} model.}

\end{figure*}

\begin{figure*}
 \centering
 \vspace*{1in}
 \includegraphics[keepaspectratio=true,scale=0.4]{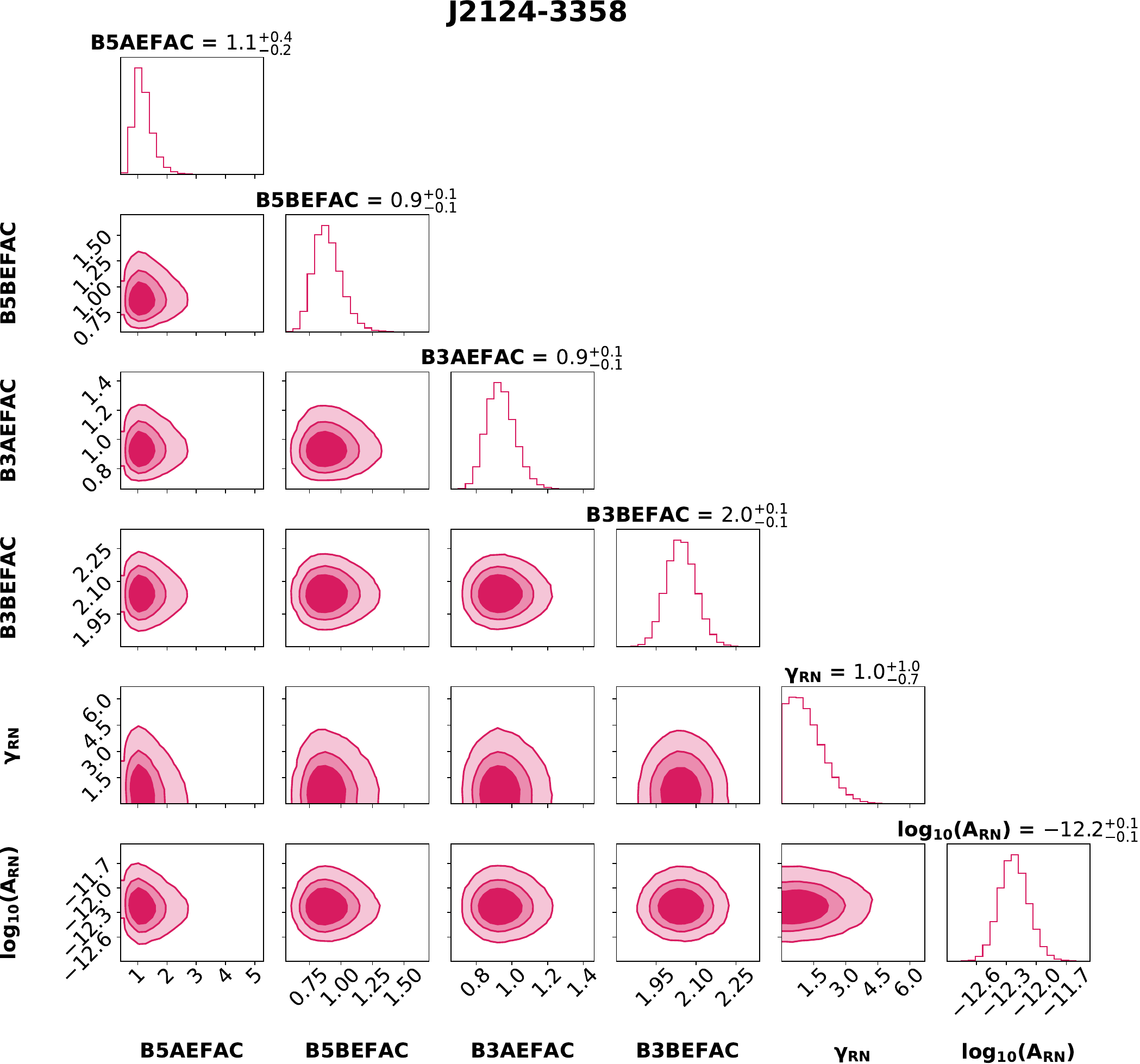}
   \captionsetup{labelformat=empty}
  \caption{FIG. 4(M): J2124$-$3358 posterior distributions with 68\%,90\%,99\% credible intervals for white noise and achromatic red noise for \textit{WR} model.}
\end{figure*}

\begin{figure*}
 \centering
 \vspace*{1.25in}
 \includegraphics[keepaspectratio=true,scale=0.23]{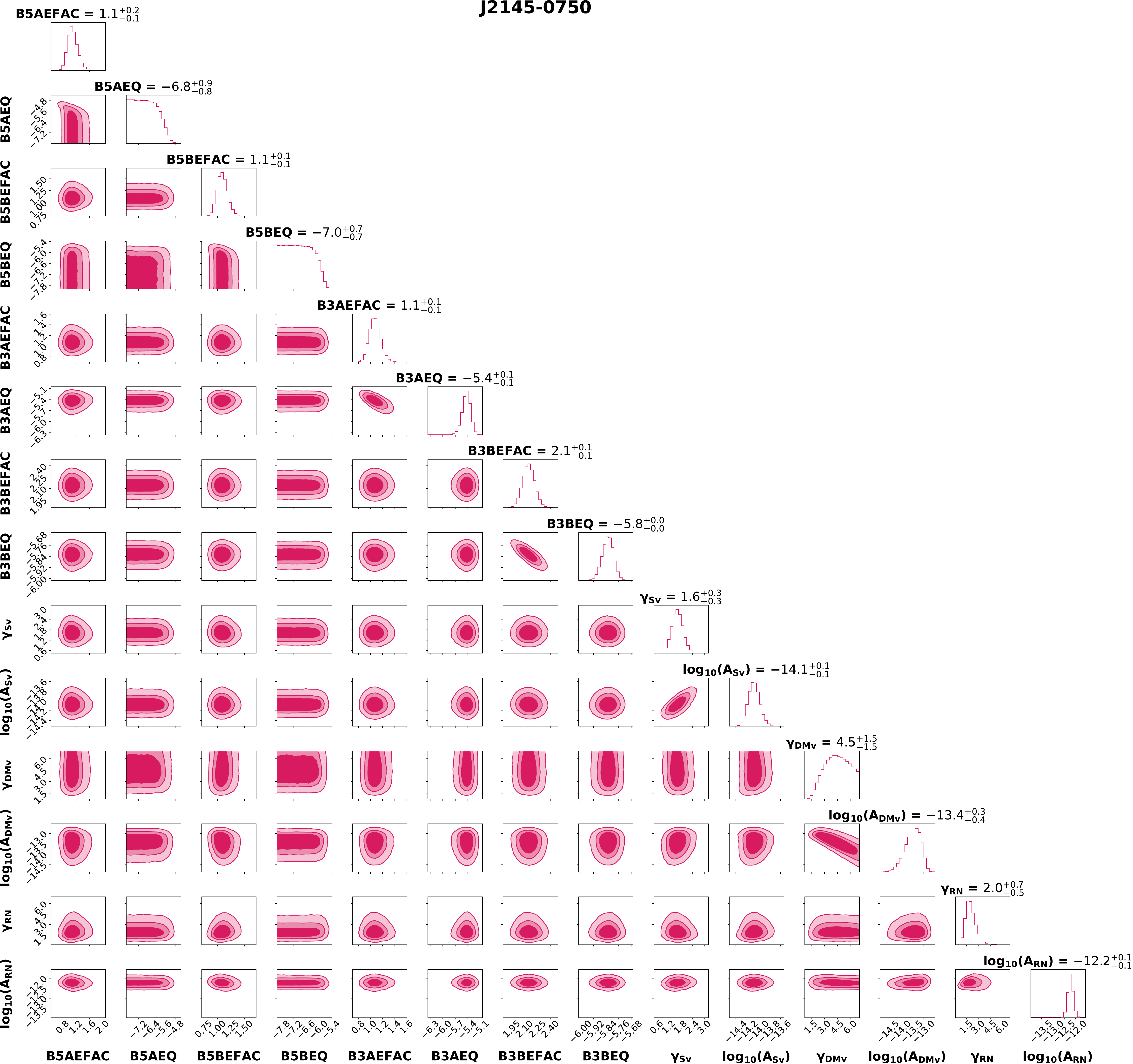}
   \captionsetup{labelformat=empty}
  \caption{FIG. 4(N): J2145$-$0750 posterior distributions with 68\%, 90\%, 99\% credible intervals for white noise, achromatic red noise, DMv and Sv for \textit{WRDS} model.}
\end{figure*}

\end{document}